\documentclass{article}

\usepackage{PRIMEarxiv}

\usepackage[utf8]{inputenc} % allow utf-8 input
\usepackage[T1]{fontenc}    % use 8-bit T1 fonts
\usepackage{hyperref}       % hyperlinks
\usepackage{url}            % simple URL typesetting
\usepackage{booktabs}       % professional-quality tables
\usepackage{amsfonts}       % blackboard math symbols
\usepackage{nicefrac}       % compact symbols for 1/2, etc.
\usepackage{microtype}      % microtypography
\usepackage{lipsum}
\usepackage{fancyhdr}       % header
\usepackage{graphicx}       % graphics

\usepackage{amsmath}
\usepackage{xspace}
\usepackage{hologo}
\usepackage{longtable}
\usepackage{makecell}
\usepackage[most]{tcolorbox}
\usepackage{tabularx,booktabs}
\usepackage{ltablex}
\usepackage{colortbl}
\usepackage{array}
\usepackage{geometry} 
\usepackage{tikz} 
\usepackage{ragged2e}
\newcolumntype{P}[1]{>{\RaggedRight\hspace{0pt}}p{#1}}

\newtheorem{definition}{Definition}

\newcommand{\dbtfull}{diversity-based testing\xspace}
\newcommand{\Dbtfull}{Diversity-based testing\xspace}

\newcommand{\dbttfull}{\dbtfull technique\xspace}
\newcommand{\dbttsfull}{\dbtfull techniques\xspace}

\newcommand{\dbt}{DBT\xspace}

\newcommand{\Dbttxfull}{\Dbtfull (\dbt) technique\xspace}

\newcommand{\etal}{et~al.\xspace}
\newcommand{\ie}{i.e.\xspace}

\newcommand{\artefact}{artefact\xspace}
\newcommand{\artefacts}{artefacts\xspace}
\newcommand{\artefactCAPS}{ARTEFACT\xspace}

\newcommand{\tdg}{test data generation\xspace}
\newcommand{\Ctdg}{Test data generation\xspace}
\newcommand{\tcp}{test case prioritisation\xspace}
\newcommand{\Ctcp}{Test case prioritisation\xspace}
\newcommand{\tcs}{test case selection\xspace}
\newcommand{\Ctcs}{Test case selection\xspace}
\newcommand{\tsr}{test suite reduction\xspace}
\newcommand{\Ctsr}{Test suite reduction\xspace}
\newcommand{\tse}{test suite quality evaluation\xspace}
\newcommand{\Ctse}{Test suite quality evaluation\xspace}

\newcommand{\numsearchenginepapers}{2,837\xspace}

\newcommand{\numscreened}{2,593\xspace}
\newcommand{\numnoncore}{62\xspace}
\newcommand{\numart}{58\xspace}
\newcommand{\numsnowballed}{20\xspace}
\newcommand{\numdbtinclart}{182\xspace}
\newcommand{\numdbtexclart}{144\xspace}
\newcommand{\numpapers}{\numdbtexclart\xspace}

\newcommand{\numtdg}{55\xspace}
\newcommand{\numtcp}{32\xspace}
\newcommand{\numtcs}{21\xspace}
\newcommand{\numtse}{11\xspace}
\newcommand{\numtsr}{10\xspace}
\newcommand{\numtfl}{9\xspace}

\newcommand{\numartefacts}{24\xspace}
\newcommand{\nummetrics}{70\xspace}
\newcommand{\numproblems}{11\xspace}

\newcommand{\code}[1]{
  \texttt{#1}
}

\newcommand{\lessthantenspace}{\phantom{1}}
\newcommand{\nonequalrankspace}{\phantom{=}}
\newcommand{\rankspace}{\hspace{.5em}}

\newcommand{\rqtrends}{RQ1: ORIGINS, TRENDS, AND PROMINENT PAPERS}
\newcommand{\rqtrendsexpanded}{What were the first papers in \dbt? What are the publication trends, and most prominent venues and authors in the
\dbt field?}

\newcommand{\rqproblems}{RQ4: PROBLEMS}
\newcommand{\rqproblemsexpanded}{What are the software testing problems to which \dbt techniques have been applied?}

\newcommand{\rqmetrics}{RQ2: SIMILARITY METRICS}
\newcommand{\rqmetricsexpanded}{What similarity metrics have been used in the literature? Which ones have been used
the most, and why?}

\newcommand{\rqartefacts}{RQ3: \artefactCAPS{S}}
\newcommand{\rqartefactsexpanded}{What \artefacts have been used (e.g., test inputs, software requirements,
abstract models) as the basis for applying similarity metrics to generate
diverse test suites?}

\newcommand{\rqsubjects}{RQ5: SUBJECT DOMAINS}
\newcommand{\rqsubjectsexpanded}{To what subject domains have \dbt techniques been applied?}

\newcommand{\rqtools}{RQ6: TOOLS}
\newcommand{\rqtoolsexpanded}{What \dbt tools have been developed by researchers?}

\title{A Survey of the Metrics, Uses, and Subjects of Diversity-Based Techniques in Software Testing
}

\author{
  Islam T. Elgendy, \\
  University of Sheffield \\
  UK\\
   \And
  Robert M. Hierons  \\
  University of Sheffield \\
  UK\\
   \And
  Phil McMinn  \\
  University of Sheffield \\
  UK\\
}

\begin{document}
\maketitle

\begin{abstract}
There has been a significant amount of interest regarding the use of
    \dbttsfull in software testing over the past two decades. 
    \Dbttxfull uses similarity metrics to leverage the dissimilarity between
    software artefacts --- such as requirements, abstract models, program
    structures, or inputs --- in order to address a software testing problem.
    \dbt techniques have been used to assist in finding 
    solutions to several different types of problems including
    generating test cases, prioritising them, and reducing very large test
    suites.
    This paper is a systematic survey of \dbt techniques that summarises the key
    aspects and trends of \numpapers~papers that report the use of
    \nummetrics~different similarity metrics with \numartefacts~different types
    of software artefacts, which have been used by researchers to tackle
    \numproblems~different types of software testing problems. 
    We further present an analysis of the recent trends in \dbt techniques
    and review the different application domains to which the techniques have
    been applied, giving an overview of the tools developed by researchers 
    to do so. Finally, the paper identifies some \dbt challenges that
    are potential topics for future work.
\end{abstract}

% keywords can be removed

\section{Introduction}
\label{sec:introduction}

One of the key challenges associated with finding software failures is the
intractability of testing each possible behaviour of a piece of software,
since the number of behaviours a non-trivial software system is capable of
exhibiting tends to be very large, if not infinite.
For this reason, software testing researchers have devised many techniques
to help a tester decide what to test. Either explicitly or implicitly, one
can argue that all software testing techniques embed the notion of {\it
        diversity}. That is, since it is not possible to know where or how software
failures will happen a priori, it is neither effective nor efficient to
keep looking in the same or similar places --- i.e., to have tests that
are overly-concerned with certain aspects of the software to the neglect of others.
Different code coverage metrics, for example, dictate all
statements or all branches of the software are executed, ensuring the
diversity of code structures exercised by tests~\cite{Ammann2016,Myers2011}.
Model-based techniques ensure diversity of tests by ensuring distinct
abstract states of the system are tested along with the transitions between
them~\cite{Ammann2016,Utting2010}.
Even boundary-value analysis~\cite{Ammann2016,Hamlet1990}, which advocates
testing potentially similar inputs to encourage exploration around the
thresholds of predicates in a program, does so in the knowledge that small
mistakes in the conditions of predicates (for example, the use of ``\code{>}''
instead of ``\code{>=}'' etc.) can trigger unexpected, divergent software
behaviours.

However, the problem facing the tester is what to do when even the test
requirements generated by these techniques result in too many tests. For these
reasons, software testing researchers of late have become increasingly
interested in a class of techniques that we refer to in this survey as
    {\it diversity-based testing (DBT) techniques}.

At a high level, a \dbt technique selects tests based on their
    {\it dissimilarity} from others already used or that exist in the test suite.
A \dbt technique measures dissimilarity using a {\it similarity metric} based on some aspect
or set of {\it \artefacts} related to the test; for example,
the inputs it uses (e.g.,~\cite{Bueno2007,Bueno2008,Feldt2016,Shahbazi2014,Shahbazi2015B,Shahbazi2018}),
the outputs it obtains given its inputs (e.g.,~\cite{Alshahwan2012,Alshahwan2014,Benito2022,Matinnejad2015A,Matinnejad2016,Matinnejad2019}),
the program structures it executes (e.g.,~\cite{Beena2014,Wang2015,You2013,Zhang2019});
or on the basis of more abstract artefacts, such as the requirements the test exercises (e.g.,~\cite{Arafeen2013,Masuda2021}),
or parts of a model representation that it covers (e.g.,~\cite{Cartaxo2007,Cartaxo2009,Coutinho2016,Hemmati2010A,Hemmati2010B,Hemmati2010C,Hemmati2013}).
Examples of similarity metrics used by researchers include, for example, the
Euclidean distance between two numerical inputs (e.g.,~\cite{Bueno2007,
    Bueno2008}) or signals of a Simulink model
(e.g.,~\cite{Matinnejad2016,Matinnejad2019}), or the Edit distance between two
regular expressions (e.g.,~\cite{Cao2022}),
or the Jaccard distance between two sets of selected product line features
(e.g.,~\cite{Abd2019,Henard2013}).
More formally, we define a \dbttfull as follows:

\begin{definition} \label{def:DBT-definition}
    A {\bf \dbttfull} $D$ is a function that takes as input a
    software system $S$, a set of \artefacts $A$ representing some aspect of the system
    (such as all of its possible inputs or outputs, its requirements etc.),
    and
    a similarity metric $\mathit{sim}$ that may be applied to two or more elements of $A$
    to measure their similarity. $D$ uses $\mathit{sim}$ in conjunction with $A$ to
    output a set of tests $T$ for testing $S$.
\end{definition}

\dbt techniques have been used to tackle many different problems in software
testing; for example,
test data generation (e.g.,~\cite{Alshahwan2014,Henard2014,Matinnejad2016,Matinnejad2019}),
test case prioritisation (e.g.,~\cite{Arafeen2013,Ledru2012,Matinnejad2019,Miranda2018,Yoo2009}),
test suite reduction (e.g.,~\cite{Chetouane2020,Coutinho2016,Coviello2018B,Cruciani2019}),
and test suite quality evaluation (e.g.,
\cite{Cao2013,Neto2018,Shi2015,Xie2006B}).
They have also been applied to a wide range of subject domains, including
deep learning models (e.g.,~\cite{Aghababaeyan2023,Mosin2022}),
compilers (e.g.,~\cite{Chen2019C,Tang2022}),
web applications (e.g.,~\cite{Alshahwan2014,Biagiola2019,Marchetto2009}),
as well as general code libraries (e.g.,~\cite{Fang2014,Feldt2016,Noor2015,Miranda2018}).
The advantages of \dbt techniques have been reported in several studies about
test case analysis, generation, and
optimization (e.g.,~\cite{Feldt2008,Hemmati2015,Noor2015,Feldt2016,Coutinho2016,Henard2016,Miranda2018}).
A common finding is that on the whole, the more diverse a test suite, the more
likely it is to trigger failures and thereby detect faults (e.g.,
\cite{Benito2022,Biagiola2019,Shahbazi2015B,Feldt2016}).
Several researchers have also shown that reordering test case execution order,
based on diversity, or removing similar test cases is more likely to result in
better test suites in terms of fault-finding than those re-ordered or reduced
using other metrics, for example
coverage~\cite{Arafeen2013,Hemmati2015,Cartaxo2009}.
Finally, \dbt techniques have been applied to different types of
testing at different levels, including unit testing (e.g.,~\cite{Feldt2008,Noor2015}),
integration testing (e.g.,~\cite{De2018}) and
system testing (e.g.,~\cite{De2016,Tahvili2018}).

Given the wide range of types of \dbt techniques, and the different problems and
application domains to which they have been applied, we conducted a systematic
literature review of the field, collecting results from multiple search engines
including IEEE Xplore, the ACM Digital Library, Scopus, and SpringerLink.
\dbt appears to have originated out of the work on Adaptive Random Testing
(ART). ART was recently surveyed by Huang et al.~\cite{Huang2019}, who summarise
ART as aiming to
{\it ``enhance RT's [Random Testing's] failure-detection ability by more evenly
spreading the test cases over the input domain''.}
Some ART techniques satisfy our definition of a \dbt technique because they use a
(dis)similarity metric that measures the ``distance'' between test cases to
accomplish this spread. Others, however, do {\it not} fit our definition, because
achieve spreading by partitioning up the input domain instead --- that is, {\it without}
the use of a similarity metric (e.g.,~\cite{Chan2002,Chen2004B,Mayer2006}),
and potentially with the help of a human expert instead (e.g.,~\cite{Sabor2015}).

The purpose of this survey is to explore how \dbt techniques in general have
developed and broadened since its beginnings with ART. We do not, therefore,
include ART itself as part of our review, referring the reader to the
comprehensive survey of 140 ART papers by Huang \etal~\cite{Huang2019} instead.
ART differs significantly in ethos from \dbt. ART has traditionally focused on
one problem --- test input generation; and therefore one type of artefact --- a
program's input domain; while the definition of a similarity metric is not a
necessary component of the technique. In contrast, \dbt is about tackling any
problem in software testing where intractability is a key limiting
characteristic, for which researchers have utilised many different software
artefacts to assist in determining solutions, and have employed a wide variety
of similarity metrics to measure their diversity in order to do so.
In total, our survey finds that researchers have tackled \numproblems~different
types of testing problems using \dbt techniques, applying \nummetrics~different
metrics to \numartefacts~different types of software \artefacts.

This paper, the first 
to define and survey the wider area of \dbt techniques, summarises the work of
\numpapers~articles, which we list in a \hologo{BibTeX} file and make
available for other researchers to use via a
public GitHub repository~\cite{Bibliography2022}. Our survey makes the following
contributions:

\begin{enumerate}
    \item The first definition what constitutes a {\it \dbtfull (\dbt) technique} (Definition~\ref{def:DBT-definition}).
    \item An analysis of the trends for \dbt techniques and the prominent venues,
          papers, and authors in the research area (Section \ref{sec:rqtrends}).
    \item A compilation of the different metrics used to measure the similarity
          between testing \artefact{s} (Section \ref{sec:rqmetrics}).
    \item A summary of the \artefacts used as a basis for \dbt techniques (Section
          \ref{sec:rqartefacts}).
    \item A summary of the different software testing problems where diversity has
          been applied (Section \ref{sec:rqproblems}).
    \item An overview of the different subject domains in which diversity-based
          testing has been utilised (Section \ref{sec:rqsubjects}), and the tools available for
          addressing them (Section \ref{sec:rqtools}).
    \item A discussion of possible future research directions for \dbt techniques
          (Section \ref{sec:futre-research-directions}).
\end{enumerate}

We begin in Section \ref{sec:methodology} by presenting the survey's scope and
research questions, and the methodology we used to collect papers.

\section{Methodology}
\label{sec:methodology}

Our systematic survey aims to answer the following research questions:

\smallskip \noindent \textbf{\rqtrends} (Section \ref{sec:rqtrends}).
\rqtrendsexpanded

\smallskip \noindent \textbf{\rqmetrics} (Section \ref{sec:rqmetrics}).
\rqmetricsexpanded

\smallskip \noindent \textbf{\rqartefacts} (Section \ref{sec:rqartefacts}).
\rqartefactsexpanded

\smallskip \noindent \textbf{\rqproblems} (Section \ref{sec:rqproblems}).
\rqproblemsexpanded

\smallskip \noindent \textbf{\rqsubjects} (Section \ref{sec:rqsubjects}).
\rqsubjectsexpanded

\smallskip \noindent \textbf{\rqtools} (Section \ref{sec:rqtools}).
\rqtoolsexpanded

\subsection{Collection Approach}
\label{sec:collection-approach}

We collected papers for our survey on the 12th of April, 2023 using searches on
IEEE Xplore\footnote{\url{https://ieeexplore.ieee.org/Xplore}}, the ACM Digital
Library\footnote{\url{https://dl.acm.org}}, Scopus\footnote{\url{https://www.scopus.com/}}, and
SpringerLink\footnote{\url{https://link.springer.com}}.
The search query we used for IEEE Xplore, Scopus, and ACM was \{``{\tt software
test*}'' AND (``{\tt divers*}'' OR ``{\tt similar*}'' in title or keywords)\}, where
``{\tt *}'' is a wildcard matching any string. 
This means looking for the text ``{\tt software test*}'' anywhere in the paper and a
variation of the terms ``{\tt divers*}'' (i.e., ``diversity'', ``diversify'', or
``diversification'', etc.) or ``{\tt similar*}'' (i.e., ``similar'' and ``similarity'') in
the title or the keywords.
We restricted the latter terms to the title or keywords due to their
commonality, and hence ability to match a very large number of papers that would
(a) be intractable to screen manually, and (b) have, on average, a low
probability of being related to \dbt techniques. For example, a paper may claim
to use a ``diverse'' set of subjects for an empirical study and would be
returned in our search results without this constraint.
SpringerLink did not allow us to search for certain terms in specific places
like the title and keywords, so we used the closest search query possible, which
was \{``{\tt software test*}'' AND (``{\tt similar*}'' OR ``{\tt divers*}'')\}
anywhere in the paper.
We did not limit our searches to any particular year range.

\begin{figure}[t]
    \centering
    \includegraphics[width=\columnwidth]{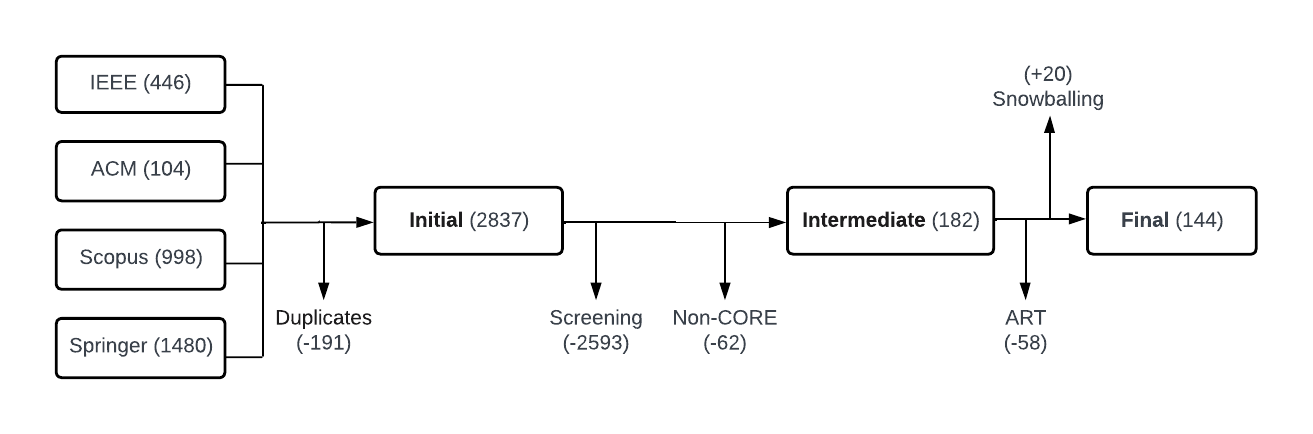}
    \vspace{-3em}
    \caption{
        \label{fig:collection-approach}
        Overview of our paper collection methodology, with the number of pages
        collected at each stage
    }
\end{figure}

Figure \ref{fig:collection-approach} gives an overview of the collection
approach and the number of papers at each stage, reporting the number of papers
retrieved from each search engine. The number of papers returned by SpringerLink
was the most, due to the aforementioned difficulty in restricting the search
terms. After we removed duplicate papers returned by more than one search
engine, the set of papers collated numbered \numsearchenginepapers~in size, that we
label the {\bf Initial} set of papers in the figure.

We then applied a screening process to this set of papers. The first and last
authors manually reviewed the contents of each paper to check that its contents
were on the subject of \dbt techniques. That is, the paper proposed, discussed,
or evaluated at least one approach fitting Definition~\ref{def:DBT-definition},
and thereby fell in the scope of this survey.
Each author made a separate judgment to include or exclude each paper, reading
paper titles and abstract, and further studying its contents in detail if
required. Following this, the two authors reviewed their decisions together,
discussing each paper that they had judged differently, coming to an agreement
to include or exclude each individual paper.
We removed \numscreened~papers from further consideration as part of this
process. These included papers that were not focused on software testing or did
not focus on software testing techniques; for example, being instead concerned with
diversity in the sense of inclusion or representation of different types of
people in a software engineering team. We further removed papers discussing software testing
techniques purporting to be ``diverse'', but did not satisfy our definition of a
\dbt given in Definition~\ref{def:DBT-definition}; for example, the technique
did not use a similarity metric, or, we could not find whether the
technique made use of a similarity metric in the text of the paper.
Following these decisions, we then removed papers that did not appear in
journals or conferences ranked ``C'' or above by the CORE Rankings
Portal~\cite{CORE2022}, with the aim of imposing a level of quality of the
papers considered by our survey. We included papers at any workshops with a
history of co-locating with conferences ranked as ``C'' or above, but excluded
papers appearing in doctoral symposia.
We removed a further \numnoncore~papers because they did not fulfil our CORE
requirements. This left \numdbtinclart~papers, which we mark as the {\bf
        Intermediate} set of papers in Figure \ref{fig:collection-approach}.

We then removed a further \numart~papers that related to Adaptive Random Testing
(ART), which is not a focus of this survey as discussed in
Section~\ref{sec:introduction}. We refer the interested reader techniques to the
existing comprehensive survey of 140 papers on ART by
Huang~\etal~\cite{Huang2019}.
Finally, in case any \dbt papers were missed by our original searches, we
applied backward snowballing~\cite{Wohlin2014}, whereupon we manually studied
the references of each of the papers in the intermediate set of papers. If a
reference involved a \dbt technique, was in a CORE ``C''-ranked venue or better,
and was not an ART paper  --- i.e., it met all of our aforementioned criteria
--- we added it to the set of papers featured in this survey. We found
\numsnowballed~additional papers via this method which we added to form the
collection of papers marked as the {\bf Final} set of papers in
Figure~\ref{fig:collection-approach}.
Each paper in the final set was read by the first author, who collated
the statistics and information needed to answer our research questions as
answered in the following sections.

\subsection{Threats to Validity}
\label{sec:threats-to-validity}

We hereby discuss validity threats to our methodology, and hence its results,
and the steps we took to mitigate these threats.

\begin{enumerate}
    \item {\bf Some relevant papers might not have been included in the searches.}
          We mitigated this risk by using a variety of search engines, including IEEE
          Xplore, ACM, Scopus, and SpringerLink with the aim of finding all papers
          relevant to our survey. Also, we picked our search terms carefully to include
          different variations of the terms ``diverse'' or ``similar'' to include terms
          like ``diversity'', ``diversify'', ``diversification'', ``similarity'', and so
          on. Finally, we used backward snowballing to find papers that the search engines
          may not have indexed.
    \item {\bf The manual screening process may have eliminated some relevant
          papers.} We mitigated this risk by having two authors do the screening
          process separately and independently, then coming together to have a final
          decision on the papers where they had different opinions.
    \item {\bf Some papers may be of low quality and their results might be
          questionable}. We attempted to mitigate this risk by including papers that are
          only published in peer-reviewed journals and conferences ranked ``C'' and above
          by the CORE ranking portal.
\end{enumerate}

While we acknowledge that it is possible that we may have missed some \dbt
papers, we are confident that we have included the majority of relevant
articles, and that our survey provides an accurate account of the key trends and
the state of the art of \dbt.
To support the replication of our study and as a useful resource for other
researchers, we have made the final set of papers we collected available in a
\hologo{BibTeX} file in a publicly-available GitHub
repository~\cite{Bibliography2022}.

\section{\rqtrends}
\subsection*{\textbf{\rqtrendsexpanded}}
\label{sec:rqtrends}

\begin{figure}[t]
    \centering
    \includegraphics[width=\columnwidth]{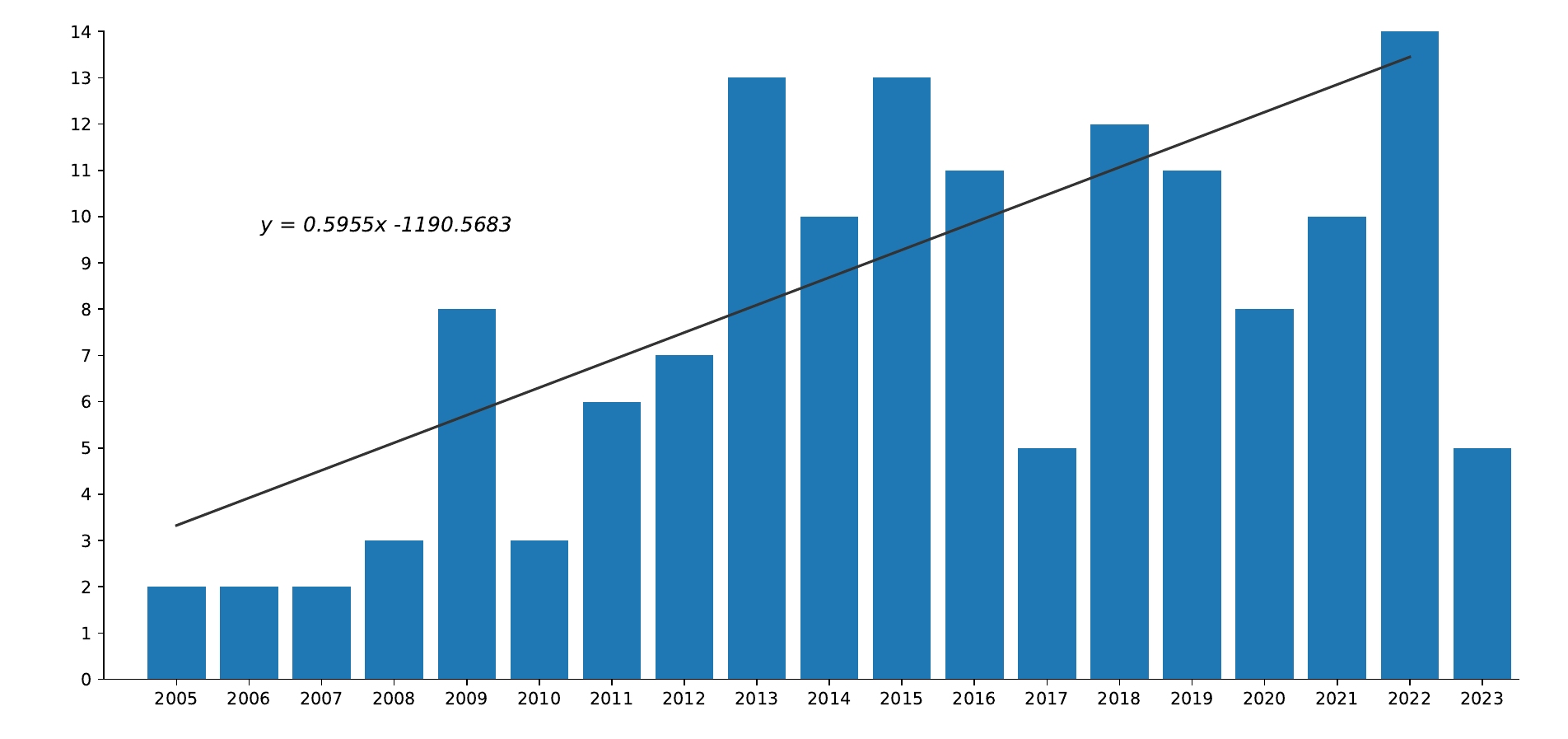}
    \vspace{-3em}
    \caption{
        \label{fig:papers-per-year}
        Papers published per year for the final set of papers collected using
        our methodology. Since the data was collected partway through 2023, the
        bar for that year is incomplete and as such we plot the trendline
        through 2005--2022 only. 
    }
\end{figure}

In this section we study the origins of \dbt, studying the first papers. We also
present and study trends in the number of \dbt publications per year. Finally,
we analyse the most cited \dbt papers, the most prolific \dbt authors, and the
venues with the highest frequency of publishing \dbt works.

\subsection{First \dbt Papers}
\label{sec:origins}

The first \dbt papers in our final collected set of papers appeared in 2005,
are two papers by Xie~\etal~\cite{Xie2005} and Hao~\etal~\cite{Hao2005}.
These first two papers are notable in that they apply the principles of \dbt to
assist two existing approaches, namely search-based test data generation and
fault localisation respectively, rather than being standalone \dbt techniques in
their own right.
Xie~\etal's \dbt work was part of a search-based strategy for maintaining
population diversity so as to avoid premature convergence of the search on
suboptimal test cases~\cite{Xie2005}. The approach monitors the similarity
between the individuals in the population, and once it exceeds a certain
threshold, their technique significantly increases the mutation probability to
increase population diversity. We discuss this paper further in
Section~\ref{subsec:population-diversity}
(p.\pageref{subsec:population-diversity}).
Hao~\etal's work~\cite{Hao2005} was a short paper proposing to detect similarity
in test cases to remove bias from suspiciousness scores for fault localisation
techniques. We discuss this paper in more detail in
Section~\ref{subsec:fault-localization}
(p.\pageref{subsec:fault-localization}).

However, peering into the intermediate set of papers collected by our
methodology (Section~\ref{sec:methodology}) shows that authors started writing
about ART techniques in terms of diversity and similarity in the year
before these two papers appeared, in 2004~\cite{Chen2004, Chan2004C, Chen2004B, Chen2004D}.
ART itself, however, can be traced back earlier to 2001 by a paper by
Chen~\etal~\cite{Chen2001}, as part of a review of proportional
sampling strategies. Not all techniques discussed require a similarity
metric.
The first papers to evaluate ART using a similarity metric --- Euclidean distance with
numerical inputs --- were various formulations by Chen~\etal~\cite{Chen2004,
  Chen2004C, Chen2004D} in 2004, which were built on partition strategies.

Other than to track the lineage of \dbt techniques in this section, however, ART
is not among the \dbt approaches considered in the remainder of this paper, for
the reasons discussed in
Sections~\ref{sec:introduction}~and~\ref{sec:methodology}. We refer the
interested reader to the comprehensive survey of ART by
Huang~\etal~\cite{Huang2019}. The remainder of this paper focuses exclusively on
non-ART approaches to \dbtfull.

\subsection{Number of Papers Year-By-Year}
\label{sec:collection-results}

Figure~\ref{fig:papers-per-year} shows the number of \dbt publications per year.
(Since the literature search was done in April of 2023 --- see our methodology in
Section~\ref{sec:methodology} for more details --- the bar for 2023 is
incomplete, and the trend line is only fitted for the years 2005--2022.)
The figure shows a trend of an increasing number of papers focussed on \dbt
techniques year on year.
The year with the highest number of publications was the most recent complete
year --- 2022. All in all, these data show that the field of \dbt is growing.

\begin{table}[t]
    \centering
    \caption{
        \label{table:papers-per-venue}
        Most frequent publishing venues
    }
    % \vspace{2em}
    %
    \begin{tabular}{@{}lr@{}}
        \toprule
            {\bf Rank / Venue} & {\bf \# Papers}
            \\

        \midrule

        \nonequalrankspace 1 \rankspace Transactions on Software Engineering (TSE) & 12  \\
                  =2 \rankspace International Conference on Software Testing, Verification and Validation (ICST) & 10 \\
                  =2 \rankspace International Conference on Automated Software Engineering (ASE) & 10 \\
\nonequalrankspace 4 \rankspace Information and Software Technology (IST) & 9  \\
\nonequalrankspace 5 \rankspace International Symposium on Search Based Software Engineering (SSBSE) & 8 \\
                  =6 \rankspace International Symposium on Software Testing and Analysis (ISSTA) & 6\\
                  =6 \rankspace International Conference on Software Engineering (ICSE) & 6 \\
\nonequalrankspace 8 \rankspace International Workshop on Automation of Software Test (AST) & 5 \\
                  =9 \rankspace Software Quality Journal (SQJ) & 4 \\
                  =9 \rankspace International Workshop on Mutation Analysis (MUTATION) & 4 \\

        \bottomrule
    \end{tabular}
\end{table}
\begin{table}[t]
    \centering
    \caption{
        \label{table:prominent-papers}
        Papers with the highest numbers of citations
    }
    %
    % \vspace{-1em}
    % 
    \begin{tabular}{@{}l@{\hspace{.5em}}P{6cm}rP{1.45cm}rr@{}}
        \toprule

        {\bf Rank / Authors} & {\bf Title} & {\bf Year} & \bf{\makecell{Section\\(Page)}} & {\bf \# Cites} & {\bf \makecell{Mean cites\\per year}}  
        \\

        \midrule

        % Bypassing the combinatorial explosion: Using similarity to generate and prioritise t-wise test configurations for software product lines
\lessthantenspace 1 \rankspace Henard \etal~\cite{Henard2014} & Bypassing the combinatorial explosion: Using similarity to generate and prioritise t-wise test configurations for software product lines  & 2014 & \ref{subsec:spl-configuration-diversity} (p.\pageref{subsec:spl-configuration-diversity}) & 226 & 22.6 \\
\arrayrulecolor{lightgray}\hline
% Achieving Scalable Model-Based Testing through Test Case Diversity
\lessthantenspace 2 \rankspace Hemmati \etal~\cite{Hemmati2013} & Achieving Scalable Model-Based Testing through Test Case Diversity   & 2013 & \ref{subsubsec:mbt-selection} (p.\pageref{pap:hemmati2013}) & 245 & 22.3 \\
\hline
% Test case prioritisation using requirements-based clustering 
\lessthantenspace 3 \rankspace Arafeen \etal~\cite{Arafeen2013} & Test case prioritisation using requirements-based clustering  & 2013
& \ref{subsubsec:clustering-prioritisation} (p.\pageref{subsubsec:clustering-prioritisation}), \ref{subsec:requirements-diversity} (p.\pageref{subsec:requirements-diversity}) & 189 & 17.2\\
\hline
% Clustering test cases to achieve effective and scalable prioritisation incorporating expert knowledge 
\lessthantenspace 4 \rankspace Yoo \etal~\cite{Yoo2009} & Clustering test cases to achieve effective and scalable prioritisation incorporating expert knowledge  & 2009
& \ref{subsubsec:clustering-prioritisation} (p.\pageref{subsubsec:clustering-prioritisation}), \ref{subsec:execution-diversity} (p.\pageref{subsec:execution-diversity}) & 254 & 16.9 \\
\hline
% FAST Approaches to Scalable Similarity-Based Test Case prioritisation 
\lessthantenspace 5 \rankspace Miranda \etal~\cite{Miranda2018} & FAST Approaches to Scalable Similarity-Based Test Case Prioritisation & 2018
& \ref{subsec:test-case-prioritisation-diversity} (p.\pageref{subsec:test-case-prioritisation-diversity}), \ref{subsec:test-script-diversity} (p.\pageref{subsec:test-script-diversity}) & 95 & 15.8 \\
\hline
% Test Set Diameter: Quantifying the Diversity of Sets of Test Cases 
\lessthantenspace 6 \rankspace Feldt \etal~\cite{Feldt2016} & Test Set Diameter: Quantifying the Diversity of Sets of Test Cases  & 2016 & \ref{subsec:input-diversity} (p.\pageref{par:Feldt2016}) & 126 & 15.6 \\
% Test Generation and Test prioritisation for Simulink Models with Dynamic Behavior 
\hline
\lessthantenspace 7 \rankspace Matinnejad \etal~\cite{Matinnejad2019} & Test Generation and Test prioritisation for Simulink Models with Dynamic Behaviour & 2019
& \ref{subsubsec:model-based-testing} (p.\pageref{subsubsec:model-based-testing}), \ref{subsec:output-diversity} (p.\pageref{subsec:output-diversity}) & 74 & 14.8 \\
\hline
\lessthantenspace 8 \rankspace Matinnejad \etal~\cite{Matinnejad2016} & Automated Test Suite Generation for Time-Continuous Simulink Models  & 2016
& \ref{subsubsec:model-based-testing} (p.\pageref{subsubsec:model-based-testing}), \ref{subsec:output-diversity} (p.\pageref{subsec:output-diversity}) & 104 & 13.0 \\
\hline
% Improving multi-objective test case selection by injecting diversity in genetic algorithms
\lessthantenspace 9 \rankspace Panichella \etal~\cite{Panichella2015B} & Improving multi-objective test case selection by injecting diversity in genetic algorithms  & 2015 & \ref{subsubsec:multi-objective} (p.\pageref{subsubsec:multi-objective}) & 129 & 12.9 \\
\hline
10 \rankspace Noor \etal~\cite{Noor2015} & A similarity-based approach for test case prioritisation using historical failure data  & 2015 & \ref{subsubsec:prioritisation-test-cases} (p.\pageref{par:Noor2015}) & 100 & 12.5 \\
\arrayrulecolor{black}

        \bottomrule
    \end{tabular}

\end{table}

\subsection{Most Frequent Publishing Venues}

Table~\ref{table:papers-per-venue} shows the ten most-frequent publishing venues.
The table shows that \dbt papers have frequently appeared in both the top
journals and conferences in the field of Software Engineering and Testing.
IEEE Transactions on Software Engineering (TSE) is the journal with the most
\dbt papers (12), and the most publications overall. The two conferences with
the highest number of publications, which are ranked second on the list are the
International Conference on Software Testing, Verification and Validation
(ICST), and the International Conference on Automated Software Engineering (ASE)
both with 10 papers each.
Conferences dominate the table, appearing in five of the ten ranked positions,
compared to three different journals, and two workshops.

\subsection{Papers with the Highest Number of Citations}
We collected citation counts from Google Scholar\footnote{\url{https://scholar.google.com}},
with counts correct to 11th May 2023.

The paper with the most citations (254) is that of Yoo \etal~\cite{Yoo2009} on
clustering test cases for test case prioritisation.
However, this paper was also published in 2009, which is relatively old for a
\dbt paper, and therefore have more time than other works to gain citations.
For this reason, we calculated the mean number of citations per year and list
the top ten papers in terms of this metric in
Table~\ref{table:prominent-papers}.

The papers with the highest mean citations per year are by Henard
\etal~\cite{Henard2014} and Hemmati \etal~\cite{Hemmati2013}.
Henard \etal~\cite{Henard2014} use a similarity-based approach for software
product lines (SPL), using diversity between configurations to prioritise the
set of configurations that should be tested.
The paper is an important reference for other studies in
testing SPL systems. We discuss it in more detail in Section
\ref{subsec:spl-configuration-diversity}
(p.\pageref{subsec:spl-configuration-diversity}).
Hemmati \etal~\cite{Hemmati2013} focussed on test case selection in the context
of Model-Based Testing. We discuss this paper in Section
\ref{subsubsec:mbt-selection} (p.\pageref{pap:hemmati2013}).

Test case prioritisation is the focus of five of the ten papers appearing in
Table~\ref{table:prominent-papers}, with the other five covering a variety of
other topics including % 
test case selection, test generation, model-based testing, and new similarity
metrics for measuring the diversity of test suites. We discuss the most
prominent problems to which \dbt has been applied more generally in
Section~\ref{sec:rqproblems}.

\subsection{Prominent Authors}

The top five most prolific authors in the papers collected in our study are
listed in Table~\ref{table:prominent-authors}.
The author publishing the most papers is Lionel Briand, with 12~publications,
focussed on applying \dbt in
model-based testing, and Simulink models in particular (see
Section~\ref{subsubsec:model-based-testing},
p.\pageref{subsubsec:model-based-testing}; and
Section~\ref{subsubsec:mbt-selection}, p.\pageref{subsubsec:mbt-selection}).

Ranked second in the list with eight papers each are Francisco Gomes de Oliveria
Neto and Hadi Hemmati. Francisco Gomes de Oliveira Neto has mainly used \dbt
approaches to solve the problem of test case selection, discussed in
Section~\ref{subsubsec:mbt-selection} (p.\pageref{subsubsec:mbt-selection}),
while Hadi Hemmati has a
research interest in automated software testing and applying diversity in
\tcp and model-based testing (discussed in
Section~\ref{subsec:test-case-prioritisation-diversity},
p.\pageref{subsec:test-case-prioritisation-diversity};
and Section~\ref{subsubsec:mbt-selection}, p.\pageref{subsubsec:mbt-selection}).

Ranked fourth is Robert Feldt, who has applied his work in
Search-Based Software Testing to diversity, with seven papers.
This work is discussed in Section~\ref{subsec:test-data-generation}
(p.\pageref{subsec:test-data-generation}). He also proposed a
similarity metric named ``Test Set Diameter'' (TSDm)  % DONE add name of metric
that can be applied to measure diversity of the whole test suite
(discussed in Section~\ref{subsec:input-diversity}, p.\pageref{subsec:input-diversity}).

\begin{table}[tbp]
    \centering
    \caption{
        \label{table:prominent-authors}
        Authors with the highest numbers of papers
    }
    %
    % \vspace{-1em}
    %
    \begin{tabular}{@{}P{9cm}rP{5cm}}
        \toprule

        {\bf Rank / Author} & {\bf \# Papers} & 
        \\

        \midrule

        \nonequalrankspace 1 \rankspace Lionel Briand & 12 & \cite{Hemmati2010C,Hemmati2010A,Hemmati2010B,Hemmati2011,Hemmati2013,Matinnejad2015A,Matinnejad2016,Matinnejad2019,Aghababaeyan2023,Rogstad2013,Liu2017,Liu2019} \\
\nonequalrankspace \phantom{1} \rankspace (\url{https://www.lbriand.info}) & & \\
\arrayrulecolor{lightgray}\hline
=2 \rankspace Francisco Gomes de Oliveira Neto & 8 & \cite{Cartaxo2007,Cartaxo2009,De2016,De2018,De2020,Neto2018,Dobslaw2020,Mosin2022} \\
\nonequalrankspace \phantom{3} \rankspace (\url{https://www.gu.se/en/about/find-staff/franciscodeoliveiraneto}) & & \\
\hline
=2 \rankspace Hadi Hemmati & 8 & \cite{Hemmati2010A,Hemmati2010B,Hemmati2010C,Hemmati2011,Hemmati2013,Hemmati2015,Noor2015,Mondal2015} \\
\nonequalrankspace \phantom{3} \rankspace (\url{https://lassonde.yorku.ca/users/hhemmati}) & & \\
\hline
\nonequalrankspace 4 \rankspace Robert Feldt & 7 & \cite{Feldt2008,Feldt2016,Neto2018,De2020,Dobslaw2020,Marculescu2016,Poulding2017} \\
\nonequalrankspace \phantom{1} \rankspace (\url{http://www.robertfeldt.net}) & & \\
\hline
=5 \rankspace Antonia Bertolino & 5 & \cite{Miranda2014,Bertolino2015,Greca2022,Miranda2018,Cruciani2019} \\
\nonequalrankspace \phantom{5} \rankspace (\url{http://bertolino.isti.cnr.it}) & & \\
\hline
=5 \rankspace Zhenyu Chen & 5 & \cite{Wu2012,Chen2011,Shi2015,Fang2014,Feng2015} \\
\nonequalrankspace \phantom{5} \rankspace (\url{https://software.nju.edu.cn/zychen}) & & \\
\hline
=5 \rankspace David Clark & 5 & \cite{Feldt2016,Joffe2019,Menendez2021,Menendez2022,Benito2022} \\
\nonequalrankspace \phantom{5} \rankspace (\url{http://www0.cs.ucl.ac.uk/staff/D.Clark}) & & \\
\hline
=5 \rankspace Shiva Nejati & 5 & \cite{Matinnejad2015A,Matinnejad2016,Matinnejad2019,Liu2017,Liu2019} \\
\nonequalrankspace \phantom{5} \rankspace (\url{https://engineering.uottawa.ca/people/nejati-shiva}) & & \\
\hline
=5 \rankspace Paolo Tonella & 5 & \cite{Marchetto2009,Kifetew2013,Biagiola2019,Menendez2021,Yoo2009} \\
\nonequalrankspace \phantom{5} \rankspace (\url{https://www.inf.usi.ch/faculty/tonella}) & & \\
\arrayrulecolor{black}

        \bottomrule
    \end{tabular}

    \vspace{-2em}
\end{table}

Finally, a number of authors are ranked fifth in the list, with five papers
each. They are Antonia Bertolino, Zhenyu Chen, David Clark, Shiva Nejati,
and Paolo Tonella who have worked on a handful of topics in \dbt including
\tcp (discussed in Section~\ref{subsec:test-case-prioritisation-diversity}, p.\pageref{subsec:test-case-prioritisation-diversity}),
test report prioritisation (discussed in Section~\ref{subsec:test-report-diversity}, p.\pageref{subsec:test-report-diversity}),
test data generation using output diversity (discussed in Section~\ref{subsec:output-diversity}, p.\pageref{subsec:output-diversity}),
testing cyber-physical systems (discussed in Section~\ref{subsubsec:model-based-testing}, p.\pageref{subsubsec:model-based-testing}),
and web testing (discussed in Section~\ref{subsubsec:web-testing}, p.\pageref{subsubsec:web-testing}), respectively.

\begin{tcolorbox}[title=Conclusions --- \rqtrends]
    The first paper on \dbt techniques in our collected results was published in
  2005, and the number of publications has a strong upward trend year on year since then.
  The venue featuring the most papers to date (12) is the
  Transactions on Software Engineering (TSE) journal with the
  International Conference on Software Testing,
  Verification and Validation (ICST)
  and the International Conference on Automated Software Engineering (ASE)
  appearing next, with 10~papers each.
  The paper with the most citations (254) is that of Yoo \etal~\cite{Yoo2009}
  on clustering test cases for test case prioritisation, published in 2009.
  The papers with the highest mean number of citations per year are by Henard
  \etal~\cite{Henard2014}, on diversity for configurations of software product
  lines for testing, published in 2014; and Hemmati \etal~\cite{Hemmati2013}
  on test case selection for model-based tests, published in 2013.
  Finally, the most prolific author in the field is Lionel
  Briand, having published 12~papers on the topic of \dbt.
\end{tcolorbox}
\section{\rqmetrics}
\subsection*{\textbf{\rqmetricsexpanded}}
\label{sec:rqmetrics}

All studies that apply diversity in their approaches use some metric to measure the
level of similarity or diversity. The similarity can be calculated for inputs, outputs, or any other
testing \artefact{s}. One such \artefact{} can be the test scripts or test cases themselves.
It is simple to calculate the distance between any two strings, but for multiple strings, usually,
researchers concatenate each set of strings into a single string, and then the two concatenated
strings can be compared directly.

There are many similarity metrics used in the literature, and we found \nummetrics metrics in the collected papers.
Some of these metrics are well-known, like Euclidean distance, Hamming distance, and so on, while others
are more specific to certain types of subject domains or new metrics.

\begin{figure}[t]
    \centering
    \includegraphics[width=0.6\columnwidth]{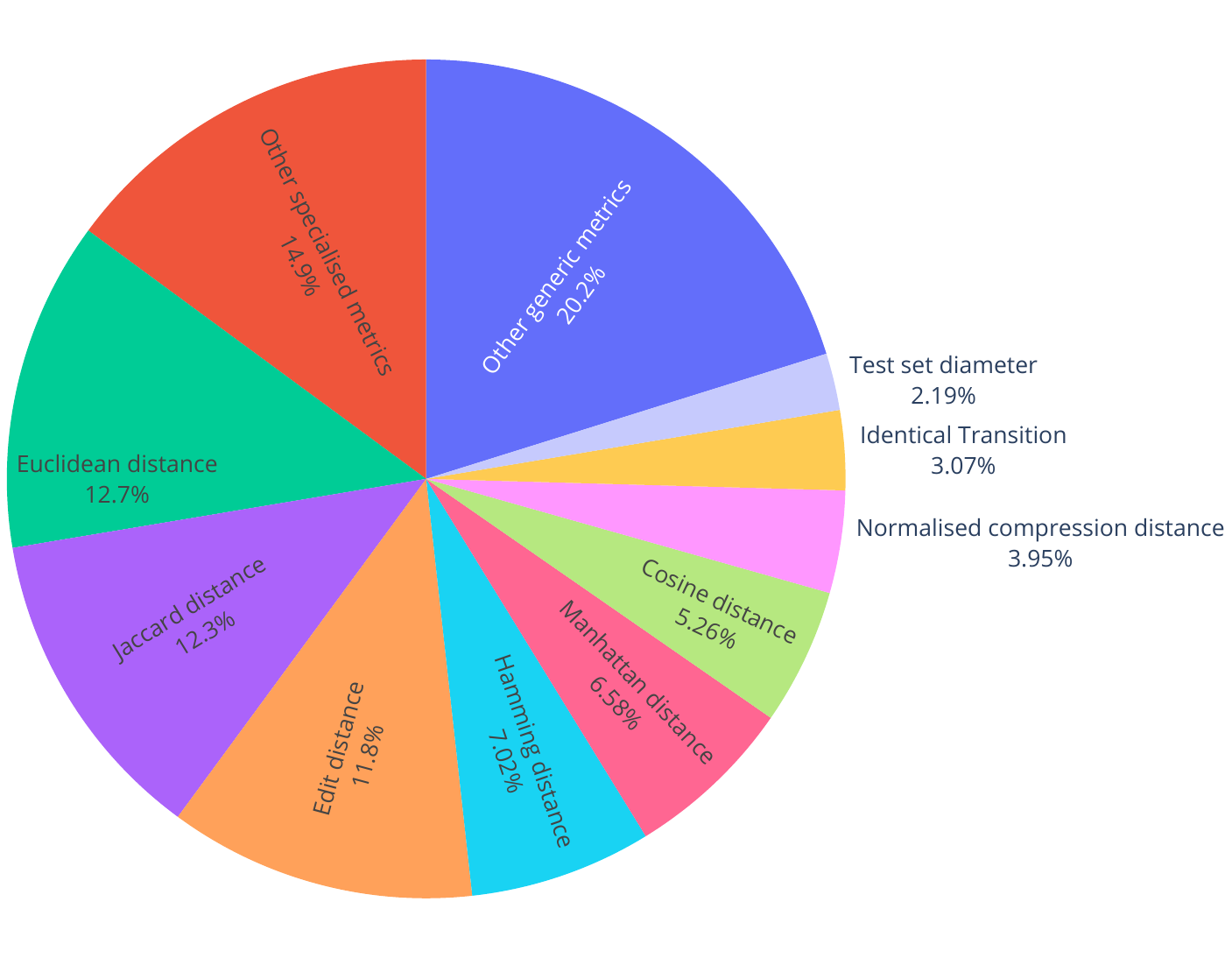}

    \vspace*{-2em}\caption{
        \label{fig:metrics-distribution}
        Distribution of similarity metrics used in \dbt techniques.
    }
\end{figure}

We categorized the similarity metrics into two groups. The first group consists of the generic similarity
metrics that originated from other fields. The second group
are specialised metrics in Software Engineering to measure similarity based on information
acquired from software programs, or metrics proposed to solve a Software Engineering problem.
Figure~\ref{fig:metrics-distribution} shows the distribution of similarity metrics used in \dbt papers.

Table~\ref{table:metrics-summary} presents the generic similarity metrics, while Table~\ref{table:other-metrics-summary}
lists the specialised similarity metrics in software engineering. For both tables, records are
ordered by usage popularity in the literature and then by alphabetical order. For each similarity
metric, we provide a citation for more details about the metric,
a short description, how many papers used that metric, and all papers using the metric in our collection.
As shown in Figure~\ref{fig:metrics-distribution}, $80.2\%$ of the \dbt papers used generic
similarity metrics, while only $19.8\%$ used specialised metrics. The largest portion is the
``other generic metrics'', which makes up $20.2\%$, but it contains 32 generic metrics. These
metrics are coded G8 to G40 in Table~\ref{table:metrics-summary}. The second-largest portion
in the pie chart is the ``other specialised metrics'' which contains 28 specialised metrics
coded S3 to S30 in Table~\ref{table:other-metrics-summary}.

The three most popular similarity metrics are Euclidean distance, Jaccard distance, and Edit 
distance used in 29, 28, and 27 papers, respectively. Numeric programs are used by many 
researchers to evaluate their techniques and Euclidean distance is a natural choice for 
such programs. Also, Jaccard distance is widely used when the testing artefacts can be 
represented as sets, with an example being software product lines in which a product can 
be seen as being a set of features. Furthermore, if inputs are strings then one might use 
Edit distance.

\begin{longtable}[t]{lP{4cm}@{~~~}p{6.5cm}@{~~~}rp{2cm}}
  \caption[Caption]{A list of the generic similarity metrics with a citation of the source, a brief description,
  the number of papers using them, and citation these papers.} \label{table:metrics-summary} \\
  \toprule
  {\bf ID} & {\bf Metric \& Source} & {\bf Description} & {\bf Total} & {\bf Papers}
  \\

  \midrule

  % The data in this file should be automatically output
% from code that processes raw results data.
% Rank & Metric & Def Ref. & Description & Total & Papers \\
G1 & Euclidean distance~\cite{AlphaEuclidean} & The square root of the sum of the squared differences between the vectors $X$ and $Y$. The formula is: \newline  \nonequalrankspace
\(\displaystyle
\mathit{Euc}(X,Y) = \sqrt{\sum_{i=1}^{n} (x_{i} - y_{i})^{2}}
\)
& 29    &
\cite{Albunian2020,Arafeen2013,Arrieta2019,Bueno2007,Bueno2008,Bueno2014,Carlson2011,Chetouane2020,Coviello2018A,Coviello2018B,Cruciani2019,Farzat2010,Greca2022,Guarnieri2022,Haghighatkhah2018B,Hemmati2015,Ledru2009,Ledru2012,Mahdieh2022,Matinnejad2015A,Matinnejad2016,Matinnejad2019,Nunes2015,Pei2014,Rogstad2013,Shahbazi2015B,Shimari2022,Wang2015,Zhao2015}
\\
\arrayrulecolor{lightgray}\hline
G2 & Jaccard distance~\cite{Jaccard1901} & The ratio of intersection over union between two sets $A$ and $B$ of values. The formula is:  \newline
\(\displaystyle
\mathit{Jac}(A,B) = 1- \dfrac{|A \cap B|}{|A \cup B|} = 1 - \dfrac{|A \cap B|}{|A| + |B| - |A \cap B|}
\) \newline \newline
Sometimes expressed, (e.g.~\cite{Wang2015}) as: \newline
\(\displaystyle
\mathit{Jac}(A,B) = 1 - \dfrac{A . B}{A . B + \omega(\parallel A \parallel^{2}+\parallel B \parallel^{2} - 2(A . B))}
\) \newline
with with $\omega=1$.
& 28    &
\cite{Cao2022,Coutinho2016,Coviello2018A,Coviello2018B,De2018,Fischer2016,Guarnieri2022,Haghighatkhah2018A,Hemmati2010B,Hemmati2011,Henard2013,Henard2014,Liu2017,Liu2019,Mariani2021,Mei2013A,Mei2013B,Miranda2014,Miranda2018,Neto2018,Tang2022,Viggiato2023,Wang2015,Xiang2021,You2013,Zhang2019,Zhao2011,Zhao2013}
\\
\hline
G3 & Edit distance~\cite{Levenshtein1966} & The minimum number of edits \emph{(insertions, deletions or substitutions)} required to change one string into the other. It takes into consideration that parts of the strings can be similar even if not in corresponding places, and can work with strings of different sizes.
& 27    &
\cite{Almulla2020,Asoudeh2013,Biagiola2019,Boussaa2015,Cao2022,Coutinho2016,Coviello2018A,Coviello2018B,De2018,Fang2014,Flemstrom2016,Guarnieri2022,Hemmati2010A,Hemmati2010B,Hemmati2010C,Hemmati2013,Ledru2009,Ledru2012,Ma2018,Mariani2021,Mondal2015,Noor2015,Shahbazi2015B,Wang2023,Wu2012,Zhang2012,Shimmi2022}
\\
\hline
G4 & Hamming distance~\cite{Hamming1950} & The number of times when the corresponding characters in two strings are different. Some (e.g.~\cite{Huang2017B}) refer to this as ``Overlap'' distance. & 16    &
\cite{Abd2019,Cao2013,Coviello2018A,Coviello2018B,Feng2015,Hemmati2010B,Huang2017B,Ledru2009,Ledru2012,Mondal2015,Noor2015,Shahbazi2015B,Wang2009,Wang2014,Xie2009,Yoo2009}
\\
\hline
G5 & Manhattan distance~\cite{AlphaManhattan} & The sum of the absolute differences between two vectors $X$ and $Y$. The formula is: \newline \nonequalrankspace
\(\displaystyle
\mathit{Man}(X,Y) = \sum_{i=1}^{n} |x_{i} - y_{i}|
\)
& 15    &
\cite{Boussaa2015,Cao2013,Chen2019C,Chen2021B,Guarnieri2022,Haghighatkhah2018A,Hemmati2015,Ledru2009,Ledru2012,Liu2017,Liu2019,Mahdieh2022,Nunes2015,Rogstad2013,Shahbazi2015B}
\\
\hline
G6 & Cosine similarity~\cite{AlphaCosine} & The cosine of the angle of two vectors $X$ and $Y$. The formula is: \newline \nonequalrankspace
\(\displaystyle
\mathit{Cosine}(X,Y) = \frac{\sum_{i=1}^{n} x_{i} \times y_{i}}{\sqrt{\sum_{i=1}^{n} x_{i}^{2}} \times \sqrt{\sum_{i=1}^{n} y_{i}^{2}} }
\)
& 12    &
\cite{Azizi2021,Brooks2009,Coviello2018A,Coviello2018B,Lin2017,Mahdieh2022,Semerath2018,Semerath2020,Shahbazi2015B,Sondhi2022,Viggiato2023,Wang2015}
\\
\hline
G7 & Normalised compression distance~\cite{Cilibrasi2005} & An approximation of the Kolmogorov complexity using real-world compressors. & 9    &
\cite{Aghababaeyan2023,De2018,De2020,Dobslaw2020,Feldt2008,Feldt2016,Guarnieri2022,Haghighatkhah2018A,Rogstad2013}
\\
\hline
G8 & Tree edit distance~\cite{Zhang1989} &  The minimum number of edit operations required to change one tree into the other. & 4 &
\cite{Mei2013A,Mei2013B,Shahbazi2014,Shahbazi2018}
\\
\hline
G9 & Locality-sensitive hashing~\cite{Shakhnarovich2006} & A technique that maps similar strings or inputs to the same hash code with high probability to get a fast estimation of the dissimilarity between two subjects. & 3 &
\cite{Haghighatkhah2018A,Miranda2018,Shahbazi2015B}
\\
\hline
G10 & Crowding distance~\cite{Mahfoud1995} &  A measure of how far a chromosome or an individual is from the rest of the population. & 2 &
~\cite{Panichella2015B,Scalabrino2016}
\\
\hline
G11 & Geometric diversity~\cite{Kulesza2012} &  The measurement of feature similarity between two feature vectors given an input sample. & 2 &
\cite{Aghababaeyan2023,Jiang2023}
\\
\hline
G12 & Gower-Legendre distance~\cite{Gower1986} &  A variant of the Jaccard Index (G2) where the weight $\omega$ is $1/2$. & 2 &
~\cite{Mondal2015,Wang2015}
\\
\hline
G13 & Isolated subTree distance~\cite{Tanaka1988} &  A variation of tree edit distance (G8), where disjoint subtrees are mapped to similar disjoint subtrees of another set. & 2 &
\cite{Shahbazi2014,Shahbazi2018}
\\
\hline
G14 & Jaro-Winkler distance~\cite{Winkler1999} &  A variation of the Jaro distance (G23) that adds more weight in strings starting with the exact match characters. & 2 &
~\cite{Abd2019,Coutinho2016}
\\
\hline
G15 & L2-test~\cite{Goldreich2000} & The distance between a uniform distribution and a sampled distribution by checking if the sampled distribution is $\epsilon$-far from uniformity. & 2 &
~\cite{Menendez2021,Menendez2022}
\\
\hline
G16 & Mahalanobis distance~\cite{Chandra1936} & The distance between a point and a distribution. & 2 &
~\cite{Nunes2015,Rogstad2013}
\\
\hline
G17 & Needleman-Wunsch distance~\cite{Needleman1970} &  Originally used in bioinformatics to align protein or nucleotide sequences, and can be used to identify similarities between two test cases by encoding them. & 2 &
~\cite{Hemmati2010B,Hemmati2011}
\\
\hline
G18 & Canberra distance~\cite{AlphaCanberra} &  A weighted version of the Manhattan distance (G5), in which each term in the sum is normalised. & 1 &
~\cite{Nunes2015}
\\
\hline
G19 & Chebyshev distance~\cite{AlphaChebyshev} &  The greatest difference between two points in two vectors along any coordinate dimension. & 1 &
~\cite{Nunes2015}
\\
\hline
G20 & Fractional distance~\cite{Aggarwal2001} &  A variation of the Euclidean distance (G1) to deal with multi-dimensional space. & 1 &
~\cite{Marculescu2016}
\\
\hline
G21 & Hellinger distance~\cite{AlphaHellinger} &  The difference between two distributions with Hellinger integral~\cite{Hellinger1909}. & 1 &
~\cite{Xie2022}
\\
\hline
G22 & Hill-numbers~\cite{Hill1973} &  A measure originally used in ecology that considers both species richness and species abundances in a sample. & 1 &
~\cite{Nguyen2022}
\\
\hline
G23 & Jaro distance~\cite{Jaro1989} &  The number of matching characters and the number of transpositions (i.e. matching characters but not in order) between two strings. & 1 &
~\cite{Coutinho2016}
\\
\hline
G24 & Jeffrey divergence~\cite{Jeffreys1998} &  A derived distribution from the Kullback-Leibler Divergence (G27) that is symmetric and more robust to noises. & 1 &
~\cite{Nunes2015}
\\
\hline
G25 & Jensen-Shannon distance~\cite{Menendez1997} &  An improved version of Kullback-Leibler Divergence (G27) to measure the similarity of two probability distributions. The metric is symmetric and always has a finite value. & 1 &
~\cite{Xie2022}
\\
\hline
G26 & Kronecker delta~\cite{AlphaKronecker} & A discrete function of two variables that is one if they are equal, 0 otherwise. & 1 &
~\cite{Kichigin2009}
\\
\hline
G27 & Kullback-Leibler divergence~\cite{kullback1951} &  The expected value of the logarithmic difference between two probability distributions, but it is not symmetric. & 1 &
~\cite{Xie2022}
\\
\hline
G28 & Mean-square-error~\cite{BritannicaMSE} & The average squared difference between the values predicted from a model and the actual values. & 1 &
~\cite{Mosin2022}
\\
\hline
G29 & Modified trigonometric distance~\cite{Li2006} &  A modified version of the trigonometric distance (G38) with a greater degree of accuracy for points of larger magnitude of values. & 1 &
~\cite{Nunes2015}
\\
\hline
G30 & N-gram models~\cite{Broder1997} &  A contiguous sequence of $n$ items from a given sample of text or speech. & 1 &
~\cite{Leveau2020}
\\
\hline
G31 & Proportional distance~\cite{Dickinson2001} &  The sum of squares of the difference between two vectors over the difference between the maximum and minimum values. & 1 &
~\cite{Wang2015}
\\
\hline
G32 & Singular value decomposition~\cite{Stewart1993} & An estimate of where the evolution is going in search-based approaches, by monitoring the movements of individuals across different generations. & 1 &
~\cite{Kifetew2013}
\\
\hline
G33 & Smith-Waterman distance~\cite{Smith1981} &  The alignment of local sequences for determining similar regions between two strings of nucleic acid sequences or protein sequences. & 1 &
~\cite{Hemmati2010B}
\\
\hline
G34 & Sokal-Sneath distance~\cite{Zalewski2014} &  A variant of the Jaccard Index (G2) where the weight $\omega$ is $2$. & 1 &
~\cite{Wang2015}
\\
\hline
G35 & Statistic salue $X^2$~\cite{Bugatti2008} &  A distance function that emphasizes large absolute difference existing between the feature values. & 1 &
~\cite{Nunes2015}
\\
\hline
G36 & String-Kernels~\cite{Leslie2003} &  The inner product between two strings by counting the occurrences of common substrings in the two strings. & 1 &
~\cite{Wang2015}
\\
\hline
G37 & Sellers algorithm~\cite{Sellers1980} &  A variation of the edit distance (G3) to find a sub-string in another string with at most $k$ edit operations. & 1 &
~\cite{Coutinho2016}
\\
\hline
G38 & Trigonometric distance~\cite{Li2006} &  A normalised distance between two points used in image matching. The distance between two vectors $X$ and $Y$ is
$\sum_{i=1}^{n} \sin (\arctan |x_{i} - y_{i}|)$
& 1 &
~\cite{Nunes2015}
\\
\hline
G39 & Wasserstein distance~\cite{Arjovsky2017} &  The difference between two frequency distributions over a region, which is also known as the earth mover's distance. & 1 &
~\cite{Xie2022}
\\
\hline
G40 & Word mover's distance~\cite{Kusner2015} &  The minimum amount of distance that the embedded words of one document need to be moved to reach the embedded words of another document. & 1 &
~\cite{Viggiato2023}
\\
\arrayrulecolor{black}

  \bottomrule
\end{longtable}

\begin{longtable}[t]{lP{4cm}@{~~~}p{6.5cm}@{~~~}rp{2cm}}
  \caption[Caption]{A list of the specialised Software Engineering similarity metrics.} \label{table:other-metrics-summary} \\
  \toprule
  % use a new line for each column if needed
  {\bf ID} & {\bf Metric \& Source} & {\bf Description} & {\bf Total} & {\bf Papers}
  \\

  \midrule

  % The data in this file should be automatically output
% from code that processes raw results data.
S1 & Identical transition distance~\cite{Cartaxo2007} & The number of identical transitions between two finite state machines divided by the average length of paths. & 7    &
\cite{Cartaxo2007,Cartaxo2009,De2016,Hemmati2010A,Hemmati2010B,Hemmati2010C,Hemmati2013}
\\
\arrayrulecolor{lightgray}\hline
S2 & Test set diameter (TSDm)~\cite{Feldt2016} & An extension of the pairwise normalised compression distance (G7) to multisets. & 5    &
\cite{Feldt2016,Haghighatkhah2018A,Haghighatkhah2018B,Shahbazi2014,Shahbazi2018} \\
\hline
S3 & Approach level~\cite{Mcminn2004} &  The number of mismatched branch predicates to reach the target branch. & 3 &
\cite{Alshraideh2011,Cai2021,Panchapakesan2013}
\\
\hline
S4 & Identical state distance~\cite{Hemmati2010A} & The number of identical states between two paths of finite state machines divided by their average number of states. & 3    &
\cite{Hemmati2010A,Hemmati2010C,Hemmati2013} \\
\hline
S5 & Trigger-based distance~\cite{Hemmati2010A} & An extension of identical transition similarity (S1) to account for triggers in the transitions. & 3    &
\cite{Hemmati2010A,Hemmati2010C,Hemmati2013} \\
\hline
S6 & Average population diameter~\cite{Vogel2019} & The average distance between all vectors in a population, where the distance between two vectors is the difference of their lengths. & 2 &
\cite{Vogel2019,Vogel2021} \\
\hline
S7 & Distinguishing mutation adequacy~\cite{Shin2016} & An assessment of the diversity of mutants' behaviour based on the mutants' killing information. & 2 &
\cite{Shin2016,Shin2018} \\
\hline
S8 & Extended subTree distance~\cite{Shahbazi2014} & A variation of isolated subtree distance (G12) with different mapping conditions. & 2 &
\cite{Shahbazi2014,Shahbazi2018} \\
\hline
S9 & Path distance~\cite{Buttler2004} & The size of the intersection between the two paths of multisets of trees. & 2 &
\cite{Shahbazi2014,Shahbazi2018}
\\
\hline
S10 & [GUI] State similarity~\cite{Feng2012} & The difference between the values of two GUI states using the widgets of the GUI. & 2 &
\cite{Feng2012,He2015}
\\
\hline
S11 & [Graph model diversity] Symmetric distance~\cite{Semerath2018} & The difference between two models in a domain-specific language, where it is calculated as the number of ``shapes'' contained exclusively in one of the models but not both. & 2 &
\cite{Semerath2018,Semerath2020}
\\
\hline
S12 & Test diversity~\cite{Nikolik2006} & A hybrid measure calculating the difference between two test cases in terms of branches covered,
variation of the data inputs, and standard deviation between conditions covered. & 2 &
\cite{Nikolik2006,Marchetto2009} \\
\hline
S13 & Text uniqueness~\cite{Alshahwan2012} & Text matching between two strings, where a string is unique if no other string matches it. & 2 &
\cite{Alshahwan2012, Alshahwan2014} \\
\hline
S14 & Achieved coverage of pools~\cite{Yatoh2015} & The number of items selected from a pool of values for a program's variables over the time spent using that pool of values. & 1 &
\cite{Yatoh2015} \\
\hline
S15 & Accuracy-based performance measure~\cite{Zhao2022} & The proportion of correctly predicted test inputs to all the test inputs for a DNN. & 1 &
\cite{Zhao2022} \\
\hline
S16 & [Test behavioural similarity] Accuracy (acc)~\cite{Metz1978} &  The percentage of tests that fail or pass together, calculated as the number of correct predictions divided by the total number of predictions in a confusion matrix. & 1 &
\cite{De2020}
\\
\hline
S17 & Average cyclomatic complexity per method (ACCM)~\cite{Chidamber1994} & Cyclomatic complexity is the number of independent paths in a program or method. ACCM is calculated by computing the number of independent paths within each method and then taking the sum, over all methods, of these values. & 1 &
\cite{Pizzoleto2020}
\\
\hline
S18 & Basic counting~\cite{Noor2015} & The overlapping occurrences of method calls between two failing sequences of method calls extracted from execution traces of tests. & 1 &
\cite{Noor2015} \\
\hline
S19 & Code complexity (cm)~\cite{Arafeen2013} & Consists of three types of information (Lines of code, Nested Block Depth, and Cyclomatic Complexity) derived from the source code to measure similarity. & 1 &
\cite{Arafeen2013} \\
\hline
S20 & [GUI similarity] $\mathit{CONTeSSi}(n)$~\cite{Brooks2009} & The differences of the frequencies between the past $n$ executed events of one test suite to another test suite. & 1 &
\cite{Brooks2009} \\
\hline
S21 & Distance entropy~\cite{Shi2015} & The distribution of tests in a set represented in a graph using the minimum weight set (\ie the set of vertices or edges in a weighted graph that collectively has the smallest sum of weights). & 1 &
\cite{Shi2015} \\
\hline
S22 & Enhanced Jaro-Winkler~\cite{Abd2019} & A hybrid metric between Jaro-Winkler (G14) and Hamming Distance (G4) that considers the deselected features from Hamming distance combined into the Jaro distance equation. & 1 &
\cite{Abd2019} \\
\hline
S23 & Graph edit distance~\cite{Sanfeliu1983} &  The minimum number of edit operations required to make two graphs identical. & 1 &
\cite{Tang2022}
\\
\hline
S24 & Matthew's correlation coefficient (\emph{MCC})~\cite{Matthews1975} &  A more accurate measure of tests behavioural similarity than accuracy (S16) that accounts for both true positives and true negatives. & 1 &
\cite{De2020}
\\
\hline
S25 & Probabilistic type tree~\cite{Poulding2017} & A tree structure to represent a probability distribution over the types. & 1 &
\cite{Poulding2017} \\
\hline
S26 & Response for class (RFC)~\cite{Chidamber1994} &  The sum of the number of methods inside the class and the number of external methods used by the class. & 1 &
\cite{Pizzoleto2020}
\\
\hline
S27 & Syntax-tree similarity~\cite{Masuda2021} & The structural similarity between two sentences represented as ``syntax trees'' by comparing the tree topologies, node positions, and the types of grammatical relationships. & 1 &
\cite{Masuda2021}
\\
\hline
S28 & Traces~\cite{Zalewski2014} &  The difference between two execution paths, that takes into account the branches covered and the number of times these branches were covered. & 1 &
\cite{Reddy2020}
\\
\hline
S29 & Weighted distance function~\cite{Beena2014} & The number of statements covered by one test case but not the other, and the difference of the execution times between the two test cases.  & 1 &
\cite{Beena2014}
\\
\hline
S30 & Extensible access control markup language (XACML) Similarity~\cite{Bertolino2015} & The distance between the requests attributes' values of two XACML test cases and the difference between their policies. & 1 &
\cite{Bertolino2015} \\
\arrayrulecolor{black}

  \bottomrule
\end{longtable}

\begin{tcolorbox}[title=Conclusions --- \rqmetrics]
  Many similarity metrics were used in the literature, and we found \nummetrics metrics in this study. The most used similarity
  metrics in the literature are generic metrics, where the most popular similarity metrics were Euclidean distance,
  Jaccard distance and Edit distance were found in 29, 28 and 27 papers, respectively.
  The survey results are \textbf{not indicative} of which metric would be best to use, as different factors can
  affect this choice such as the nature of the data. As will be shown in Section \ref{sec:rqartefacts} and 
  Section \ref{sec:rqproblems}, there is no clear indication of which metric performs the best, as different authors
  reported different results based on their empirical studies and the similarity metrics used in their studies.
\end{tcolorbox}

\section{\rqartefacts}
\subsection*{\textbf{\rqartefactsexpanded}}
\label{sec:rqartefacts}

Different software \artefact{s} have been used as a basis for \dbt techniques. Some of these
\artefact{s} include the input domain, output domain, etc. We discuss how these \artefact{s} were used in
the \dbt techniques.

\begin{figure}[t]
    \centering
    \includegraphics[width=0.7\columnwidth]{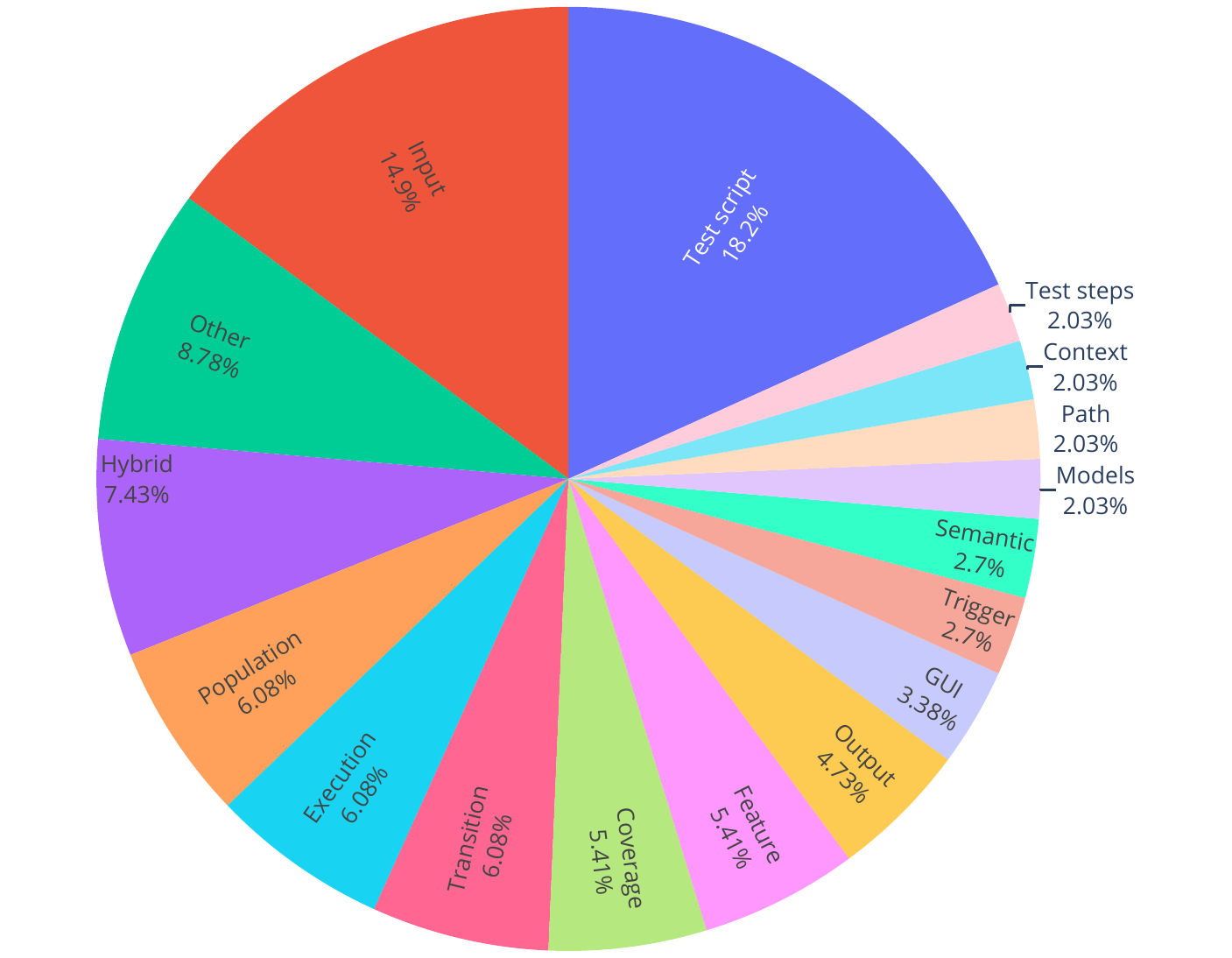}

    \vspace*{-1em}\caption{
        \label{fig:artefacts-distribution}
        Distribution of testing \artefacts used in \dbt techniques.
    }
\end{figure}

Table \ref{table:artefact-summary} presents a list
of the diversity \artefact{s} used, with a short description, the total number of papers using them,
and citations of all these papers. The most used diversity \artefact{s} are test scripts and inputs,
while test report diversity, social diversity, function, and running time diversity were used in one study each.
Figure~\ref{fig:artefacts-distribution} shows the distribution of testing \artefacts used in \dbt papers.
Papers that use test scripts or inputs as diversity \artefacts constitute $33.1\%$ of the total \dbt papers
in our study. Also, \dbt papers that use a hybrid of diversity \artefacts form $7.4\%$ of the collected papers.

\begin{longtable}[t]{p{3cm}@{~~~}p{7cm}@{~~~}rp{3cm}}
    \caption[Caption]{A list of the testing \artefact{s} used as a basis for DBT techniques.} \label{table:artefact-summary} \\
    \toprule
    {\bf Rank/Artefact} & {\bf Description} & {\bf Total} & {\bf Papers}
    \\

    \midrule

    % The data in this file should be automatically output
% from code that processes raw results data.
\nonequalrankspace\nonequalrankspace 1 \rankspace Test Script & Information from the actual text of the test cases which may include for example setup, inputs, and assertions. & 27 &
\cite{Azizi2021,Bertolino2015,Burdek2015,Carlson2011,Chen2021B,Chetouane2020,Coviello2018A,Coviello2018B,Cruciani2019,Gong2012,Greca2022,Haghighatkhah2018A,Haghighatkhah2018B,Hao2005,Hao2008,Hemmati2011,Hemmati2015,Huang2017B,Ledru2009,Ledru2012,Miranda2018,Mondal2015,Neto2018,Rogstad2013,Shi2015,Tang2022,Xia2016}
\\
\arrayrulecolor{lightgray}\hline
\nonequalrankspace\nonequalrankspace 2 \rankspace Input & Data provided to the software, which can be numbers, strings, objects, or images. & 22 &
\cite{Almulla2020,Boussaa2015,Bueno2007,Bueno2008,Bueno2014,Chen2011,De2012,Feldt2016,Mei2013A,Mei2013B,Menendez2021,Menendez2022,Mosin2022,Noor2015,Pei2014,Poulding2017,Reddy2020,Shahbazi2014,Shahbazi2015B,Shahbazi2018,Shimmi2022,Yatoh2015}
\\
\hline
\nonequalrankspace\nonequalrankspace 3 \rankspace Hybrid & Approaches that use a combination of different \artefact{s} as targets for applying diversity. & 11 &
\cite{Albunian2017,Albunian2020,Arrieta2019,Biagiola2019,Feldt2008,Liu2017,Liu2019,Ma2018,Marchetto2009,Nikolik2006,Singh2013}
\\
\hline
\nonequalrankspace =4 \rankspace Population & In search-based software testing, the collection of individuals or solutions represents the population. & 9 &
\cite{Alshraideh2011,Kifetew2013,Marculescu2016,Panichella2015B,Scalabrino2016,Vogel2019,Vogel2021,Wang2023,Xie2005}
\\
\hline
\nonequalrankspace =4 \rankspace Execution  & The actual runtime execution path of the software under test after running the test cases. & 9 &
\cite{Cao2013,Fang2014,Guarnieri2022,Pizzoleto2020,Wu2012,Xie2022,Yoo2009,Zhao2011,Zhao2013}
\\
\hline
\nonequalrankspace =4 \rankspace Transitions & The edge connecting a source state to a target state. In model-based testing, two transitions are said to be similar if they both have the same source and target states, and trigger. & 9 &
\cite{Asoudeh2013,Cartaxo2007,Cartaxo2009,Coutinho2016,De2016,Hemmati2010A,Hemmati2010B,Hemmati2010C,Hemmati2013}
\\
\hline
\nonequalrankspace =7 \rankspace Coverage & Information about statement or branch coverage achieved from running the test cases. & 8 &
\cite{Beena2014,De2018,Mahdieh2022,Wang2015,You2013,Zhang2019,Zhao2015,Zhao2022}
\\
\hline
\nonequalrankspace =7 \rankspace Feature & The specific functionalities, capabilities, or characteristics that a software product or application offers to its users. & 8 &
\cite{Abd2019,Aghababaeyan2023,Chen2019C,Farzat2010,Henard2013,Jiang2023,Joffe2019,Nunes2015}
\\
\hline
\nonequalrankspace\nonequalrankspace 9 \rankspace Output & The results produced by the software under test from running the test cases. & 7 &
\cite{Alshahwan2012,Alshahwan2014,Benito2022,Dobslaw2020,Matinnejad2015A,Matinnejad2016,Matinnejad2019}
\\
\hline
\nonequalrankspace 10 \rankspace GUI & The visual components and their properties' values that represent the interface of the software. & 5 &
\cite{Feng2012,He2015,Leveau2020,Mariani2021,Xie2006B}
\\
\hline
=11 \rankspace Trigger & A specific event or condition that causes the system to move from one state to another. In model-based testing, trigger similarity does not require the source and target states to be identical, unlike transition similarity. & 4 &
\cite{Hemmati2010A,Hemmati2010B,Hemmati2010C,Hemmati2013}
\\
\hline
=11 \rankspace Semantic & Statements that have different wordings but carry similar meanings are said to be \emph{semantically} similar. & 4 &
\cite{Cai2009,Cao2022,Lin2017,Zhang2012}
\\
\hline
=13 \rankspace Models & Graph models represent and analyze relationships between variables in a system, where the nodes of the graph are the variables and the edges are the relationships or dependencies. & 3 &
\cite{Kichigin2009,Semerath2018,Semerath2020}
\\
\hline
=13 \rankspace Path  & The possible flow of control and data within a program without executing it. & 3 &
\cite{Cai2021,Panchapakesan2013,Xie2009}
\\
\hline
=13 \rankspace Context & Data from software and environment to guide the responses and actions of the system. & 3 &
\cite{Brooks2009,Wang2009,Wang2014}
\\
\hline
=13 \rankspace Test Steps & A test step corresponds to some manual action and consists of a stimulus and an expected reaction. & 3 &
\cite{Flemstrom2016,Flemstrom2018,Viggiato2023}
\\
\hline
=13 \rankspace Software product line configuration & A combination of specific features or components from a software product line to create a customized software product. & 3 &
\cite{Fischer2016,Henard2014,Xiang2021}
\\
\hline
=18 \rankspace Mutant & Small variations introduced to a software program using mutation operators that represent faults in the software. & 2 &
\cite{Shin2016,Shin2018}
\\
\hline
=18 \rankspace Requirements & The system specifications and requirements, where test cases linked to requirements are specified through traceability matrix. & 2 &
\cite{Arafeen2013,Masuda2021}
\\
\hline
=18 \rankspace Behavioural & Tests that differ in their outcome (\ie when one test pass, another fail e.g.~\cite{De2020}), or inputs that cover branches evenly (e.g.~\cite{Nguyen2022}). & 2 &
\cite{De2020,Nguyen2022}
\\
\hline
=21 \rankspace Test Report & Reports written in natural language and contain screenshots that show the results of tests. & 1 &
\cite{Feng2015}
\\
\hline
=21 \rankspace Social & Coverage data collected from similar users who utilise a web service. & 1 &
\cite{Miranda2014}
\\
\hline
=21 \rankspace Function & A process to perform a certain functionality given some inputs and possibly producing an output. & 1 &
\cite{Sondhi2022}
\\
\hline
=21 \rankspace Running Time & The runtime execution of test cases. & 1 &
\cite{Shimari2022}
\\
\arrayrulecolor{black}

    \bottomrule
\end{longtable}

\subsection{Input}
\label{subsec:input-diversity}

Input diversity measures use the distances between the test data to calculate a diversity value for test sets.
Inputs can be program inputs, or arguments passed to methods. The inputs can be numbers, strings, or complex
objects like tree-structured data. Several authors used input diversity to guide the testing
process~\cite{Bueno2007, Bueno2008,Shahbazi2015B,Boussaa2015,Yatoh2015,Almulla2020,Biagiola2019}.

Input diversity is used in search-based software testing techniques to guide the search process.
Bueno \etal~\cite{Bueno2007, Bueno2008} used Euclidean distance as a measure that characterizes the diversity of
a test set. The diversity of test set $T$ composed of $m$ test data is:
\begin{equation}
  Div(T) = \sum_{i=1}^{m} \sum_{j=1(j \neq i)}^{m} min(D(t_{i},t_{j}))
\end{equation}
where $min(D(t_{i},t_{j}))$ is the nearest test data $j$ to $i$, and the diversity-value of a test
set $T$ is the sum of distances between each test data in the set and its nearest neighbour.
Boussaa \etal~\cite{Boussaa2015} used the novelty score algorithm and suggested the use of diversity between methods' arguments of the
individuals of a population. The Novelty metric measures the distance between test suites in the population. They used
Manhattan distance for primitive types, and Levenshtein distance for string data types. The distance between two test
suites $t_{1}$ and $t_{2}$ is:
\begin{equation}
  \mathit{distance}(t_{1},t_{2}) = \sum_{i=1}^{m}\sum_{j=1}^{p} |t_{1}(M_{i}, P_{j}) - t_{2}(M_{i}, P_{j})|
\end{equation}
where $m$ is the number of methods in the test suite, and $p$ is the number of parameters in a method. The Novelty metric
for a solution ($S$) is given by:
\begin{equation}
  \mathit{NM}(S) = \frac{1}{k} \sum_{i=1}^{k} \mathit{distance}(S, \mu_{i})
\end{equation}
where $\mu_{i}$ is the $i^{th}$ nearest neighbour of the $S$ in the population, and $k$ is set to 20.

There has also been work that has used diversity metrics for strings. Shahbazi and Miller~\cite{Shahbazi2015B}
used string inputs as a basis for diversity and used six different string distance functions, which are
Levenshtein distance, Hamming
distance, Manhattan distance, Cartesian Distance, Cosine Distance, and Locality-Sensitive Hashing (LSH).
The experiments showed that the Levenshtein distance was the best in terms of fault-detection, followed by
Hamming distance and then Cosine distance. In terms of efficiency with respect to string size, the
runtime of Hamming, Manhattan, and Euclidean are linear and much better than Levenshtein or LSH. However, in
terms of efficiency with respect to different test sizes, LSH has a lower runtime with test set size larger than
$30$ and comparable runtime with test sizes lower than $30$. Also, the more the test size increases, the runtime
of LSH gets better than other metrics. In another study, that deals with numeric and string inputs,
Yatoh \etal~\cite{Yatoh2015} extended Randoop and used multiple pools simultaneously rather than a single pool to
increase input diversity. With different pools, three basic operations occur throughout the process. The basic operations
are selecting a pool to use, adding new pools, and deleting pools. The direction of the search is determined by the
selected pool, which is based on a scoring function. The score function is based on the achieved coverage of the pool
divided by the time consumed by that pool. Adding new pools with scores of $\infty$ strengthens the search by introducing
new content. Deleting old pools gives more time budget to the remaining pools when the number of pools exceeds a threshold.
The pool selected for deletion is based again on the coverage of the pool and time spent.

For some DL systems, images of the dataset are considered input to the DNN. Mosin \etal~\cite{Mosin2022} used input diversity
(i.e. diversity between images) for a similarity-based test input prioritisation approach. The similarity between inputs
(images) is calculated using the root mean square error. In their empirical study, they reported that similarity-based
approach significantly outperformed white-box and data-box approaches. However, in terms of APFD, the similarity-based
approach was less effective than white-box, and had a similar effectiveness to data-box approach.

\label{par:Feldt2016}
A new similarity metric was proposed by Feldt \etal~\cite{Feldt2016}, which used input diversity to evaluate the proposed approach.
They proposed a metric to calculate the diversity of sets of test cases considered
as a whole, rather than a metric just to calculate the diversity between a pair of test inputs, outputs, or
traces. They called it ``test set diameter'' (TSDm), and it extends the pairwise normalised compression distance to multisets. The term
``multiset'' is used rather than set because the same string might occur more than once in the collection.
However, this metric requires a higher computational cost to calculate compared to other similarity metrics.

Input diversity is used in fuzz testing to control the large number of tests generated~\cite{Pei2014},
to generate inputs that exercise different traces~\cite{Reddy2020}, or to create a diverse set of tests~\cite{Menendez2022}.
Also, input diversity was employed for multiple dispatch languages. Poulding and Feldt~\cite{Poulding2017} proposed use
of diverse test inputs in terms of values and types.

\subsubsection{Tree structured data}
\label{subsec:tree-data-diversity}

Tree-structured data are used in many systems like HTML and XML. Therefore, there is a need to have a similarity distance
function for such data structures. Shahbazi and Miller~\cite{Shahbazi2014} proposed a similarity metric for trees and named
it \emph{Extended Subtree} (EST). EST uses the edit base distance and generalizes it using new rules for subtree mapping.
They made an empirical study on three real data sets and one synthetic data set, and found that the proposed similarity
distance function was better than other distance functions in terms of runtime, accuracy (the number of correctly
clustered/classified trees over the total number of trees), and Weighted Average of F-measures (measured in terms of
recall and precision). In a later study,
Shahbazi \etal~\cite{Shahbazi2018} proposed a black-box test case generation technique of tree-structured programs (e.g. XML
parsers or web browsers) based on the diversity of a tree test set, where each test case is a tree. The proposed approach used
a Multi-objective genetic algorithm (MOGA) using a diversity-based fitness function and tree size as objectives. The
diversity-based fitness function is:
\begin{equation}
  F_{D} = \sum_{i=1}^{n} dist(t_{i}, \beta (t_{i}))
\end{equation}
where $n$ is the test set size, $t_{i}$ is the $i$-th test case in the test set, $\beta$ is the nearest test case to $t_{i}$, and
$dist(t_{i}, \beta (t_{i}))$ is the distance function. They used six tree distance functions in their experiments: Tree
edit distance (TED), Isolated subtree (IST) distance, Multisets distance, Path distance, and Extended subtree (EST) distance.
EST was reported to outperform the other distance functions.

Similarly, in another study, Shimmi and Rahimi~\cite{Shimmi2022} used the similarity of tree-structured data
to generate test cases for Java applications. The source code is transformed into an XML structured format
using a tool called srcML \footnote{https://www.srcml.org/\#home}, then used another tool called Triang
\footnote{https://relaxng.org/jclark/trang.html} to generate a meta-model of XML Schema Definition (XSD).
A group of methods with the same exact XSD structure are similar to each other. The approach would recommend
test cases of such similar methods to be adapted by the developers to cope with the changes in the software.
They used the Levenshtein distance between the test case recommended and the developer's changes to estimate
the effort for adapting the generated test cases. The evaluation showed that only $11\%$ distance between the
generated test cases and those written by developers, which suggests low efforts are spent to make the
required changes.

\subsection{Test Script}
\label{subsec:test-script-diversity}

Many studies applied \dbt techniques using the test script as a target for measuring the diversity. This is achieved
by considering the actual text of the test cases and measuring similarities between them. Mainly, many of these
papers which use test script diversity perform \tcp~\cite{Ledru2009,Ledru2012,Hemmati2015,Azizi2021,
  Miranda2018,Greca2022,Huang2017B,Carlson2011,Bertolino2015,Chen2021B,Haghighatkhah2018A,Haghighatkhah2018B} or
\tsr~\cite{Coviello2018A,Coviello2018B,Cruciani2019,Chetouane2020}. Since the test suite already exists,
and the target is to reorder them or minimise the test suite, it is intuitive to use the actual
text of the test cases. Ledru \etal~\cite{Ledru2009,Ledru2012}
suggested the use of string distances on the text of test cases. They considered the use of four distance metrics, which are
Euclidean distance, Manhattan distance, Levenshtein or Edit distance, and Hamming distance. They reported that Manhattan distance
was the best choice for the similarity metric for test case prioritisation. Hemmati \etal~\cite{Hemmati2015} implemented the
same algorithm used in~\cite{Ledru2012}, using the Euclidean and Manhattan distances, to work on manual black-box system testing.
Rogstad \etal~\cite{Rogstad2013} utilised test script diversity for test case selection in large-scale database applications.
They used four similarity metrics in the study: Euclidean distance, Manhattan distance, Mahalanobis distance, and normalised compression distance.
Burdek \etal~\cite{Burdek2015} proposed a white-box test suite generation approach for software product lines based on the reuse
of reachability information by means of similarity among test cases. Mondal \etal~\cite{Mondal2015} used a few similarity
measures in diversity-based test case selection, such as Hamming distance, Levenshtein distance, and Dice diversity formula.

The concept of program diversity was used by Tang \etal~\cite{Tang2022}, where the distance between two test Programs
$P_{1}$ and $P_{2}$ is:
\begin{equation}
  Dist(P_{1}, P_{2}) = GED(G_{1}, G_{2}) + d(P_{1},P_{2})
\end{equation}
where $G_{1}$ and $G_{2}$ are the control flow graphs of the test programs, $GED$ is the graph edit distance, and $d(P_{1},P_{2})$
is the Jaccard distance between the statements of the test programs. The calculations of $GED$ and $d(P_{1},P_{2})$ are given in
\cite{Tang2022}.

Miranda \etal~\cite{Miranda2018} used the string representations of the test cases for similarity-based black-box
prioritisation, and used Jaccard distance to measure the similarity between two test cases. While Cruciani \etal
\cite{Cruciani2019} applied test script diversity for test suite reduction. They modelled each test case as a point
in some D-dimensional space, and used Euclidean distance to select evenly spaced test cases. Shi \etal~\cite{Shi2015}
represented the test cases as nodes in a complete graph where the weights of the edges are the distance between each
pair or nodes.

\subsection{Output}
\label{subsec:output-diversity}

Output diversity or uniqueness is a black-box technique that minimises similarities between produced outputs. The key
idea is that two test cases that produce different kinds of output have a higher probability of executing two different paths.
Outputs derived from complete or partial execution of the program capture semantic information, which potentially makes
output diversity more effective in discovering more faults.  Alshahwan and Harman~\cite{Alshahwan2012, Alshahwan2014}
proposed the use of output diversity for web application testing aiming to minimise similarities between produced outputs.
Output uniqueness can be measured by calculating the difference between a new page and the previously visited pages
in terms of one or more of the web page components, which can be the appearance and structure of the web page or the
content (i.e. the textual data that the user can view). In their empirical evaluation, they reported a strong correlation
between output uniqueness and structural coverage, a strong correlation to fault-finding, and found many faults that white-box
techniques left undetected. In another study, Benito \etal~\cite{Benito2022} utilised output diversity for unit test data
generation. However, they repurposed the
XORSample' algorithm from sampling input diversity to sampling output diversity. The output domain is randomly partitioned into
cells of equal size and randomly selects one cell. Then an input is selected that is able to produce an output in the selected
cell, which in turn that input is the new test case.

Also, output diversity is used in Model-based testing in general and in Simluink models in particular.
Matinnejad \etal~\cite{Matinnejad2015A} proposed an algorithm for test case selection of state flow models based on output diversity.
The output-diversity algorithm aims to build a new test suite from the original test suite by maximising the output signal diversity
using the Euclidean distance to measure the diversity between two output signals. A signal $sg$ is defined as a function
$sg : \{ 0, \triangle t, 2.\triangle t, \dots , k.\triangle t \} \rightarrow R_{sg}$, where $\triangle t$ is the simulation step,
$k$ is the number of steps, and $R_{sg}$ is the signal range. Each signal is represented in a vector, and the Euclidean distance
between two vectors is the similarity measure between the two of them. They made a comparison between the proposed output-based
diversity and traditional coverage-based and input-based diversity approaches. They reported that output-based diversity selection
was better than other approaches in revealing faults. In a later study, Matinnejad \etal~\cite{Matinnejad2016} proposed a new
\emph{feature-based} diversity measure. The feature of an output signal is a distinguishing shape in the signal, and a set of these
features is stored in a feature vector. The Euclidean distance calculates the diversity between two feature vectors of two
output signals. Test data generation using feature-based diversity detected more faults than vector-based diversity. Later,
Matinnejad \etal~\cite{Matinnejad2019} used normalised Euclidean distance to calculate the diversity between output
signals in terms of vector-based and feature-based diversity.

Furthermore, output diversity is used in the automatic detection of boundaries. Dobslaw \etal~\cite{Dobslaw2020} discussed
Boundary Value Analysis (BVA) and proposed an algorithm for automatic detection
of boundaries using distance functions for detection, especially when the specifications are incomplete or vague. Also, they
proposed a Boundary Value Exploration (BVE) technique that utilise the software visualization to explore the behaviour of the
SUT and identify its boundaries. The idea behind boundary detection is to use the output diversity. Given two inputs $x_{1}$
and $x_{2}$ for a program $P$, running the $P$ with $x_{1}$ and $x_{2}$ will produce two outputs $o_{1}$ and $o_{2}$. The
program derivate is $\frac{d_{\mathit{output}}(o_{1},o_{2})}{d_{\mathit{input}}(x_{1},x_{2})} $, where $d$ is the normalised
compression distance and high-value
derivatives indicate boundaries. Dobslaw \etal used a component for handling dates in the Julia programming language as a case
study and were able to automatically detect boundary behaviour.

\subsection{Requirements}
\label{subsec:requirements-diversity}

Test cases can be linked to requirements through a traceability matrix and some researchers investigated using
software requirements as the target of diversity. Arafeen \etal~\cite{Arafeen2013} used requirements' diversity
by using $k$-means clustering~\cite{Hartigan1979} algorithm,
which uses Euclidean distance, to cluster the requirements, and using requirements-test cases traceability matrix, they
linked the test cases with each cluster. They used the code complexity metric (\emph{cm}) (explained in Table
\ref{table:other-metrics-summary}) to prioritise the test case in each cluster. Masuda \etal~\cite{Masuda2021} proposed a
syntax-tree similarity method based on natural language processing for
test-case drivability in software requirements. They pre-process the requirements to select test-case derivable sentences
using the syntax-tree similarity technique, then calculate the similarity between each sentence in the requirements and test-
case-derivable sentences. Based on the selected requirements, conditions and actions are derived. They evaluated the
method using three types of requirements written in different styles, and found an increase of $16\%$ in the accuracy of
the derived conditions and actions, from a similar method based on the tree Kernel technique.

\subsection{Program Internal Information}
\label{subsec:path-diversity}

Internal information about the structure of the software can be used as a diversity artefact, such as code, branches, or paths.
An interesting idea is to make use of the similarity of a tested program to new ones. Pizzoleto \etal~\cite{Pizzoleto2020}
introduced a framework to reduce the cost of testing a program using mutation testing. The basic idea is that given
a new program $p$ similar to a group of tested programs $G$, apply the same cost reduction strategy from $G$ on $p$.
Pizzoleto \etal~\cite{Pizzoleto2020} used syntactic similarity to determine program similarities and to identify which
program group is closest to a new program. The
similarity metrics used were ACCM (Average of Cyclomatic Complexity per Method) and RFC (Response for Class).
Guarnieri \etal~\cite{Guarnieri2022} implemented the framework, experimented with 20 possible configurations and used
Euclidean distance, Manhattan distance, Expectation Maximisation (EM) algorithm, Levenshtein distance, Jaccard Index,
normalised compression distance,
and a plagiarism detection tool called JPlag. Guarnieri \etal evaluated the framework on 35 small Java programs and were
able to predict the most effective mutation operators for untested classes.

\subsubsection{Coverage}
\label{subsec:coverage-diversity}

Code coverage can be the target for diversity. Coviello \etal~\cite{Coviello2018B} proposed a clustering-based
approach for test suite reduction based on their
statement coverage. They used Euclidean distance, Hamming distance, cosine similarity, Jaccard distance, Levenshtein
distance, and string Kernels-based dissimilarity to measure the similarity between test cases. Later,
Coviello \etal~\cite{Coviello2018A} made an experimental study using many test suite reduction approaches
where one of the approaches is based on the coverage similarity between test cases. They used  the same six
similarity metrics to cluster all similar test cases into one group, and then select the test case
covering the largest amount of test requirements. Zhang and Xie~\cite{Zhang2019} conducted an empirical study
for testing deep learning systems, where one of the diversities measured is the diversity in Neuron Coverage (i.e. the
number of activated neurons). Mahdieh \etal~\cite{Mahdieh2022} used coverage information of test cases as
a basis for grouping test cases together in clusters for test case priotrisation. The similarity
between test cases is calculated using the Euclidean distance, Manhattan distance, and Cosine similarity of
the coverage information of each test case. The empirical evaluation made on five projects from the Defects4J
dataset reported that the Euclidean distance performed better than Manhattan and Cosine in terms of average
first fail detected.

\subsubsection{Path}

A program path is the possible flow of control and data within a program without
executing it. The analysis is typically based on examining the program's source code, without considering actual
runtime behaviour or specific input values.
Xie \etal~\cite{Xie2009} proposed three algorithms for calculating the similarity between the test path and the
target path for path-oriented testing. The first algorithm is based on path matching, where it finds a sub-string
of the execution track that matches the longest prefix of the target path uninterrupted and returns the length of
the matching prefix as the similarity value. The second algorithm is based on node sequence matching, where it needs
to cover the statements in the target in order without regard to interruption. The length of the longest prefix in
the target path covering the statements is the similarity value. The third algorithm is based on the scatter graph method,
where the statements of the target path and execution track must match at the same positions. This is very close to the
concept of Hamming distance, where the similarity value is the number of matching statements between the two paths.
Panchapakesan \etal~\cite{Panchapakesan2013} used path diversity combining evolutionary strategy with evolutionary algorithm.
The control flow graph (CFG) of the software under test is given, and the similarity
between two paths $p_{1}$ and $p_{2}$ are calculated from the CFG using the following similarity function:
\begin{equation}
  s(p_{1}, p_{2}) = \frac{k - 1}{\max(|p_{1}|,|p_{2}|)}
\end{equation}
where $k$ is the largest common substring such that $1 \leq k \leq \max(|p_{1}|,|p_{2}|)$. Another fitness function used is the
average relative similarity defined as:
\begin{equation}
  s_{R}(p_{1}, p_{2}) = \frac{s(p_{1}, p_{2})}{\sum_{k=0}^{|\mathit{PATHS}|} s(p_{1}, p_{2}) + s(p_{1}, p_{2})}
\end{equation}
Cai \etal~\cite{Cai2021} used path diversity to guide a search-based algorithm. The path similarity is
determined by the approach level (AL) value. The smaller the AL value, the more similar two paths can be. The AL value between two
paths is the number of mismatched branch predicates between them.

\subsection{Execution}
\label{subsec:execution-diversity}

An execution path refers to a specific sequence of statements or instructions that are executed during the runtime of a program.
It represents the flow of control within a program, outlining the order in which statements are executed, branches are taken,
and conditions are evaluated.
Yoo \etal~\cite{Yoo2009} used test execution diversity for test case prioritisation by clustering test cases. The clustering
is performed based on the similarity of the execution traces of test cases. Each test case execution generates a string of
binary bits, where each bit represents a line in the software under test, with a value of one if the corresponding statement
has been executed or zero if it was not executed. Then, the distance between two traces of execution is calculated using the
Hamming distance. Wu \etal~\cite{Wu2012,Fang2014} aimed to use test execution diversity in similarity-based test case
prioritisation techniques.
An execution profile is defined as the ordered sequence of program elements sorted based on the execution count. They used
Levenshtein distance as a similarity metric to measure the similarity between two ordered execution sequences.

Another way of utilising execution diversity is in metamorphic testing.
Cao \etal~\cite{Cao2013} investigated the relation between the fault-finding capabilities of metamorphic relations (MRs) and
the diversity of test case executions in the hope of identifying good MRs (i.e. MRs that can reveal faults). MRs are some expected
properties of the target program. The dissimilarity of the test case executions is calculated by measuring the distance using
coverage Manhattan distance (CMD), frequency Manhattan distance (FMD), and frequency Hamming distance (FHD). In order to calculate
CMD, the branches are represented in a vector with values zero (uncovered) or one (covered), and then Manhattan distance is applied.
Whereas in FMD and FHD, the vector representing the branches contains the number of times these branches were executed, then
Manhattan or Hamming distances are applied. Their empirical study asserted the importance and significance of diversity of test
case executions to effective MRs to reveal faults. Xie \etal~\cite{Xie2022} proposed a \dbt prioritisation technique
of Metamorphic test case Pairs (MPs) for Deep Learning (DL) software. The diversity between MPs is determined
based on the execution of the internal states of DNNs using new diversity metrics suitable for DNNs. The diversity
metrics used are Kullback-Leibler (KL) divergence, Jensen-Shannon (JS) divergence, Wasserstein distance (WD), and
Hellinger distance (HD). The empirical evaluation showed that JS and HD were superior to other diversity metrics producing
the highest NAPVD.

\subsection{Semantic}
\label{subsec:semantic-diversity}

Software semantics refers to the meaning and interpretation of software \artefact{s}, such as programs, libraries,
frameworks, and other software components. Semantics can be static, which refers to the rules, constraints, and
specifications that govern the structure and correctness of software \artefact{s}. Also, semantics can be dynamic,
that focus on the runtime behaviour and execution of software components.
Lin \etal~\cite{Lin2017} presented a natural-language approach based on semantic similarity to test web applications.
The natural-language approach extracts the feature vector of the DOM elements, and then applies some transformations to represent
the feature vector with multidimensional
real-numbers vectors. Then each input field in the training dataset is labelled by topic. For a new DOM element, its feature
vector is extracted, the same transformations are applied, and the most similar vectors in the training dataset are determined
using the Cosine similarity metric. The topic of the new DOM element is inferred from the assigned labels. If the similarity of the
new DOM element is less than a threshold with the top five most similar vectors, a voting process (i.e. the majority of the topics
between the top five similar vectors) determines the topic. In case of multiple candidates still exist after the voting process,
one topic is selected randomly. They made experiments in input topic identification using 100 real-world forms, splitting
them into testing and training data, and found that using large training data, the performance is close to the rule-based approach,
while the accuracy improved by
$22\%$ when integrated with their approach. Also, they performed experiments for GUI state comparison using five datasets
from the industry. Their approach outperformed other approaches in three out of the five datasets.

Zhang \etal~\cite{Zhang2012} argued that a path of executed statements represents the semantics of
the program. Therefore, if two test inputs execute the same sequence of statements, then they are semantically equivalent, and
if not, they are semantically different. Therefore, semantic diversity or test execution diversity can be used as a way to guide
\dbt techniques. Zhang \etal~\cite{Zhang2012} suggested a new similarity metric
and referred to it as \emph{test case similarity} metric to measure the semantic similarity between test cases. It is based on the
normalised edit distance between two execution paths.

Error-revealing test cases can be stored in a library to be reused in the future. Retrieval methods can be based on
keywords, attributes-values, and other syntactic measures. However, these methods lack important semantic information.
Cai \etal~\cite{Cai2009} suggested the use of semantic diversity as a basis for test case reuse and retrieval. Semantic
similarity can be measured through semantic distance computed by four difference sets, which are super-concept difference
set $D_{Sup}$ (i.e. concept's ancestors), sub-concept difference set $D_{\mathit{Sub}}$ (i.e. concept's descendants), intension
difference set $D_{\mathit{Int}}$ (i.e. difference between data properties) and extension difference set $D_{\mathit{Ent}}$ (i.e. the difference
between concept and its instance). The semantic distance is the sum of the four difference sets factored by weights.

Machine translation systems are systems that do automatic translation of text and speech from a
source language to another target language, and like any other system, they must be tested properly. Usual
approaches detect mistranslations by examining the textual (e.g., Levenshtein distance) or syntactic
similarities between the original sentence and the translated sentence, but ignoring the semantic similarity.
Cao \etal~\cite{Cao2022} proposed an automatic testing approach for machine translation systems based
on semantic similarity. The sentences are transformed into regular expressions and use Levenshtein distance to
calculate the semantic similarity between two regular expressions, or transformed into deterministic finite automata
where Jaccard distance is used to calculate the semantic similarity between their regular languages.
Also, they proposed combining the two distance metrics into a hybrid one to make use of their advantages.
The proposed semantic-based similarity approach was shown to be $34.2\%$ more accurate than other machine
translation approaches.

\subsection{Context}
\label{subsec:context-diversity}

Context-aware Pervasive Software (CPS) collects and analyzes data from software and environment to guide the responses and actions of
the system. Test cases for CPS consist of sequences of context values, such as $\langle location, activity \rangle$ in smartphones
\cite{Wang2014}. Context diversity measures the number of changes in contextual values of test cases. A context stream is a
series of context instances taken at different time intervals. Each context instance is a series of variables, which consists of
\emph{value}, \emph{type}, and \emph{time}. Wang and Chan~\cite{Wang2009} introduced the use of context diversity paired with
coverage-based criteria for constructing a test suite. The context diversity (CD) of a context stream (cstream(C)) is defined as:
\begin{equation}
  \mathit{CD}(\mathit{cstream}(C)) = \sum_{i=1}^{n-1} \mathit{HD}(\mathit{ins}(C)_{i},\mathit{ins}(C)_{i+1})
\end{equation}
where $n$ is the number of context streams, $\mathit{ins}(C)_{i}$ is the $i$-th field of a context instance, and $\mathit{HD}$ is the Hamming
distance between two context instances. In a later study, Wang \etal~\cite{Wang2014} proposed three techniques to use context
diversity to improve data-flow testing criteria for CPS. Once more, context diversity is measured using Hamming distance. The
three proposed Context-Aware Refined Strategies (CARS) techniques are referred to as \emph{CARS-H}
(\textbf{H}igh context diversity), \emph{CARS-L} (\textbf{L}ow context diversity), and \emph{CARS-E} (\textbf{E}venly
distributed context diversity). They made a case study including 8,097 lines of code, 30,000 test cases, 959
mutants, and 43,200 adequate test suites for data analysis. The results showed that higher context diversity significantly
increased the effectiveness of data-flow testing for CPS. Effectiveness was improved by $10.6\%$ to $22.1\%$ using
CARS-H, while it decreased by $2.0\%$ to $22.2\%$ using CARS-L and CARS-E had a marginal effect.

In event-driven software (EDS), test cases are a sequence of events. Brooks and Memon~\cite{Brooks2009} presented a new
similarity metric, \emph{CONTeSSi(n)} (\textbf{CON}text \textbf{Te}st \textbf{S}uite \textbf{Si}milarity), that uses the
context of $n$ preceding events in test cases. Thus, $\mathit{CONTeSSi}(n)$ is a context-aware similarity metric, which is adapted
from the cosine similarity metric, with values in the range $[0,1]$, where the closer the value is to $1$, the more similarity
there is. $\mathit{CONTeSSi}(n)$ can be used in the contexts of test case prioritisation, test suite reduction, or test case
selection. Brooks and Memon used $\mathit{CONTeSSi}$ to reduce the size of a test suite as a case study.

\subsection{GUI}
\label{subsec:gui-state-diversity}

A GUI state is represented as a set of widgets that make up the GUI $W=\{ w_{1}, w_{2}, \dots, w_{n} \}$,
a set of properties for the widgets $P=\{ p_{1}, p_{2}, \dots, p_{m} \}$, and a set of values for the properties $V=\{ v_{1}, v_{2}, \dots, v_{n} \}$.
Feng \etal~\cite{Feng2012} used GUI state similarity as a basis for GUI testing. The GUI state similarity is the sum
of all similarity values of its windows. A window similarity is the sum of all similarities of its widgets. A widget similarity
is the similarity of its properties values. Suppose a widget $i$ has $n$ properties, the value of the $k$th property is $I_{k}$
with a weight $w_{k}$. If two states have the same widget $i$, then the similarity is:
\begin{equation}
  \mathit{Sim}_{\mathit{wid}}(\mathit{widget}_{i},\mathit{widget}_{i}^{'}) = \frac{1}{n} \sum_{k=1}^{n} w_{k} \times \mathit{Sim}_{p_{k}}(I_{k},I_{k}^{'})
\end{equation}
\begin{equation}
  \mathit{Sim}_{p_{k}}(I_{k},I_{k}^{'}) =
  \begin{cases}
    0 & I_{k} \neq I_{k}^{'} \\
    1 & I_{k} = I_{k}^{'}
  \end{cases}
\end{equation}
The similarity of a window $j$ consisting of $m$ widgets in two states is:
\begin{equation}
  \mathit{Sim}_{\mathit{win}}(\mathit{win}_{j},\mathit{win}_{j}^{'}) = \frac{1}{m} \sum_{i=1}^{m} \mathit{Sim}_{\mathit{wid}}(\mathit{widget}_{i},\mathit{widget}_{i}^{'})
\end{equation}
Finally, the similarity between the two states is:
\begin{equation}
  \mathit{Sim}_{\mathit{state}}(S,S^{'}) = \frac{1}{t} \sum_{j=1}^{t} \mathit{Sim}_{\mathit{win}}(\mathit{win}_{j},\mathit{win}_{j}^{'})
\end{equation}
where $t$ is the maximum of windows between $S$ and $S^{'}$. If a window $l$ does not exist in either one of the two states,
then $\mathit{Sim}(\mathit{win}_{l},\mathit{win}_{l}^{'})=0$.

Mariani \etal~\cite{Mariani2021} used GUI semantic diversity in their study. The semantic matching consists of four components:
Corpus of Documents (Component 1), Word Embedding (Component 2), Event Descriptor Extractor (Component 3), and Semantic Matching
Algorithm (Component 4)~\cite{Mariani2021}. They used the Jaccard Distance and Edit (Levenshtein) Distance to compute the semantic
similarities between source events and a set of target events, and reported that Component 4 is the most impactful on the quality
of the results. Xie and Memon~\cite{Xie2006B} investigated GUI state diversity for evaluating the GUI test suite. In this study,
the diversity of GUI states simply means using different GUI states.

\subsection{Model}
\label{subsec:model-diversity}

Graph models represent and analyze relationships between variables in a system. A graph model consists of a graph
structure composed of nodes and edges, where nodes represent variables or entities, and edges represent relationships
or dependencies between them. Some papers considered applying diversity on graph models, either diversity within a single
model, or diversity between a set of models. In the case of a single model, internal diversity is measured through the
number of shape nodes covered by the model. In the case of a set of models, the external diversity (i.e. diversity between
pairs of models) is measured through graph distance functions.
Semeráth \etal~\cite{Semerath2018,Semerath2020} proposed a number of shape-based graph diversity metrics to measure the model
diversity. Three distance functions used in the study are Pseudo-distance, Symmetric distance, and Cosine distance. They used
mutation testing to evaluate the test suite of the models in terms of mutation scores. They empirically evaluated the proposed
diversity metrics on $23439$ models in the context of test case prioritisation and found that there is a positive correlation
between diversity and mutation score. The study reported that using the proposed diversity metrics as a guide to generate diverse
test cases for model testing would yield a test suite with higher fault detection.

\subsection{Transition \& Trigger Diversity}
\label{subsec:transition-diversity}

In the context of model-based testing, the
test cases are considered to be paths traversing a Labelled Transition System (LTS)~\cite{Keller1976}. An LTS is a directed
graph defined in terms of states and labelled transitions between states to describe system behaviour. Formally, LTS is defined
as a 4-tuple $\langle S, L, T, s_{0}\rangle$, where $S$ is the set of states, $L$ is the set of labels, $T$ is the transition
relation, and $s_{0}$ is the start state such that $s_{0} \in S$. Using an LTS model, all possible test cases are
generated using depth-first search according to the \emph{all-one-loop-paths}, which means that paths with no loops, or paths
that loop once are considered. Coutinho \etal~\cite{Coutinho2016} proposed using similarity-based approaches to reduce
test suites in the context
of model-based testing and focused on LTS. The inputs used were 12 LTS specifications with unique configurations.
Test cases were generated for each specification, and a set of random failures were defined. The failures were
$10\%$ of the test suite size. They used all-transition-coverage as a test requirement. The reduction approach
calculates a \emph{similarity matrix}, which is the similarity degree between each pair of the test cases. They made
an analysis of the effectiveness of six different string distances to compute the similarity between the test
cases and applied them to three test subjects. The distance functions they used were the Similarity function,
Levenshtein distance, Sellers algorithm, Jaccard index, Jaro distance, and Jaro-Winkler distance. They concluded
that the choice of the similarity metric does not affect the size of the reduced test suite, but it affects the
fault coverage.
Cartaxo \etal~\cite{Cartaxo2007, Cartaxo2009} and Gomes de Oliveira Neto \etal~\cite{De2016} used transition diversity in similarity-based
test case selection. The similarity is calculated using the \emph{Similarity function} or identical transition similarity
(\emph{``It''}) measure mentioned in Table \ref{table:other-metrics-summary}. In a later study, Hemmati \etal
\cite{Hemmati2010A} developed the similarity-based test case selection using trigger diversity rather than only transition
diversity. They developed trigger-Based similarity (\emph{``Tb''}) measure explained in Table \ref{table:other-metrics-summary}.

Another paper used transition diversity in search-based software testing.
Asoudeh and Labiche~\cite{Asoudeh2013} used search-based GA and represented the entire test suite of the model under test as an
individual. The fitness function was defined based on transition diversity and calculated using the Levenshtein distance to measure
the similarity of the individual solutions.

\subsection{Mutants}
\label{subsec:mutants-diversity}

An interesting idea in mutation testing is to include the diversity of the mutants and tests in addition to number of mutants
killed. Shin \etal~\cite{Shin2016} proposed a new mutation adequacy criterion called \emph{distinguishing mutation adequacy
  criterion}, where the idea is that
fault-detection can be improved using stronger mutation adequacy criteria. Mutants can be distinguished by their killing
tests, and the proposed mutation adequacy criteria aim to distinguish as many mutants as possible plus killing mutants.
For example, if there exists three mutants $\{m_{1},m_{2},m_{3}\}$ in a program, and three test cases where $T1$ kills all
three mutants, $T2$ kills $m_{1}$ and $m_{2}$, and $T3$ kills $m_{3}$. Traditional criteria can select only $T1$ and this
can be adequate, while the \emph{distinguishing mutation adequacy criterion} will select $T2$ and $T3$ because it has more
diverse test cases. They made an empirical study using 45 real faults from $\mathit{Defects4J}$ showing that test suites adequate
to their proposed adequacy criterion have a $76.8\%$ increase in fault-detection capabilities than traditional criterion.
However, the size of the test suites is $1.67$ times more tests and $5.74$ times more selection time than test suites
generated for traditional criterion. This study used the distinguishing mutation adequacy criterion in the context of
test case selection, but it can be used in other contexts as well such as test data generation or test case prioritisation.
Shin \etal~\cite{Shin2018} made a more comprehensive empirical study using 352 real faults from $\mathit{Defects4J}$ on the
distinguishing mutation adequacy criterion. The results reported were similar to the previous study and showed $1.33$ times
more faults detected than the traditional mutation adequacy criteria with an increase of size by $1.56$ time more.

\subsection{Feature}
\label{subsec:feature-diversity}

Features of a software refer to the specific functionalities, capabilities, or characteristics that a software product or
application offers to its users. Features differ depending on the type of the application or the software under test.
Many researchers used the idea of feature diversity in many areas, such as testing Content-Based Image Retrieval (CBIR)~\cite{Nunes2015},
constructing meaningful search spaces~\cite{Joffe2019}, compiler testing~\cite{Chen2019C}, and SPL~\cite{Henard2013,Abd2019}.

CBIR is the process of finding content in a database most similar to an object image, which
extracts features from images and calculates the similarities of these features with the images stored in the database using
traditional distance functions. Nunes \etal~\cite{Nunes2015} made an empirical study of 10 similarity functions in CBIR in
the context of software testing systems with graphical outputs. The similarity functions used in the study were Euclidean,
Manhattan, Chebyshev, Canberra, Mahalanobis, Trigonometric, Modified Trigonometric, Statistic Value $\mathcal{X}^{2}$, Jeffrey
Divergence and Cosine. The similarity of two images $A$ and $B$ can be calculated using feature vectors of length $n$ extracted,
where $a_{i}$ and $b_{i}$ are the $i$-th feature of $A$ and $B$. The results of the empirical study showed that some similarity
functions are sensitive to a large difference in value in one of the extracted features. Another study that also uses image
datasets used feature diversity to test DNNs. Aghababaeyan \etal~\cite{Aghababaeyan2023} used a feature extraction model
to extract features from images and calculated the feature diversity using three metrics, which are the Geometric diversity,
normalised compression distance (NCD), and Standard Deviation. Their empirical study showed that NCD did not perform well
in measuring data diversity and
that Geometric diversity had the best correlation to fault-detection, and it significantly outperformed neuron coverage.
Similarly, Jiang \etal~\cite{Jiang2023} suggested an approach to generate test inputs by maximising feature diversity
calculated by Geometric diversity and evaluated it on four image datasets.

In compiler testing, a feature vector of a test program refers to a representation that captures the characteristics or features
of the program. It is a multidimensional vector that encodes specific attributes or properties of the program, which are relevant
for testing and evaluation purposes. Chen \etal~\cite{Chen2019C} used the idea of feature diversity for compilers and proposed
HiCOND. HiCOND uses the Manhattan Distance to measure the diversity between the feature vectors that represent the test programs.
HiCOND uses particle swarm optimization (PSO) to measure the diversity of the generated test programs based on their feature vectors.

Constructing a meaningful search space (fitness landscape) refers to the process of defining and organizing a set of possible
solutions or configurations in a way that effectively represents the problem domain and facilitates the search or exploration
of potential solutions. Joffe and Clark~\cite{Joffe2019} proposed an approach to construct meaningful search space using neural
networks (NNs). A set of execution traces and various properties are used to train the NN, and the outputs of the NN construct the proper
search space that is representative of the properties of interest. They used an auto encoder for diversity based on the similarity of
the data points salient features (i.e. features that are most important to differentiate one data point from another). They used
five real-world programs in their experiments and showed that the resulting search spaces are continuous, large, and representative
of various properties of interest.

In software product lines engineering, a feature model (FM) is a structured representation of the common and variable features
within a family of related software products. The model defines the presence or absence of features and allows for configuration
and customization of products by selecting or deselecting specific features.
Henard \etal~\cite{Henard2013} used mutation testing on FMs of a software product lines to
confirm that diverse test cases are more capable of detecting the mutants. They introduced erroneous variants of FMs using
mutation operators and then used a similarity measure to compare between two products evaluating the similarity degree of
a test suite ($S$). In this
context, a product consists of $n$ features of FM, and Jaccard Index can be used to measure the similarity degree between two
products based on selected features. A result of $1$ indicates that the two products are entirely different, while a result of
$0$ means they are identical. Then, the $S$ is ordered according to the similarity degree, such that the top test cases are the
most dissimilar to all other test cases in $S$. They made experiments using 12 real feature models of different sizes on
similar and dissimilar test suites and found that dissimilar test suites have more capability in detecting the FMs' mutants.
Abd-Halim \etal~\cite{Abd2019} proposed an enhanced string distance function for similarity-based test case prioritisation
in software product lines. The modified string distance is based on a hybrid between Jaro-Winkler and Hamming distance referred
to as Enhanced Jaro-Winkler (EJW). They used their modified string distance on multiple of different prioritisation and evaluated
them in terms of average percentage fault detection.

\subsection{Behavioural}
\label{subsec:behavioural-diversity}

Two tests that fail or pass together are said to have similar behaviour~\cite{De2020}.
Gomes de Oliveira Neto \etal~\cite{De2020} introduced measures to calculate the behavioural diversity of test cases, and referred to it as
\emph{b-div}. They used mutation testing to compare the executions and failure outcomes of the test cases to capture the
behavioural diversity. The focus of their study is the test outcomes and their patterns, which are collected
in a \emph{test outcome matrix} (TOM). The rows of the matrix represent the test cases, and the columns represent the
different mutants. The individual cells of the matrix are the test outcomes $(pass/fail)$ of the test cases when they are
executed against the mutants. The $b$-div is calculated
between the rows of the TOM in order to determine how much the tests differ in terms of their outcome. A confusion matrix of the
outcomes is made for each pair of tests. Two distance measures are calculated from the confusion matrix, which is accuracy
(acc) and Matthew's Correlation Coefficient (MCC). TP (true positive) occurs when a pair of tests fail together, TN (true negative) occurs when a pair of tests pass together,
FP (false positive) occurs when the first of the pair of tests pass and the second of the pair of tests fail, and FN (false
negative) occurs when the first of the pair of tests fail and the second of the pair of tests pass. For example, assume a pair
of tests $(t_{1}, t_{2})$, where $t_{1} = \{ 1, 1, 0, 0, 1 \}$ and $t_{2} = \{ 0, 0, 1, 0, 1 \}$, the classification is
$\{ FN, FN, FP, TN, TP \}$ ($0$ is a pass and $1$ is a fail). The MCC distance yields a value between $-1$ and $+1$,
where $+1$ represents an exact match and $-1$ represents a total difference. The
normalised distance is calculated by $d_{\mathit{mcc}} = \frac{(1-d_{\mathit{mcc}})}{2}$ to get values between $0$ and $1$.
Nguyen and Grunske defined behavioural diversity as covering many branches evenly with diverse inputs~\cite{Nguyen2022}.
They proposed a new metric to measure behavioural diversity, which is based on the Hill numbers ecology diversity metric.

\subsection{Test Report}
\label{subsec:test-report-diversity}

Crowdsourced testing tasks are performed, and the results are given in reports referred to as ``\emph{test reports}'', which are
written in natural language and contain screenshots. The test reports are inspected manually to determine their fault-finding value.
Feng \etal~\cite{Feng2015} proposed a diversity-risk-based prioritisation approach of the test reports for manual inspection and
called it \emph{DivRisk} strategy. The diversity aspect helps in removing duplicates and ensuring a wide range of reports are inspected,
while the risk aspect ensures that the more likely fault-revealing reports are inspected first. Keywords are extracted from the test
reports and placed in a \emph{Keyword vector} (KV) and a \emph{Risk vector} (RV), which serve as the basis for the risk assessment
and diversity calculations. Based on the KV, a distance matrix is created to represent the distance between each pair of the test
reports using Hamming distance. They conducted an empirical study on three projects evaluating the effectiveness of DivRisk, and found
that it greatly outperforms random prioritisation by $14.29\%$ to $34.52\%$ in terms of APFD. Also, the drop of APFD of DivRisk to the
ideal theoretical results was $7.07\%$.

\subsection{Test Steps}
\label{subsec:test-steps-diversity}

In system integration level testing, tests are represented as \emph{``test steps''}, where each test step correspond
to some manual action and a test step consists of a stimulus and an expected reaction~\cite{Flemstrom2018}. The similarity
between test steps can be calculated using any of the similarity distance functions. Flemstrom \etal~\cite{Flemstrom2016}
used the Levenshtein distance to measure the overlap between test steps. In a later study, Flemstrom \etal~\cite{Flemstrom2018}
used similarities between test steps to guide the ordering of test cases.
Viggiato \etal~\cite{Viggiato2023} used test steps similarity for test suite reduction. In order to group similar
test steps together, they used five text embedding techniques (Word2Vec, BERT, Sentence-BERT, Universal Sentence Encoder, and
TF-IDF), two similarity metrics (Word Mover's Distance and cosine similarity), and two clustering techniques (Hierarchical
Agglomerative Clustering and $k$-Means). Finally, they used the grouped test steps to find similar test cases using three different
similarity metrics (simple overlap, Jaccard, and cosine metrics).

\subsection{Software Product Lines Configuration}
\label{subsec:spl-configuration-diversity}

Software product lines (SPL) configuration refers to the process of selecting and combining specific features or components from
a software product line to create a customized software product that meets specific requirements or user preferences. It
involves choosing the desired features, their variations, and their configurations to create a tailored software solution.
Diversity between SPL configurations is used in many papers~\cite{Henard2014,Fischer2016,Xiang2021}.
In MBT, diverse test cases are found to be more effective in finding faults, and therefore it can be used to reduce the size
of the test suite, which is very suitable to use in the context of SPL due to its huge set of feature
models (FMs)~\cite{Henard2013}. Henard \etal~\cite{Henard2014} proposed a search-based approach for generating product
configurations for large SPL and a prioritisation algorithm for the set of product configurations. They used
the Jaccard distance to measure the similarity between two SPL configurations. In this approach, two configurations with a
distance value of $1$ are completely different, while a distance value of $0$ means they are identical. They made an empirical
study on 100 artificially generated and 14 real feature models from $t = 2$ to $t = 6$, comparing their approach with other
tools and showed that other tools failed in dealing with large SPLs, while their approach was effective and scalable. Also, they
reported that the similarity heuristic can replace $t$-wise.
Fischer \etal~\cite{Fischer2016} assessed using similarity in Combinatorial Interaction Testing
(CIT) for Software Product Lines with real faults. They used the approach by Henard \etal~\cite{Henard2014} on an open-source
web content management system called \emph{Drupal} as an industry case study. They analyzed how faults distribute among
features, and the effectiveness of similarity in detecting actual faults in comparison to classical $t$-wise coverage. Fischer \etal
reported that the similarity-based CIT approach is competitive with $t$-wise coverage, indicating that similarity is a
valid alternative metric for $t$-wise coverage. In another study, Xiang \etal~\cite{Xiang2021} investigated the relationship
between similarity measures and the t-wise coverage in SPL testing.
They reported a significant positive relationship and proposed the use of a novelty search (NS) algorithm for similarity-based
SPL testing to sample a set of products as dissimilar as possible. The novelty score was reported to have a higher positive
correlation with t-wise coverage than other traditional fitness functions used in GAs. They used the Jaccard distance to calculate
the novelty score of an SPL configuration. They made an empirical study on 31 feature models and showed how NS outperforms GA in
$77\%$ of the cases.

\subsection{Social}
\label{subsec:social-diversity}

Miranda \etal~\cite{Miranda2014} proposed a different coverage metric referred to as \emph{Social Coverage}, that uses coverage
data collected from similar users to customize coverage information. The goal is to identify a group of entities that are of interest
to the user automatically without the need to provide such information a priori. Each user has a list of services performed, and
the data coverage of these services is obtained through code instrumentation. In this work, the similarity between the target
user and all other system users is calculated using the Jaccard Index measure. Given a similarity threshold, any service below that
threshold is discarded, and the union of the remaining similar services serves as the targeted entities and the Social Coverage is
calculated as follows:
\begin{equation}
  \mathit{Social} \: \mathit{Coverage} = \frac{\text{number of covered entities}}{\text{number of targeted entities}}
\end{equation}
In traditional coverage, the denominator would be the total number of entities, which can contain a lot of unrelated entities.
They evaluated the Social coverage metric on a real-world application, and they were able to predict the entities of interest to
the user with an average percentage of $97\%$ and an average recall of $75\%$.

\subsection{Function}
\label{subsec:function-diversity}

If two libraries implementing the same functionality, and have two test suites, they can serve each other by
applying one test suite on the other library or adapting it to reveal undetected bugs. Sondhi \etal~\cite{Sondhi2022}
suggested exploiting the fact that many library projects overlap and have similar functionalities to use the associated
test suites on similar projects with some adaption lowering the cost.  They used cosine similarity distance to
measure the similarity between library functionalities. The empirical study showed that using this approach revealed
67 defects across 12 libraries.

\subsection{Running Time}
\label{subsec:runtime-diversity}

The idea of monitoring the runtime execution of test cases, and applying diversity metrics on runtime information
is only used by Shimari \etal~\cite{Shimari2022}. They proposed a clustering-based test case selection using
the execution traces of test cases for an Industrial Simulator. Runtime information of test cases was encoded
into vectors of integers and clustered together into groups using \emph{k}-means clustering algorithm. The Euclidean
distance was used to measure the similarity between these vectors. A case study was provided to compare the
proposed clustering-based approach against random selection, a variation of the approach that ignores the number
of occurrences, and the method used already in the industrial simulator. They found that the proposed approach
selects test cases with high coverage and a high diversity of execution time.

\subsection{Population}
\label{subsec:population-diversity}

Some studies implemented \dbt techniques in an Evolutionary algorithm, and in their
studies they managed to increase population diversity.

One issue found in evolutionary algorithms used for test data generation is the \emph{genetic drift} phenomena. The genetic drift
problem occurs when individuals become very similar to one another, limiting the diversity of the population, and causing early
convergence into suboptimal solutions.
Xie \etal~\cite{Xie2005} proposed a dynamic self-adaptation strategy for evolutionary structural testing referred to as \emph{DOMP}
(\textbf{D}ynamic \textbf{O}ptimization of \textbf{M}utation \textbf{P}ossibility). The approach analyzes the evolution process and
checks for premature convergence, then sharply increases the mutation probability to increase the diversity of the population. DOMP
has a parameter to control the frequency of observations to analyze the population. In every observation, another parameter controls the
number of individuals to check whether the proportion of gene-type exceeds a threshold value, and the mutation probability is adjusted
to an enhanced mutation probability controlled by the tester. The new mutation probability will persist for a number of generations
defined by the tester until the diversity of the population is restored.
Alshraideh \etal~\cite{Alshraideh2011} proposed the use of multiple populations rather than a single
population keeping the solutions diverse, and preventing the process of falling into a local optima.
Kifetew \etal~\cite{Kifetew2013} proposed an orthogonal exploration of the input space
approach to introduce diversity in the population to counter the genetic drift problem. The diversity of the population is
increased by adding individuals in orthogonal directions via Singular Value Decomposition (SVD). Three variants of SVD-based GA
have also been introduced, which are history-aware orthogonal, reactive exploration, and a combination of the two. They evaluated
their approach on 17 Java classes and compared it with traditional GAs, and showed that coverage has increased in $47\%$ of the
cases, and when effectiveness was the same, the efficiency has increased in $85\%$ of the cases.
Panichella \etal~\cite{Panichella2015B} proposed DIV-GA (\textbf{DIV}ersity based \textbf{G}enetic \textbf{A}lgorithm), that
inject new orthogonal individuals to increase diversity during the search process for test case selection. DIV-GA uses an orthogonal
design to generate an initial diversified population and then uses SVD to maintain diversity through the search process.
Almulla and Gay~\cite{Almulla2020} proposed an approach to increase diversity in test data
generation referred to as adaptive fitness function selection (AFFS). AFFS changes the fitness functions used during the generation
process to increase diversity. They used the Levenshtein distance as a fitness function and as a target for reinforcement learning.
They evaluated AFFS on 18 real faults from a real-world project called Gson and found AFFS created more diverse test suites than
using static fitness functions. However, the increase in fault-finding capabilities is very small and inconclusive.

An alternative way of maintaining population diversity is through using multiple subpopulations and ensure that
individuals with high diversity migrate between the subpopulations. Wang \etal~\cite{Wang2023} used this idea
for web application testing and proposed a parallel evolutionary test case generation approach.
With multiple subpopulations being evolved, the migration of individuals from one subpopulation to another
keeps the diversity high. The diversity of an individual is calculated using the Levenshtein distance to all
individuals in the target subpopulation, and the individuals with the farthest distances are selected for migration.

\subsection{Hybrid}
\label{subsec:hybrid-diversity}

Approaches that use a combination of different \artefact{s} as targets for applying diversity are
referred to as hybrid diversity approaches. Feldt \etal~\cite{Feldt2008} proposed a model for
test variability referred to as VAT (VAriability of Tests), where they identified $11$ variation
points in which tests might differ from each other including ``test setup'', ``test invocation'',
``test execution'', ``test outcome'', and ``test evaluation''. They suggested using normalised compression distance to measure
the diversity of the VAT trace. In another study, Albunian \etal~\cite{Albunian2017,Albunian2020}
defined three diversity measures based on fitness entropy, test executions,
and Java statements. The first two referred to as \emph{phenotypic diversity} are behavioural or semantic
diversity, and the last one is referred to as \emph{genotypic diversity} is a structural or syntactic diversity. They
measure behavioural similarity by constructing a vector for each test case representing the number of execution times
of each statement. Then, Euclidean distance is used to calculate the similarity between two vectors. In order to measure
the syntactic similarity between test cases, statements are normalised first by replacing the variable name with a placeholder,
then constructing a vector representing the number of occurrences of each statement. The Euclidean distance is again used
to calculate the syntactic similarity.

Liu \etal~\cite{Liu2017} used three test objectives in a search-based algorithm to increase test suite diversity
and minimise size. The fitness functions used to guide the search process are referred to as coverage dissimilarity,
coverage density, and number of dynamic basic blocks. The \emph{coverage dissimilarity} increases diversity between
test executions and uses Jaccard distance to measure it. The \emph{coverage density} is an adaptation of the test
coverage density which is the average size of test execution slices related to every output over the static backward
slice of the output. High-density values indicate that large parts of the model are covered.
In a later study,
Liu \etal~\cite{Liu2019} added a new test objective based on output diversity. The \emph{Output Diversity}
aims to maximise the diversity between the output signals and uses a normalised Manhattan distance to measure the output
diversity between two output signals. Also, Arrieta \etal~\cite{Arrieta2019} proposed using input and output diversity
for Simulink models. They used the Euclidean distance as a basis for the two similarity metrics on input and output signals.

Ma \etal~\cite{Ma2018} used hybrid static and dynamic diversity metrics. The static diversity metric is based on
Levenshtein distance, which makes sure that there is diversity in the program structures of
the test cases. The other dynamic diversity metric enforces diversity in the test cases' ability to
cover untested thread schedules. This dynamic diversity value is the number of
interleaving instances that are discovered by the new test case, and not exposed by the
previously executed test cases. The static diversity metric helped in showing more interleaving
instances but with a high processing time. The dynamic diversity metric reduced the number of required
active tests for revealing interleaving instances with high performance.

Marchetto and Tonella~\cite{Marchetto2009} applied hybrid diversity for testing Ajax Web applications.
Three fitness functions based on test diversity are \emph{EDiv}, \emph{PDiv}, and \emph{TCov}. EDiv is
based on the execution frequency of each event in the finite state machine (FSM). PDiv is based on the
execution frequency of each pair of semantically interacting events. TCov is based on the FSM coverage
by the test cases. Biagiola \etal~\cite{Biagiola2019} used input diversity and navigation diversity for
web testing. A test suite that covers the web application is generated using the distance metric between
the test actions and input data, maximising the diversity of navigation sequence and input data. The distance
metric used is composed of two parts. These are the difference between the action sequences in the two test
cases, and the difference in input values utilised by the identical actions in those sequences.

\begin{tcolorbox}[title=Conclusions --- \rqartefacts]
  \dbt techniques used different software \artefact{s} as a basis for the similarity calculations.
  These \artefact{s} included input, output, test script, test execution, etc. The most used \artefact{} reported
  in the literature was test scripts with $18.2\%$ of the papers reported in this study. After that, $14.9\%$ of the
  collected papers used inputs as a basis for their approaches. Over one-third of the \dbt techniques
  used either input or test script diversity, because both of them are simple to use as they are usually
  obtained without the need to instrument, run the tests, or build
  any models of the software under test, making them very appealing to use by testers.
\end{tcolorbox}
\section{\rqproblems}
\subsection*{\textbf{\rqproblemsexpanded}}
\label{sec:rqproblems}

\dbt techniques have been used in solving software testing problems, such
as test data generation, test case prioritisation, etc. We explore the various studies made
regarding these problems and discuss the impact of \dbt techniques upon them.

\begin{figure}[tbp]
    \centering
    \includegraphics[width=0.6\columnwidth]{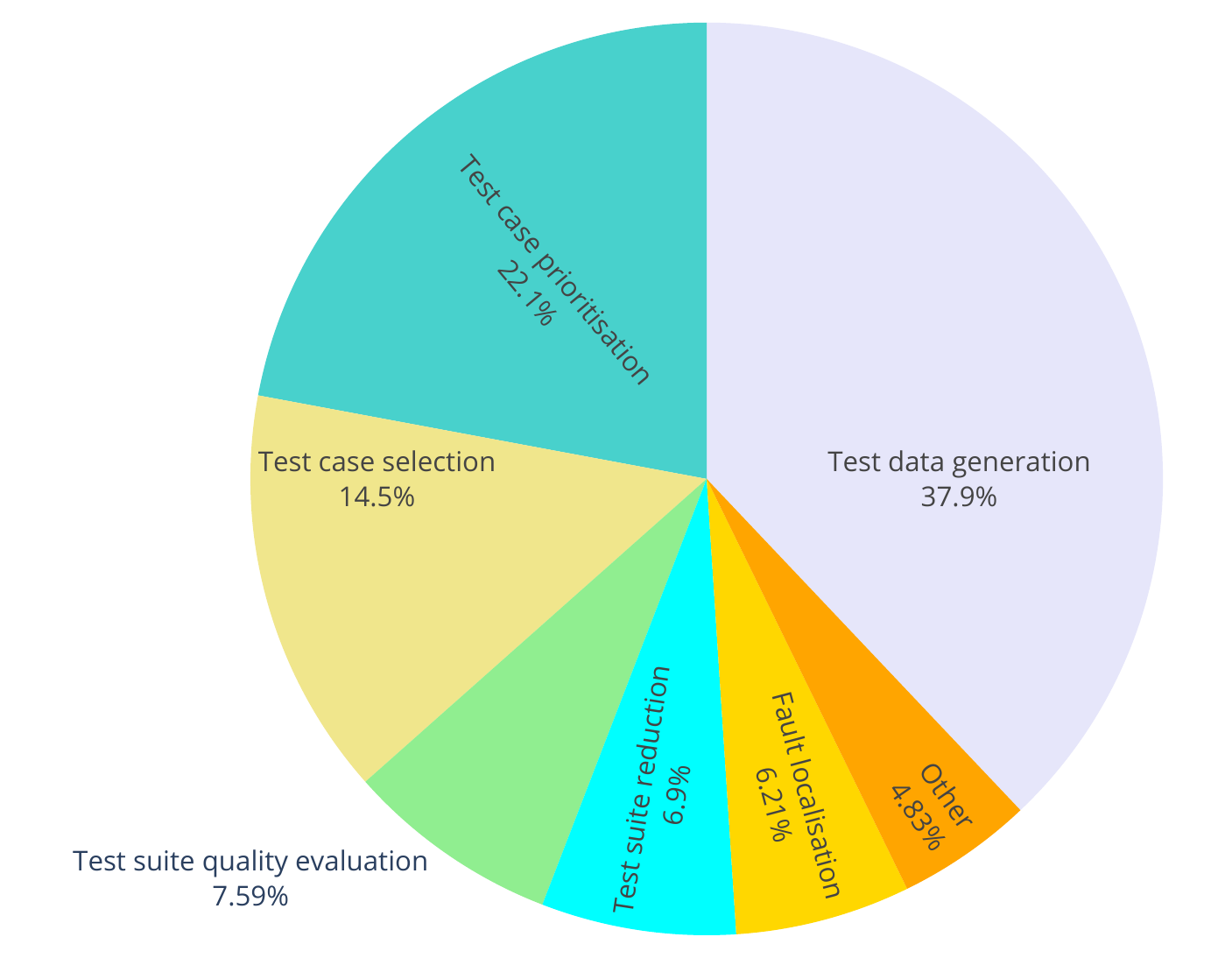}

    \vspace*{-1em}\caption{
        \label{fig:probelms-distribution}
        Distribution of \dbt papers addressing problems in software testing.
    }
\end{figure}

Table \ref{table:problems-summary} presents a list of the software testing problems addressed
using \dbt techniques, with a short description of the problem, the total number of papers solving them,
and citation of all these papers.
There are \numtdg studies focussing on \tdg, \numtcp concerning \tcp,
\numtcs studies about \tcs, \numtse studies regarding \tse, \numtsr studies on \tsr and \numtfl
studies with regard to fault localisation. Figure~\ref{fig:probelms-distribution} shows the distribution
of the papers tackling software testing problems. As it shown from the figure, $60\%$ of the collected
\dbt papers deal with \tdg and \tcp. Out of the \numproblems testing problems discussed, almost
three-quarters of the \dbt techniques are applied in \tdg, \tcp, and \tcs.
There are seven other papers that
introduced new similarity metrics~\cite{Feldt2016,Miranda2014}, investigated test reuse in software product
lines~\cite{Cai2009,Burdek2015}, utilised output diversity in boundary value testing~\cite{Dobslaw2020}, used \dbt in test adaptation~\cite{Sondhi2022},
and argued for teaching diversity principles in software testing~\cite{Chen2011}. The paper by Matinnejad \etal~\cite{Matinnejad2019}
addressed both \tdg and \tcp where it is cited in both problems in Table \ref{table:problems-summary}.

\subsection{Test Data Generation}
\label{subsec:test-data-generation}

Test data generation is the process of creating data to test a software program
according to some adequacy criteria. It is the most researched area where \dbt techniques
have been proposed.

\begin{table}[htbp]
    \centering
    \caption{
        \label{table:problems-summary}
        A list of the software testing problems addressed using DBT techniques.
    }
    \begin{tabular}{lp{5cm}@{~~~}rp{4cm}}
        \toprule
      
        Rank/Problem & Description & Total & Papers
        \\

        \midrule

        % The data in this file should be automatically output
% from code that processes raw results data.

\nonequalrankspace 1 \rankspace \Ctdg & The process of creating data to test a software program according to some adequacy criteria. & \numtdg &
\cite{Albunian2017,Albunian2020,Almulla2020,Alshahwan2012,Alshahwan2014,Alshraideh2011,Asoudeh2013,Benito2022,Biagiola2019,Boussaa2015,Bueno2007,Bueno2008,Bueno2014,Cai2021,Cao2022,Chen2019C,Feldt2008,Feng2012,Fischer2016,Guarnieri2022,Henard2013,Henard2014,Jiang2023,Joffe2019,Kifetew2013,Lin2017,Ma2018,Marchetto2009,Marculescu2016,Mariani2021,Masuda2021,Matinnejad2016,Matinnejad2019,Menendez2021,Menendez2022,Nguyen2022,Panchapakesan2013,Pizzoleto2020,Poulding2017,Reddy2020,Scalabrino2016,Shahbazi2014,Shahbazi2015B,Shahbazi2018,Tang2022,Vogel2019,Vogel2021,Wang2009,Wang2014,Wang2023,Xie2005,Xie2009,Yatoh2015,Zhang2012,Shimmi2022}
\\
\arrayrulecolor{lightgray}\hline
\nonequalrankspace 2 \rankspace \Ctcp & The process of re-ordering test cases so that the tester gets maximum benefit in the case testing is prematurely stopped at some arbitrary point due to some budget constraints. & \numtcp &
\cite{Abd2019,Arafeen2013,Azizi2021,Bertolino2015,Carlson2011,Chen2021B,De2020,Fang2014,Feng2015,Flemstrom2018,Greca2022,Haghighatkhah2018A,Haghighatkhah2018B,He2015,Hemmati2015,Huang2017B,Ledru2009,Ledru2012,Mahdieh2022,Matinnejad2019,Mei2013A,Mei2013B,Miranda2018,Mosin2022,Noor2015,Semerath2018,Semerath2020,Wang2015,Wu2012,Xie2022,Yoo2009,Zhao2015}
\\
\hline
\nonequalrankspace 3 \rankspace \Ctcs & The process of selecting a subset of the current available tests that are relevant to some set of recent changes. & \numtcs &
\cite{Aghababaeyan2023,Arrieta2019,Cartaxo2007,Cartaxo2009,De2012,De2016,De2018,Farzat2010,Hemmati2010A,Hemmati2010B,Hemmati2010C,Hemmati2011,Hemmati2013,Matinnejad2015A,Mondal2015,Panichella2015B,Rogstad2013,Shimari2022,Shin2016,Shin2018,Zhao2022}
\\
\hline
\nonequalrankspace 4 \rankspace \Ctse & The process of judging on the quality of the current test suite and giving insight to testers about certain properties which can guide the tester to improve or fix some issues in the test suite. & \numtse &
\cite{Cao2013,Flemstrom2016,Leveau2020,Neto2018,Nikolik2006,Nunes2015,Shi2015,Singh2013,Xiang2021,Xie2006B,Zhang2019}
\\
\hline
\nonequalrankspace 5 \rankspace \Ctsr  & The process of finding redundant test cases in the test suite and discarding them in order to reduce the size of the test suite. & \numtsr &
\cite{Beena2014,Brooks2009,Chetouane2020,Coutinho2016,Coviello2018A,Coviello2018B,Cruciani2019,Kichigin2009,Pei2014,Viggiato2023}
\\
\hline
\nonequalrankspace 6 \rankspace Fault localisation & The process of identifying the locations of faults in a program. & 9 &
\cite{Gong2012,Hao2005,Hao2008,Liu2017,Liu2019,Xia2016,You2013,Zhao2011,Zhao2013}
\\
\hline
=7 \rankspace New Metric & Introducing a new similarity metric that can be used to solve any of the software testing problems. & 2 &
\cite{Feldt2016,Miranda2014}
\\
\hline
=7 \rankspace Test Reuse & Resuing part or the whole of a test suite on other systems. & 2 &
\cite{Cai2009,Burdek2015}
\\
\hline
=9 \rankspace Test Adaptation & The process of identifying a test suite from another program that can be adapted to be used on the SUT. & 1 &
\cite{Sondhi2022}
\\
\hline
=9 \rankspace Boundary value analysis & A software testing technique in which tests are designed to include representatives of boundary values in a range. & 1 &
\cite{Dobslaw2020}
\\
\hline
=9 \rankspace Teach DBT & Teaching software testing methods based on diversity principles. & 1 &
\cite{Chen2011}
\\
\arrayrulecolor{black}

        \bottomrule
    \end{tabular}

\end{table}

\subsubsection{Search-Based Approaches}

Search-based approaches transform the testing objectives into search problems and solve these
problems using an evolutionary process. The evolutionary process
can use a single solution, such as hill climbing, or can use a population of individuals, such as
genetic algorithms, that go through multiple iterations until the stopping criteria are met.

Researchers have applied diversity to construct individuals with higher fitness values in search-based approaches.
Bueno \etal~\cite{Bueno2007, Bueno2008} employed a metaheuristic algorithm to
evolve ``randomly generated test sets'' (RTS) towards ``diversity oriented test sets'' (DOTS),
that aims to meticulously cover the
input domain. They identified metaheuristics that can be utilised to generate DOTS and proposed an
algorithm called ``Simulated Repulsion'' (SR). SR begins with RTS and their
diversity is enhanced to
evolve to DOTS by simulating systems of particles contingent upon electric repulsion
forces.
Bueno \etal~\cite{Bueno2014} went on to implement that measure in a test data generation technique.
They used simulated annealing (SA) and genetic algorithms (GA) along with their SR algorithm for
generating DOTS. They found that SR outperforms GA, which in turn outperforms SA. Their proposed
SR found greater diversity values for the sets than GA and SA in a lower number of iterations. However,
the computational complexity of SR is higher than both SA and GA. Further experiments were made to
compare between DOTS and RTS, and found that DOTS increased the coverage and mutation score up to
$30\%$. However, for small test sizes, RTS can perform better than DOTS.
Boussaa \etal~\cite{Boussaa2015}
used a Novelty Search (NS) algorithm for test data generation using statement coverage as an evaluation
criterion, exploring the search space without any regard for an objective. Rather than selecting test cases
based on their coverage capabilities, the test cases are selected based on their ``diversity'', or based
on a novelty score of how different the test cases are compared to all other
solutions.
This helps avoid
premature convergence that GAs suffer from. They applied their initial implementation to the
Apache Commons Math library,
but did not report any results, and no empirical study was made to evaluate the algorithm.

Other approaches use fitness functions based on diversity metrics to guide the search-based
algorithm in generating a diverse test suite. Shahbazi and Miller~\cite{Shahbazi2015B} investigated
black-box testing methods where the input value is a string and used two fitness
functions to guide the test case generation. The first fitness function was to control
the diversity of the test cases referred to as $F_{D}$, and the other fitness function was to control
the length distribution of the string test cases referred to as $F_{L}$, and they employed a Multi-Objective
Genetic Algorithm (MOGA) using both fitness functions to be minimised.
Shahbazi and Miller conducted an experimental study on 13 real-world programs and showed
that test generation sets using MOGA achieved higher mutation scores than RT for all subjects, indicating
that increasing the diversity of the test cases leads to better fault-detection rates.

Some studies investigated the effects of population diversity in order to check if it helps in generating
test suites achieving higher statement and branch coverage~\cite{Albunian2017}.
Albunian \etal~\cite{Albunian2020} investigated the effect of holding the population diversity on
the Many-Objective Sorting Algorithm (MOSA) for test data generation. Three diversity measures (two behavioural
or semantic diversity and one structural or syntactic diversity) were used to guide the generation of test data,
and used fitness sharing technique to maintain the population diversity. The experiments reported that
maintaining diversity does not affect coverage, but may help prevent the search process from stagnating.
Alshraideh \etal~\cite{Alshraideh2011} proposed using multiple-populations rather than a single population
to maintain the diversity of the solutions and to prevent the search process from being stuck in a local optima.
They made an empirical evaluation performed on eight JScript functions and showed that
the multiple-population outperformed the single-population in terms of the number of executions required for branch
coverage, execution time, efficiency, and search effectiveness.

\label{pap:Feldt2008}
An alternative approach to assessing the diversity of a set of test inputs or the population diversity
is to compare the resultant program execution traces. Feldt \etal~\cite{Feldt2008}
proposed a model for test variability referred to as VAT (VAriability of Tests), which
we discuss in Section \ref{subsec:hybrid-diversity} (p.~\pageref{subsec:hybrid-diversity}).
The generated VAT trace of a test consists of
all the values from the execution of a test with all the variation points in the
VAT model.
They implemented the tests in Ruby and used a Ruby function to attach into the program interpreter and
trace the execution of new lines. This function saves information including the line number, the code on that line,
and all the variables' values that are accessed on this line of code. The information is saved in a
file as a VAT trace, where it can be used later to calculate the diversity between the different traces.
They used the triangle classification problem to demonstrate their model and compared it with three human
subjects where they showed that their proposed model found many of the same clustering made by the
three persons.

A few more studies explored path diversity in evolutionary approaches.
Xie \etal~\cite{Xie2009} used an evolutionary testing approach for test data generation for path-oriented
testing. The fitness function is based on the similarity between the execution track and the target path. Given a
target path $\langle p_{0}, p_{1}, \dots p_{n-1} \rangle$ and an execution track of a test case $\langle t_{0}, t_{1}, \dots t_{m-1} \rangle$,
the distance between the test and the target is calculated as follows:
\begin{equation}
  \mathit{distance}(\mathit{test}, \mathit{target}) = n - \mathit{similarity}(\mathit{test}, \mathit{target})
\end{equation}
where $n$ is the length of the target path, and $\mathit{similarity}(\mathit{test}, \mathit{target})$ is the similarity between the two paths
calculated by some distance function; e.g., Hamming distance. The empirical study showed that the proposed approach
outperformed other traditional techniques for long and complicated paths, while
it did not perform well in short paths.
Panchapakesan \etal~\cite{Panchapakesan2013} proposed a white-box test data generation combining evolutionary
strategy with an evolutionary algorithm based on path diversity. The control flow graph (CFG) of the software under
test is given, and the similarity between the two paths is calculated from the CFG. They evaluated their approach on
three case studies and found that the proposed method outperformed GA in terms of runtime while covering all feasible
paths.
Cai \etal~\cite{Cai2021} proposed a search-based test data generation algorithm using path coverage criterion
called the ``binary searching iterative algorithm''. It is based on the idea that similar test cases usually cover
similar paths. The proposed algorithm finds the closest (\ie most similar) uncovered path to a discovered path
to it, and then it searches for inputs left and right of the region of the already discovered path. This kind of
``binary'' searching finds inputs to cover the target path quickly. They evaluated the proposed algorithm on 12
programs and found it outperforms other techniques generating less number of test cases in 11 out of 12 cases.

\subsubsection{Fuzz testing}
\label{sec:fuzz-testing}

Fuzz testing, also known as fuzzing, is a software testing technique that involves feeding unexpected,
random, and invalid inputs to a program in order to discover bugs, vulnerabilities, and other issues that
might not be found through traditional testing methods. Fuzz testing is reported to improve the
robustness and security of software~\cite{Klees2018, Tsankov2012}, and is widely used in industries such as cybersecurity, software
development, and quality assurance. By generating a large number of test cases, fuzz testing helps
uncover errors and weaknesses that might go unnoticed in traditional manual or automated testing,
thereby helping to enhance the overall quality and reliability of software products.
Zhang \etal~\cite{Zhang2012} proposed a fuzzing approach that starts by using a similarity-based black box
fuzzing for test data
generation. Then, a subset of the generated test data is selected based on the similarity metric. Finally, new
inputs are generated from the selected subset using combination testing.

Other researchers used fuzz testing for property-based testing, which is a software testing
approach that focuses on checking the validity of general properties or invariants of a system.
Instead of writing specific test cases, property-based testing uses properties, which are
high-level assertions or conditions that should hold true for a wide range of inputs.
Reddy \etal~\cite{Reddy2020} proposed a black-box test data
generation approach of diverse valid test inputs quickly in the context of property-based testing.
They used reinforcement learning (RL) to guide and control the random input generator. The proposed
approach was evaluated against random test data generation and evolutionary algorithms using four
real-world benchmarks. They found that the proposed approach generates more diverse inputs
given the same time budget as other approaches. In another study, Nguyen and Grunske~\cite{Nguyen2022}
aimed to increase behavioural diversity in addition to having large branch coverage. They
developed a feedback-driven fuzzing technique for generator-based fuzzers. Nguyen and Grunske
evaluated their approach to Ant, Maven, Rhino, Closure, Nashorn, and Tomcat. They found better
behavioural diversity compared with other fuzzers like Zest~\cite{Padhye2019} and RLCheck~\cite{Reddy2020}.
Menendez \etal~\cite{Menendez2022} proposed an approach called \emph{HashFuzz} that combines coverage and diversity
using hash functions to create a diverse set of tests for C programs improving system testing. The key idea is to
transform the program by adding new branches right after the input, which forces the input to be chosen from each
partition from the input space to produce near-uniform test sets. In order to find inputs from different partitions
of the input space, they used the XOR hash family in the transformation to limit the fuzzer. They evaluated their
approach on eight programs from the Google Fuzzer Test Suite using four other fuzzers: AFL, FastAFL, FairFuzz, and
LibFuzzer. They showed an improvement in diversity leading to a slight increase in branch coverage and a significant
increase in crash detection between $28\%$ to $97\%$.

\subsubsection{Model-based testing}
\label{subsubsec:model-based-testing}

Researchers have also applied diversity in model-based testing.
Asoudeh and Labiche~\cite{Asoudeh2013} proposed a genetic algorithm for
test data generation of extended finite state machines (EFSMs) to maximise coverage and test case
diversity and minimise overall cost. They represented the entire test suite as a chromosome, where each gene
represents a sequence of transitions of the state model. The fitness function guides the search, and it is defined
based on the similarity of the individual solutions.

In Cyber Physical Systems (CPSs), models have time-continuous behaviour. Therefore, test inputs and outputs are
defined as continuous signals representing values over time. Matinnejad \etal~\cite{Matinnejad2016} proposed
an approach for test data generation of Simulink models. The proposed approach aims to maximise diversity in output signals rather than maximising structural coverage,
on the basis that test cases that yield diverse output signals are more likely to reveal different types of faults~\cite{Matinnejad2016}.
They evaluated the test data generation approach on two industrial and two public-domain Simulink models. The proposed approach
is found better in terms of fault-detection than random generation. In a later study, Matinnejad \etal~\cite{Matinnejad2019}
modified the test data generation approach to be compatible with continuous behaviours, to fix unrealistic
assumptions about test oracles, and more scalable using meta-heuristic search. Also, they proposed a test case
prioritisation approach based on a combination of test coverage and test suite output diversity
\cite{Matinnejad2019}. The empirical study made on two industrial Simulink models showed that the proposed
test data generation and prioritisation approach is better in terms of fault-detection and decision coverage than random testing and
coverage-based approaches.

Deep Learning (DL) software has become widely used in many applications, and few works applied \dbt
techniques to generate test inputs for DL software. Jiang \etal~\cite{Jiang2023} proposed an approach
for generating valid and diverse test inputs for a testing task-specific DNN. The empirical evaluation
using four image datasets compared the proposed approach to greedy strategy, coverage-guided, and random
method. The experiments showed that the proposed approach increased the number of generated valid test
inputs by $36\%$ to $100\%$, increased the average diversity of generated data by $0.29\%$ to $2.8\%$,
and the average generation time decreased by $15.7\%$ to $47\%$ compared to the other generation methods.

\subsubsection{Focused testing}

Sometimes it is important to generate a test suite that focuses on a specific part of the program under test. This
part could be new code, or modified code. Also, we might want to focus on comparing new and previous functionality.
This kind of testing is referred to as \emph{focused testing}. The main issue with focused testing is that the inputs are not diversified enough.
Menendez \etal~\cite{Menendez2021} proposed an approach named \emph{Diversified Focused Testing} (DFT) that uses
a search technique generating a diverse set of tests to cover a given program point ($\mathit{pp}$) repeatedly to reveal faults
affecting the given $\mathit{pp}$. DFT analyses the program constraints required to reach $\mathit{pp}$,
selects the appropriate comparison operators for these constraints, and then incorporates parameters into
the chosen operators. Finally, DFT constructs a final formula containing the parameterized constraints.
DFT applies the search to produce inputs uniformly at random to execute $\mathit{pp}$ generating a diverse test suite. They
made an experimental study on 105 programs and compared DFT against focused testing, symbolic execution, and random
testing. They showed that mutation score improved from $3\%$ to $70\%$, and fault-detection has improved in $100\%$
of the cases.

\subsubsection{Web testing}
\label{subsubsec:web-testing}

Web testing is the process of testing web-based applications to ensure that they function correctly
and meet the intended requirements.

Some researchers used \dbt techniques in web testing. Marchetto and Tonella~\cite{Marchetto2009} applied
search-based testing maximising test case diversity to explore the large space of long interacting
sequences in Ajax Web applications. To control the size of the solution, rather than generating all
test cases up to a maximum length, the most diverse test cases are selected. Marchetto and Tonella
evaluated their approach on two real-world Ajax applications and found the generated interaction
sequences are the most effective.
Alshahwan and Harman~\cite{Alshahwan2012, Alshahwan2014} proposed a black-box test criterion based on
Output Uniqueness (OU), and used six web applications to study the white-box coverage and fault detection
achieved. Output uniqueness detected $92\%$ of the faults found by branch coverage and also complemented white-box
techniques by finding $47\%$ of other undetected faults. In a later study, Benito \etal~\cite{Benito2022} presented
an approach for unit test generation using ``Output Diversity Driven'' (ODD) rather than ``Input Diversity Driven'' (IDD)
like Alshahwan and Harman~\cite{Alshahwan2014}. Benito \etal showed that ODD test sets have higher output
uniqueness than IDD by $50\%$, and have a $63\%$ higher correlation output uniqueness score and detecting
faults than IDD.

A state flow graph (SFG) can represent the navigational model of a web application, where the nodes of the graph
represent the dynamic DOM states of the web pages, and the edges represent the events that trigger the transitions
between the pages. Biagiola \etal~\cite{Biagiola2019} presented a diversity-based system-level web test generation
algorithm that selects the most encouraging candidate test cases according to their dis-similarity from previously
generated tests. Therefore, the algorithm considers the test cases which inspect different behaviours of the application
for in-browser execution. The approach starts by extracting the navigational model of the dynamic web pages. Then, page
objects are created to model each web page, and object-oriented classes that show the actions executable in each web page
are modelled as methods. Finally, a test suite that covers the web application is generated using the distance metric between
the test actions and input data. The test generator starts with a random test. Then a set of candidate test cases is
generated, and the distances between the candidates and the tests already generated are calculated. The candidate with
the highest distance value is executed. Only when the test path of the executed test case differs from the execution
trace obtained from running the test suite increasing the coverage of the navigational model, then it
is added to the set of test cases. Biagiola \etal made an empirical evaluation of their proposed approach
using six web applications. They produced test suites better in terms of coverage, fault detection rates, and
processing speed than crawling-based and search-based test generators.

Furthermore, test data for web applications can be generated using evolutionary approaches.
Wang \etal~\cite{Wang2023} proposed a parallel evolutionary test case generation approach, that increases
population diversity, for web application testing. The parallel execution greatly decreases the time needed
to generate the tests, while maintaining population diversity prevents the premature convergence to a local
optima. The empirical evaluation made on six web applications showed a decrease of $33.43\%$ and $63.1\%$
in iterations and evolution time, respectively, while having the same coverage as other single evolutionary
algorithms.

\subsubsection{Mobile applications testing}

With the increasing popularity of mobile devices and the growing number of mobile apps, mobile application testing
has become an essential aspect of software testing. Mobile application testing is the process of testing mobile apps
to ensure that they function as intended and meet the specified requirements. Once more, diversity concepts can be
used in testing mobile applications.
Vogel \etal~\cite{Vogel2019} investigated the effect of diversity for the search problem of test data generation for mobile
applications using a test data generation tool called \textsc{Sapinez} which is based on the heuristic NSGA-II. Each individual
in the population is a test suite, which in turn consists of $n$ test cases, each is a sequence of $m$ GUI events.
The test cases are referred to as an \emph{``ordered test sequences''}. The distance between two test suites is
the sum of the distances between their ordered test sequences. The distance between two ordered test sequences is the difference
of their length increased by one for each different event. The diversity of the population is measured through the average population
diameter. \textsc{Sapinez} loses the diversity of solutions over time, which may lead to premature convergence~\cite{Vogel2021}.
Vogel \etal~\cite{Vogel2021} proposed \textsc{Sapinez}$^{\mathit{DIV}}$ to include more diversity. Four mechanisms were used to introduce
diversity into test suites. They evaluated the \textsc{Sapinez}$^{\mathit{DIV}}$ against \textsc{Sapinez} on 34 applications and found that
\textsc{Sapinez}$^{\mathit{DIV}}$ outperforms \textsc{Sapinez} in terms of coverage, or at least the same on all subjects. However, the
execution time of the generated test suite takes more time.

\subsubsection{Directed random testing}

Pacheco \etal~\cite{Pacheco2007A} proposed a technique called feedback-directed random test
generation and implemented a tool called ``Randoop''~\cite{Pacheco2007, Just2014A} that
generates unit tests for object-oriented programs. The main difference between this technique
and random testing is the use of feedback from previously generated tests. This technique
extracts public methods from the classes of the software under test (SUT), and then constructs
method sequences, with each sequence constituting a unit test. In each method sequence,
inputs are generated, modified, and used by method calls. The process begins with basic
method sequences containing either a single method call or a constant. Subsequently, it
expands these sequences by adding a newly chosen method and previously generated method
sequences. Sequences
that cause exceptions are not extended. Randoop executes the sequences it creates, using
the results of the execution to create assertions that capture the behaviour of the program, then
creates tests from the code sequences and assertions. However, one problem with Randoop,
is that code coverage stops increasing after some point, due to the inability of the
technique to find any new interesting tests that can contribute to the coverage~\cite{Yatoh2015}.
Values produced by methods in SUT are correlated with the
arguments used, and similar arguments result in similar return values, which in turn are used from the
feedback as similar arguments, causing over-directness.
Another issue is that due to Randoop's randomness, code coverage can vary considerably according
to the chosen random seed.
Yatoh \etal~\cite{Yatoh2015} extended this work and proposed a method named \emph{feedback-controlled
  random test generation}, which limits and controls the feedback in order to increase the diversity of
generated tests. They used multiple pools simultaneously
rather than a single pool, with different pools having different contents guiding the test generation
into more diverse test cases. The feedback-controlled generation algorithm employs multiple pools
simultaneously and concurrently. It dynamically determines the direction for further search and
decides when to halt the search in each direction, utilising the feedback information.
They experimented on eight
different real-world application libraries, and found that branch coverage increased by $78\%$ to
$204\%$ over the original feedback-directed algorithm.

\subsubsection{Multiple dispatch language testing}

In a multiple dispatch language, many methods can have the same name but differ in the number or type of
parameters of the methods. The correct method call is determined by the data types of the actual arguments
passed to the method. This concept is referred to as \emph{function overloading} in many programming languages.
Poulding and Feldt~\cite{Poulding2017} proposed a probabilistic approach for generating diverse test inputs for
multiple dispatch languages. They made an empirical study to evaluate the approach using 247 methods across
nine built-in functions of the Julia programming language and found three real faults.

\subsubsection{GUI testing}

A Graphical User Interface (GUI) application is an event-driven software, where the input is a sequence of events that changes
the state of the application, and then produces an output. The events can be clicks on buttons, messages in text fields, etc.
A GUI test is made up of a sequence of events interacting with the GUI and some assertions on the GUI state.
Feng \etal~\cite{Feng2012} proposed a 3-way technique for GUI test case generation based on event-wise partitioning. The event-
wise partitioning is performed according to the level of similarity between the GUI states. They use $k$-means algorithm
to cluster the states. For evaluation, they used two Java open-source applications in their experiments, and found the approach
a little better in fault-detection but with a smaller test suite size compared with Event Semantic Interaction technique.
Mariani \etal~\cite{Mariani2021} made an empirical study on the reuse of GUI tests based on
semantic similarities of GUI events. They proposed a semantic matching algorithm named \emph{SemFinder}, and showed that it
outperformed other algorithms. Events with the highest similarity score or all events above a certain threshold can be considered in the test
reuse approach. The study had 253 configurations of the semantic components, 337 distinct queries, and 8099 unique GUI events.
Mariani \etal found that all semantic matching components have an effect on the quality of the results, but the most impactful
component was the Semantic Matching Algorithm.

\subsubsection{Compiler testing}

Compilers like other software may contain bugs, and warning diagnostic messages could be faulty or missing. A specific type of
testing called, compiler testing, is used to detect compiler bugs. An important task in compiler testing is to generate test
programs that are able to reveal bugs. Chen \etal~\cite{Chen2019C} proposed an approach called \emph{HiCOND} (\textbf{Hi}story-Guided
\textbf{CON}ﬁguration \textbf{D}iversiﬁcation) for test-program generation based on historical data that are bug-revealing and
diverse in the sense that they are capable of revealing many types of bugs. HiCOND starts by generating a set of test
configurations utilising the historical data of compiler testing. Then, HiCOND generates test programs using the set of test
configurations. They evaluated the proposed approach on GCC and LLVM compilers, and found HiCOND detected $75\%$, $133.33\%$ and
$145\%$ more bugs than three random test generation and two variants of swarm testing, respectively. Also, $61\%$ of the bugs were
detected only HiCOND. In a later study, Tang \etal~\cite{Tang2022} proposed an approach to build diverse warning-sensitive
programs to detect compiler warning defects.
They named the approach DIPROM (\textbf{DI}versity-guided \textbf{PRO}gram \textbf{M}utation approach). DIPROM takes as input
a test program, and it removes the dead code and constructs the abstract syntax tree (AST). Then, DIPROM mutates the AST using
63 mutators to generate program variants using a diversity-based strategy. Finally, differential testing detects warning
defects in different compilers. They made an empirical study using GCC and Clang compilers to compare DIPROM against other approaches including
HiCOND, Epiphron, and Hermes. They showed that DIPROM reveals more bugs by $76.74\%$ than HiCOND, $34.30\%$ than Epiphron, and
$18.93\%$ than Hermes all in less time.

\subsubsection{Test case recommendation}

Similar code structures tend to have similar test cases and co-evolve together, some researchers made
use of this pattern to generate test cases for structurally similar methods. Shimmi and Rahimi~\cite{Shimmi2022}
proposed such an approach that assumes an existing test suite that is either incomplete or outdated. In the former case,
the approach recommends additional test cases to complete the test suite, or in the latter case, change
the current test cases to adapt to the changes introduced in the code. First, the structural similarity
between methods is computed, and similar methods are grouped. Then, they identify the corresponding and
missing test cases, and modify the test cases with appropriate values. Finally, new test cases for the
new methods are recommended. The empirical evaluation made on five large open-source Java applications
showed that the approach generated $72\%$ of the minimum required test cases.
The limitations of this approach are the assumption made of
an existing test suite, and assumes a certain naming convention to be held such that the name of the test
case must have the word ``test'' followed by the tested class or method name. Also, it depends on two other
open-source tools, which if they are changed or not available, the whole approach might not work.

\subsection{Test Case Prioritisation}
\label{subsec:test-case-prioritisation-diversity}

Test case prioritisation (TCP) re-orders test cases so that the tester gets maximum benefit in the case testing is prematurely
stopped at some arbitrary point due to some budget constraints~\cite{Yoo2010B}. \dbt techniques have been
used in many test case prioritisation papers.

\subsubsection{Prioritisation based on test cases}
\label{subsubsec:prioritisation-test-cases}

Some techniques use the test cases themselves as a source to guide the prioritisation process.
Ledru \etal~\cite{Ledru2009,Ledru2012} proposed a prioritisation algorithm that is based on the idea that test
diversity strengthens the test suite capability for fault detection. The fitness function used in the prioritisation
algorithm is based on the distance between each test case and the set of prior test cases in the test suite. The
distance measure is the distance between a test case $t$ and the nearest element of the set of test cases $T$, where
\begin{equation*}
  \mathit{dist}(t, T) = \min_{t_{i} \in T , t_{i} \neq t} {d(t,t_{i})}
\end{equation*}
For $N$ test cases in $T$, the fitness function is the sum of distances between each test case and the set of prior test
cases in a given permutation $P = t_{1},t_{2}, \cdots, t_{n}$.
The prioritisation algorithm used is a greedy algorithm, where in each iteration the most distant test case from the
set of ordered test cases is selected. In the case of multiple test cases with the same maximum value, one of them is
selected randomly. The ordered set initially consists of the first test case in the test suite. For each string distance,
the algorithm computes a \emph{similarity matrix}, which is an $N \times N$ matrix containing the distance values between
each test case and the other test cases in the test suite. The experimental results showed that prioritisation based on
string distances is better in finding faults than the random ordering of the test suite.
Wu \etal~\cite{Wu2012} aimed to enhance the similarity-based test case prioritisation techniques by accounting
for the use of a relative number of execution times of program elements. Test cases are selected based on their diversity
with the test cases already selected until the list of test cases is done. Fang \etal~\cite{Fang2014} expanded more
about the use of similarities of ordered execution sequences for test case prioritisation. Wang \etal~\cite{Wang2015}
proposed a similarity-based test case prioritisation technique according to the distance between pairs of test cases in
terms of branch coverage. All branches in the program can be represented as a vector $V:<v_{1}, v_{2}, \dots, v_{n}$,
where $v_{i}$ is the number of times a branch $i$ is executed. They used six different similarity measures in
their experiments, which are Jaccard Index (JI), Gower-Legendre (GL), Soka-Sneath (SS), Euclidean distance (ED), Cosine similarity
(CS), and Proportional distance (PD) metric. They evaluated their approach on four C programs and found that Euclidean distance
is the most effective.

Scalability and performance are crucial to take into consideration if any technique should be used in an industrial environment.
Miranda \etal~\cite{Miranda2018} introduced an approach for test case prioritisation based on diversity that is both
fast and scalable with big test suites sizes and referred to it as the FAST family of test case prioritisation. The approach
utilises some techniques in big data domain for similarity detection. The FAST family techniques use Shingling, Minhashing,
and LSH algorithms to find diverse test cases in a short time, and can be used for white-box or black-box prioritisation.
The input to the algorithm is a \emph{coded test suite info}, which is the code coverage information represented as a set
in the case of white-box prioritisation, or for black-box prioritisation, it is the string representation of the test cases.
Miranda \etal made a comparison between 20 techniques of the FAST family and 17 other techniques in test case prioritisation
using 10 subjects in C and Java. The evaluation was done in terms of Average Percentage of Faults Detected (APFD). They
found that in black-box prioritisation, the efficiency has improved significantly without any effect on effectiveness. In a
later study, Greca \etal~\cite{Greca2022} made an empirical study to compare between test case selection and test case
prioritisation. They used a tool called Ekstazi~\cite{Gligoric2015} for test case selection, and FAST~\cite{Miranda2018} for
test case prioritisation. Then, they used the two tools to combine file-based test case selection and similarity-based test
case prioritisation. First, test case selection is performed using Ekstazi, and then the selected subset of the test suite is
ordered using test case prioritisation FAST. They made their experiments using 12 Java projects from Defects4J, and found
that their proposed approach is more effective than Ekstazi and FAST.

Manual black-box system testing is a type of software testing in which the tester evaluates the functionality of a software
application without having any knowledge of its internal workings or source code. Hemmati \etal~\cite{Hemmati2015} modified
existing test case prioritisation techniques to work on manual black-box system testing. They implemented three different
techniques for test case prioritisation based on topic coverage, diversity, and riskiness. In order to evaluate the three
techniques, they made an empirical study on releases of Mozilla Firefox (4 old releases and 9 new releases) and found
that in a rapid-release environment, prioritisation based on riskiness is $65\%$ more effective in terms of APFD. However,
there is no clear advantage over any technique for a traditional software development environment.

Some empirical studies were conducted to evaluate the effectiveness of \dbt \tcp (TCP) techniques.
Huang \etal~\cite{Huang2017B} made an empirical study of TCP techniques using similarity measures for
abstract test cases. The study included 14 similarity measures and two techniques
(local and global techniques) for TCP. Also, the study was made on five real-world programs written
in $C$ language, and measured the coverage level and fault-finding capabilities of each similarity
measure. They reported that global TCP outperformed local TCP in coverage and fault-finding capabilities.
Also, Haghighatkhah \etal~\cite{Haghighatkhah2018A} made an empirical study to investigate whether
similarity-based test case prioritisation (SBTP) is more capable than random ordering to locate faults,
and the best implementation of SBTP. They implemented five different techniques and experimented on
six Java programs from the Defects4J dataset. The five implementations are based on
Manhattan distance, Jaccard distance, Normalised compression distance (NCD), NCD-Multisets, and Locality
sensitive hashing (discussed in Section~\ref{sec:rqmetrics}). The SBTP algorithms were greedy algorithms
based on the implementation of
Ledru \etal~\cite{Ledru2012}. They found that SBTP techniques outperform random ordering
and that no particular implementation is superior to others. However, Locality-sensitive hashing was
the fastest but the least effective of the other four implementations.

\label{par:Noor2015}
An alternative idea is based on the rationale that a previous faulty test case is more likely to
reveal faults again. Thus, test cases that are similar to these previously faulty test cases have
higher chance of detecting new faults as well.
Noor and Hemmati~\cite{Noor2015} defined a class of metrics to measure the similarity of test cases
to previously failing test cases, and proposed a test case prioritisation approach based on this
metric. The similarities between test cases are identified based on their sequences of method calls.
They used three distance functions to measure the similarity
between test cases, which are Basic Counting, Hamming distance, and Levenshtein (Edit) distance.
They made an empirical study on five real-world Java projects with real
faults, and found the proposed approach more effective than other traditional metrics.

\subsubsection{Prioritisation based on test steps}

\emph{``Test steps''} refer to the individual actions or procedures that need to be followed in order to execute a test case.
Test steps are typically written in a clear and concise manner and provide detailed instructions on how to carry out
the test case.
Flemstrom \etal~\cite{Flemstrom2018} proposed ordering test cases based on their similarities in terms of
test steps to reduce the cost of translating manual tests into executable code. The automation effort of
translating the test steps into executable code can be reduced if the executable code of a test step can be reused for
similar test steps. They made a case study on four projects with a total of 3919 integration test cases making up
35180 test steps, written in natural language. The results showed that the expected manual effort was reduced by
$12\%$, and the proposed prioritisation method was as effective as other heuristic approaches but faster and had less
computational cost.

\subsubsection{Prioritisation based on program executions}

Some researchers designed their prioritisation techniques based on program execution diversity.
Gomes de Oliveira Neto \etal~\cite{De2020} proposed a test case prioritisation approach based on the behavioural diversity of test cases, and
referred to it as \emph{b-div}. They used mutation testing to compare the executions and failure outcomes of the test
cases to capture the behavioural diversity. They compared the $b$-div measures against usual diversity metrics between
artefacts (\emph{a-div}), such as test scripts, inputs, or outputs. The study ran experiments on six different Java
projects in the context of test prioritisation and evaluated the techniques in terms of their average percentage
of fault detection (APFD). They found that $b$-div measures found more mutants than $a$-div measures in all subjects.

\subsubsection{Prioritisation for deep learning software}

Deep Learning (DL) is being used nowadays for a variety of applications and there is a need to prioritise the inputs
of the DL software to find a violation or incorrect predictions as early as possible to reduce the cost of running time
\cite{Mosin2022,Xie2022}. Some of these prioritisation approaches are \dbt techniques. Mosin \etal
\cite{Mosin2022} proposed an input prioritisation technique for DL systems, and compared between white-box, data-box,
and black-box approaches in terms of effectiveness and efficiency. They implemented surprise adequacy as a white-box
approach, autoencoder-based as a data-box approach, and similarity-based as a black-box approach. The empirical evaluation
was made on four image classification datasets, and they found that similarity-based has the highest efficiency with a total
running time of 4556-4947 and 9381-9623 times faster than autoencoder-based and surprise adequacy, respectively.
In terms of APFD, surprise adequacy was the best ranging from $0.785$ to $0.914$ APFD, while autoencoder-based, and
similarity-based ranged from $0.532$ to $0.744$ APFD and $0.579$ to $0.709$ APFD, respectively.

Since Metamorphic Testing is used to tackle the oracle problem for DL systems~\cite{Xie2022}, an alternative way of test
input prioritisation is to prioritise according to diversity between Metamorphic test case Pairs (MPs). An MP consists
of the \emph{source case} (\ie existing test case) and the \emph{follow-up case} (\ie a new test case) generated using a
specific Metamorphic Relation (MR). Xie \etal~\cite{Xie2022} proposed a \dbt prioritisation technique of MPs
for Deep Learning (DL) software.  The hypothesis is that if the source test case executes a faulty execution pattern,
then a diverse follow-up test case will bypass that faulty execution pattern and a violation would be detected, or vice
versa. Therefore, diversity between the executions' patterns of MPs will have a higher probability of detecting violations.
The approach starts by executing all MPs and records their execution states. Then, prioritisation of the MPs starts based
on the diversity of the MPs executions. The empirical evaluation made on three image classification DNN models showed
that the proposed \dbt prioritisation technique increased the detection of violations compared to the neuron
coverage-based approach in terms of Normalised Average Percentage Violations Detected (NAPVD).

\subsubsection{Prioritisation for WS-BPEL programs}

WS-BPEL (Web Services Business Process Execution Language) programs are XML-based executable processes that
describe the flow of interactions between different web services to achieve a specific business goal. WS-BPEL
programs typically consist of a series of activities, including invoking web services, manipulating data, and
performing various control structures such as loops, conditions, and exception handling. WS-BPEL programs can
be designed using a visual editor or written directly in XML.
Mei \etal~\cite{Mei2013A,Mei2013B} proposed a similarity-based test case prioritisation technique for regression
testing of WS-BPEL programs. The proposed techniques are based on the structural similarity of XML-based \artefact{s} between
test cases in three levels: Web Services Description Language (WSDL) interface specification, XML-based messages, and BPEL workflow process. Also, the
proposed approach selects pairs of test cases without replacement in turn. They used the tree edit distance to measure
the similarity between two XML documents and adapted this metric incorporating Jaccard distance to measure the similarity
between two sets of XML documents. They empirically evaluated the proposed approach and found it better in terms of
average percentage of fault detection (APFD) than random ordering and other techniques. Bertolino \etal~\cite{Bertolino2015} proposed a similarity-based test
case prioritisation for access control, which is a security mechanism that ensures that only intended users can access
the data. The approach order Extensible Access Control Markup Language (XACML) test cases using distance functions.
In XACML, a test suite consists of a number of
access control requests. They defined two distance functions to measure the similarity between two access control requests.
The first distance function is referred to as simple similarity, which is based on the lexical distance of the requests'
parameters. Simple similarity is a modified version of Hamming distance. The second distance function referred to as
XACML similarity, in addition to the values of the parameters, considers the XACML policy. They used mutants to represent
faults in their empirical study and found that the proposed approach is better than random
prioritisation in terms of APFD.

\subsubsection{Prioritisation in CI environments}

Some studies investigated diversity-based test case prioritisation in Continuous Integration (CI) environments. Haghighatkhah \etal
\cite{Haghighatkhah2018B} proposed using diversity-based (DB-TCP) and history-based test case prioritisation (HB-TCP).
They used three similarity measures in their approach, which are Manhattan distance, normalised compression distance (NCD), and NCD-Multiset. The NCD-Multiset
was the best in terms of effectiveness, but it is slower to calculate than the other two measures. They evaluated
the approach on six projects and found that DB-TCP is superior in the early stages when no history data interval is available,
but in later stages, HB-TCP becomes better. Hence, merging the two approaches gave better results than using any of them
separately. This means applying DB-TCP in the early stages until enough historical intervals are available to apply the
more powerful history-based diversity using NCD Multiset. Also, Azizi~\cite{Azizi2021} proposed a technique for regression
test case prioritisation that selects test cases based on the test case's
textual similarity to the changed parts of the code using Cosine Similarity measure. However, if a test case can reveal an
error, but does not have matching terms with the changed code, that test case can be ignored in the ordering. Therefore,
Azizi generated tags for the test cases based on their topics, and mapped the test cases with the tags. All test files are
grouped, and each term is tokenised removing mathematical operators and some other reserved keywords. The most repeated terms
are identified as initial tags, then the Latent Dirichlet Allocation (LDA) technique~\cite{Blei2003} is used to suggest tags for
test cases based on their relevance to each topic. The changes in the code between two versions are extracted using a \emph{diff}
tool and placed into a query, which is a vector of tokens. Now given a set of documents $D = \{d_{1}, d_{2}, \dots, d_{n}\}$
and a query $Q = \{q_{1}, q_{2}, \dots, q_{n}\}$, where each element in $D$ and $Q$ is the frequency of the term $i$ in $D$
and $Q$. The similarity is calculated as:
\begin{equation}
  \mathit{Sim}(D,Q) = \sum_{i=1}^{n} tf_{D}(d_{i}) . \mathit{tf}_{Q}(q_{i}) . \mathit{idf}({t_{i}})^{2}
\end{equation}
where $\mathit{tf}$ is the term frequency, and $\mathit{idf}$ is the inverse document frequency. The tagging system helps to organise
the test suite and aid testers to find relevant tests more quickly reducing the cost of regression testing.
Azizi made an empirical study to evaluate her approach using 37 versions of
six open source applications written in Java and C\#. She made a comparison with five test case prioritisation techniques
and found the proposed approach more efficient and effective.

\subsubsection{Clustering-based prioritisation approaches}
\label{subsubsec:clustering-prioritisation}

Some studies used clustering-based approaches for test case prioritisation (TCP). The general idea would be to group similar
test cases according to some criterion together, and then prioritise the test cases within each cluster. Finally,
construct the final list of test cases by selecting test cases from each cluster according to some rule, usually
in a round-robin fashion. Yoo \etal~\cite{Yoo2009} proposed a TCP technique by clustering test cases based on their dynamic
runtime behaviour. A human expert, who can provide rich domain knowledge, prioritises the cluster of similar test cases, which
is much smaller than the actual test cases. An Agglomerative Hierarchical Clustering technique is used to cluster
test cases. Test cases in each cluster are prioritised using a coverage-based greedy prioritisation algorithm and a
human expert prioritises the clusters. The final test suite is
constructed by selecting from each cluster, going by prioritisation order, the highest ranked test case and discarding it from
the cluster, and so on until all test cases from all clusters are appended in the list of final test cases. They evaluated
the approach on seven test suites and compared it against coverage-based TCP. Their proposed approach was better in terms
of fault-finding capabilities. In another paper, Carlson \etal~\cite{Carlson2011} proposed a clustering approach for TCP
that uses code coverage, code complexity, and history data on real faults. They used the Euclidean distance to measure the
similarity between test cases, then applied agglomerative hierarchical clustering to group similar test cases together.
Then, in each cluster, test cases are ordered using code coverage, code complexity, and fault history information. The final
list of test cases is constructed by selecting one test case from each cluster in a round-robin fashion. They experimented
with their approach on an industrial program with real faults and found that their approach improves the effectiveness over other
techniques with no clustering in terms of the average percentage of fault detection (APFD) metric. Zhao \etal~\cite{Zhao2015} proposed a hybrid technique for regression
test case prioritisation that incorporates code coverage
similarities with Bayesian Networks test case prioritisation, which includes source code change information, test coverage data,
and software quality metrics. The technique clusters test cases with similar code coverage in groups, and then the clustered test
cases in each group are prioritised using Bayesian Networks algorithm. Finally, the test cases from each cluster are selected in
a round-robin method to construct the final list of test cases. They used Euclidean distance to measure the distance between test
cases based on their code coverage. They evaluated their technique on two Java applications with mutation faults and one Java
application with manually seeded faults. They compared their technique against additional greedy, Bayesian Networks, and
Bayesian Networks with feedback, and found that it outperformed them in terms of APFD.

String Distance test case prioritisation (SD-TCP) uses the similarity between test cases as a basis for ordering test cases.
Although SD-TCP is simple and fast, it is vulnerable to extreme cases of tests where the test case may be very long or very
short giving it higher priority in the ordering process. Chen \etal~\cite{Chen2021B} proposed a TCP method based on
K-medoids and Similarity (KS-TCP). KS-TCP solves this problem by clustering similar test cases into groups and then prioritises
the clusters using string distance greedy-based algorithms. First, a similarity matrix is calculated using the Manhattan distance.
Second, test cases are grouped into K-clusters using the K-medoids algorithm. Third, each cluster is sorted using a greedy
algorithm from the distance matrix. Finally, test cases are selected from each cluster in order in a round-robin fashion. They
evaluated KS-TCP against SD-TCP and Random prioritisation and found that KS-TCP outperforms both in terms of APFD, and KS-TCP is
faster than SD-TCP.

An alternative method of clustering based on the properties of the test cases is to cluster test cases based on requirements.
Since different requirements have different priorities for the clients, one can build a cluster prioritisation based on
requirements' importance and whether the code for that requirement has been modified or not.
Arafeen \etal~\cite{Arafeen2013} implemented such \tcp technique based on requirements clustering. The rank of a cluster $k$ is:
\begin{equation}
  R(k) = \frac{1}{x(k)} \sum_{i=0}^{x(k)} w(i,k)
\end{equation}
where $x(k)$ is the number of requirements in cluster $k$, and $w(i,k)$ is the weight of requirement $i$ in cluster $k$.
$w(i,k) = \mathit{Priority}(i) \times \mathit{Modified}(i)$, where $\mathit{Priority}(i) = \{1(\mathit{low}), 2(\mathit{medium}), 3(\mathit{high})\}$
and $\mathit{Modified}(i) = \{1(\mathit{unmodified}), 2(\mathit{modified})\}$. Finally, according to the ranking of clusters, test cases are picked. The higher the rank
of the cluster, the more test cases are selected. This is computed by calculating the average number of test cases per
cluster $t$, and selecting $t$ test cases from the highest-rank cluster and for each subsequent cluster, the number is
decreased by $10\%$, until all test cases are selected. They evaluated their approach using two Java programs with
multiple versions and requirements documents and found the approach to be effective.

Moreover, some researchers applied clustering of test cases based on their coverage information and predicted
fault-proneness. Mahdieh \etal~\cite{Mahdieh2022} proposed such a \tcp approach and evaluated the approach on
five projects from the Defects4J dataset. They reported that the proposed clustering-based combined with
fault-proneness was better than traditional coverage-based approaches and prioritisation using only
fault-proneness in terms of average first fail detected.
Mahdieh \etal~\cite{Mahdieh2022} proposed a test case prioritisation approach using bug history and test
case similarities as guides to the reordering process. Firstly, the fault-proneness of code units is determined
through a defect prediction model.
Secondly, similar test cases are clustered into groups using a hierarchical clustering algorithm. Similarity
between test cases is calculated using the Euclidean distance, Manhattan distance, and Cosine similarity of
the coverage information of each test case and the predicted fault-proneness. Then, for each
group, test cases are prioritised according to their coverage capabilities. Finally, the final list of test
cases is constructed by selecting test cases from each cluster using both fault-proneness and the order of
test cases within each cluster. The empirical evaluation was made on five projects from the Defects4J dataset,
and reported that the Euclidean distance performed better than Manhattan and Cosine. Also, the proposed
clustering-based combined with fault-proneness was better than traditional coverage-based approaches and
prioritisation using only fault-proneness in terms of average first fail detected.

\subsection{Test Suite Reduction}
\label{subsec:test-suite-reduction-diversity}

Test suite reduction (TSR) techniques aim to find redundant test cases in the test suite and to discard them
in order to reduce the size of the test suite~\cite{Yoo2010B}. In TSR literature, Coviello \etal~\cite{Coviello2018A}
defined an adequate approach as an approach where the reduced test suite covers the same test requirements as the
original test suite, while an inadequate approach is an approach where the reduced test suite
does not cover the same test requirements as the original test suite. Coviello \etal~\cite{Coviello2018A}
made an experimental study of adequate and inadequate test suite reduction approaches in terms of the
trade-off between the reduction in size and fault-finding capability. They used statement, method,
and class coverage as test requirements on 18 approaches. One of the assessed reduction
approaches is an inadequate approach based on the coverage similarity between test cases.
They concluded that inadequate approaches offer a better trade-off between size reduction
and loss in fault-finding capability than inadequate approaches. Coviello \etal~\cite{Coviello2018B}
proposed a clustering-based approach for inadequate test suite reduction. The clustering-based approach
places similar test cases into groups based on their statement coverage, and then treats the test cases in
each group as redundant. The algorithm gets the statement coverage of each test case using the JaCoCo tool and
represented as a binary vector, where 1 means that the corresponding statement is covered, 0 otherwise.
A similarity matrix of size $n \times n$ is constructed, where $n$ is the number of test cases and each cell in the
matrix stores the similarity value of the pair of test cases corresponding to the indices of the cell. They used
Hierarchical Agglomerative Clustering (HAC) as the clustering algorithm to group test cases together. The reduced
test suite will select one test case from each cluster that covers the largest number of statements. They made an
empirical study on 19 versions of four Java programs and found that cosine similarity and Jaccard distance gave
the best results.  In order to fix scalability issues, Cruciani \etal~\cite{Cruciani2019}
proposed a family of scalable approaches based on similarity for test suite reduction, that uses techniques
from the big data domain. The proposed approach is considered inadequate TSR as it targets
a certain number of test cases, but it can be adapted to do an adequate reduction.
However, it is not suitable for large-scale test suites. The tester specifies the
desired number of test cases. Then, they model each test case as a point in some
D-dimensional space, and select evenly spaced test cases. The main contribution is to make
a reduction of a large size test suite in a short time (\emph{within seconds}) with a comparable
loss of fault detection to other techniques.

Some \tsr techniques involve applying clustering algorithms to group test cases with similar characteristics
or behaviours into clusters, then from each cluster, one or more representative test cases are selected~\cite{Beena2014,Chetouane2020,Viggiato2023}.
Beena and Sarala~\cite{Beena2014} proposed a clustering-based multi-objective test suite reduction approach aiming to maximise statement
coverage and minimise execution time. The subset of the test suite covering all test requirements is referred to as the
hitting set. Using statement coverage and execution time information, a similarity matrix is constructed and used for test case
clustering. Finally, using the clusters of test cases, the minimal hitting set is generated. The clustering approach uses Hitting
Set Directed Acyclic Graph algorithm to identify clusters through weighted distance function.
Based on the similarity of test cases, any test case similarity falling
below a certain threshold will be placed in a cluster.
Chetouane \etal~\cite{Chetouane2020} proposed an algorithm for test suite reduction that combines $k$-means clustering
with binary search. Euclidean distance has been used in the $k$-means clustering. The general idea of the algorithm is
to group test cases that are close together in clusters and then
pick a test case from each cluster, reducing the size of the test suite. They use binary search to look for the proper
number of clusters such that the coverage or mutation score of the reduced test suite does not deviate much from the
original test suite. First, the coverage of the entire test suite of size $n$ is computed. Then, the coverage is computed
after halving the test suite into $\frac{n}{2}$ clusters. If the coverage is the same as the original, the number of
clusters is decreased, otherwise, it is increased. A test suite is constructed by selecting a test case from each cluster,
and the test suite with coverage close to the original is selected. They made an experimental study on Java programs,
and used statement, branch, MC/DC coverage and mutation score as evaluation metrics. The study was able to achieve
a $95.9\%$ reduction maintaining coverage but with a drop in mutation score, and a $82.2\%$ reduction maintaining both
coverage and mutation score of the original test suites. Viggiato \etal~\cite{Viggiato2023} proposed an approach to
identify similar test cases written in natural language to reduce the
cost of manual testing performed. The proposed approach uses clustering, text similarity and embedding to identify similar test
steps and then compare test cases based on their similarity to test steps grouped in the same cluster. They evaluated the approach
using a case from an educational game company in terms of F-score (a measure based on precision and recall). The proposed approach
detected similar test cases with the performance of an F-score of $83.47\%$.

Diversity was also employed for \tsr in the context of model-based testing. Coutinho \etal~\cite{Coutinho2016}
proposed focused on labelled transition systems and used a similarity-based approach. The reduction approach
calculates a \emph{similarity matrix}, which is the similarity degree between each pair of the test cases. The
algorithm searches the similarity matrix for the highest value, which represents the most two similar test cases.
If two cells in the matrix have the same highest value, one of these cells is chosen randomly. The most similar
test cases are considered to determine the test case with the lower number of transitions, and select this test
case for removal. If the reduced test suite satisfies all the requirements with the removal of the first test
case chosen, the test case is removed from the similarity matrix. If one of the test requirements is not
satisfied, the test case is added back to the test suite, and the second test case is considered for removal.
This process continues until all the test cases that do not affect the test requirements are removed.

Another form of reduction is to \emph{``tame''} the inputs. Pei \etal~\cite{Pei2014} proposed using delta debugging
trails to ``tame'' the fuzzer. Taming the fuzzer means controlling
the large number of test cases generated, especially the tests that keep triggering the same bug. The hypothesis is that the
larger the distance between two test cases according to some distance function, the more likely they belong to different
groups. Therefore, by applying the Furthest Point First (FPF) algorithm, the most diverse test cases can be retained to
discover diverse bugs and reduce the number of test cases required. Two distance measures based on the Euclidean distance
are used to measure the distance
between any pair of test cases. The first distance measure is the \emph{Single-linkage}, where the distance between two
sets of failing test cases is the distance between the nearest pair. The second distance measure is the \emph{Average-linkage},
where the distance is the average pair-wise distance between the two sets.

Some other works defined a set of test metrics to calculate the diversity between pairs of test cases in
a regression test suite that can be based on interaction behaviour models~\cite{Kichigin2009},
or based on block coverage criteria, control flow of a program, variable definition-usage, and data values
\cite{Singh2013}.

\subsection{Test Case Selection}
\label{section:test-case-selection}

Test case selection is the process of selecting a subset of the currently available tests that are relevant to some
set of recent changes~\cite{Yoo2010B}, where \dbt techniques can be used. Rogstad \etal~\cite{Rogstad2013}
proposed a black-box regression test case
selection approach based on similarity for large-scale database applications. First, the input domain is divided
using classification tree models. Then, similarity-based test case selection is applied to select test cases from
each partition. They conducted experiments on a industrial large database application and found that improvements were
marginal against random selection. However, if the partitions contain many test cases and the variations
within each partition are large, then the approach is worthwhile.

In some cases, emergency changes have to be made to the software quickly, but running the entire test suite can be
time-consuming, and selecting a subset can be risky. Farzat~\cite{Farzat2010} proposed a heuristic approach for test
case selection aiming to maximise coverage and diversity to lower the risks posed by introducing emergency changes into
the software. The proposed approach selects a subset of the test cases that cover the most important features of the
program according to a certain feature ordering or priority. The formal model contains a reward function that calculates
an overall coverage for a test suite adjusted by a penalty function calculated using Euclidean distance. The overall
coverage is the sum of the coverage of each selected test case.

Another possible method to mitigate the risk of discarding test cases to lower the costs of running tests is
using Continuous Integration (CI) and/or Continuous Delivery (CD) pipelines. Gomes de Oliveira Neto \etal~\cite{De2018}
aimed to study the harmonies between CI and similarity-based test optimization by evaluating similarity-based
test case selection (SBTCS) on integration-level tests executed on continuous integration pipelines of two
companies. They selected test cases that maximise the diversity of test coverage and reduce feedback time to developers.
In a case study, an evaluation between three similarity functions and random test selection was conducted, in
terms of test coverage and execution time of the selected subset. The similarity functions used are normalised Levenshtein (NL),
Jaccard distance and normalised compression distance. The selection does not require any information other than the test cases themselves, and the
selection techniques are evaluated according to how much of the test requirements were covered, dependencies and steps in
relation to the reduction of the test suite. Gomes de Oliveira Neto \etal~\cite{De2018} showed that SBTCS managed to reduce the time needed
to run integration tests on CI servers while maintaining full test coverage of various criteria from test \artefact{s}.

In a recent study, Shimari \etal~\cite{Shimari2022} proposed a clustering-based test case selection using the
execution traces of test cases for an industrial simulator. The goal was to select a subset of the regression test suite
that is capable of achieving high code coverage and diverse runtime executions for the simulation (\ie short
and long simulations to reflect regular usage patterns). The case study showed that the proposed approach selects
test cases with high coverage and high diversity of execution time.

\subsubsection{Multi-objective algorithms}
\label{subsubsec:multi-objective}

Multi-objective algorithms have been used for test case selection in a number of studies~\cite{De2012,Mondal2015,Panichella2015B}.
De Lucia \etal~\cite{De2012} discussed the concept of asymmetric distance preserving, useful to improve the
diversity of non-dominated solutions produced by multi-objective Pareto efficient genetic algorithms. They
proposed a multi-objective algorithm for test case selection by increasing population diversity in the obtained
Pareto fronts. They made an empirical study on four programs from the Siemens benchmark and found that using
diversity as one of the objectives improved the convergence speed of the genetic algorithms. Mondal \etal~\cite{Mondal2015}
proposed using coverage-based and \dbt techniques using NSGA-II multi-objective algorithm maximising
both coverage and diversity of test cases in the context of test case selection.
They made an experimental study on real-world case studies and found that \dbt selection is slightly
better than coverage-based selection. They found that \dbt is much more effective than coverage-based
for some projects, but also coverage-based is much better in other cases. However, coverage-based and \dbt
can be complementary to each other. Thus, the proposed bi-objective approach is better than using only coverage-based
or \dbt by $10-16\%$. Panichella \etal~\cite{Panichella2015B} suggested an improvement in multi-objective
genetic algorithms (MOGAs) by diversifying the solutions of the population for test case selection. They proposed DIV-GA
(\textbf{DIV}ersity based \textbf{G}enetic \textbf{A}lgorithm), that inject new orthogonal individuals to increase diversity
during the search process. The selection strategy is based on the concept of crowding distance, where individuals who
have higher distances from the rest of the population have a higher probability of being selected. They made an empirical
study on $11$ real-world programs and showed that DIV-GA outperforms greedy algorithms and MOGAs in fault-detection
and also better in terms of cost.

\subsubsection{Model-based test selection}
\label{subsubsec:mbt-selection}

Model-Based Test Selection (MBTS) focuses on selecting a subset of test cases that provide adequate coverage of the system's
functionality and requirements utilising a model of the system under test. \dbt techniques are used
extensively in this context. Cartaxo \etal~\cite{Cartaxo2007, Cartaxo2009} proposed a similarity-based approach for MBTS.
A similarity path matrix of size $n \times n$ is built, in which each cell represents the similarity between a pair of
test cases. The algorithm finds the highest value in the matrix, which represents the highest similarity between a pair
of test cases and discards one of them. If one test case has fewer transitions than the other, then it is discarded. If
both test cases have the same number of transitions, then one of them is discarded randomly. This process continues until
the desired size is reached. They made a comparison between the similarity-based selection and random selection in three
case studies. They found that similarity-based selection is better than random selection for a selection goal of $20\%$ or
more. For small selection percentages, they found that random selection works better. In a similar paper, Gomes de Oliveira Neto \etal
\cite{De2016} presented a similarity-based MBTS. The proposed approach detects modifications in the code by analyzing the
similarities between test cases. The selection approach constructs a
similarity matrix, and classifies test cases into reusable, targeted, or obsolete. Then, it removes similar to obsolete test cases and
adds test cases covering modifications. Finally, it removes redundant test cases, includes dissimilar reusable test cases and
exports the resulting subset. They performed an industrial case study and an experiment, and found that the proposed approach had similar
fault-detection with other similarity-based approaches, but with a significant increase in covering the modified parts of the model.
Hemmati \etal~\cite{Hemmati2010A} proposed a similarity-based test case selection (STCS) for UML state
machines using Genetic Algorithm (GA). The fitness function is based on Trigger-Based similarity (\emph{``Tb''}). The crossover
and mutation operations can result in an invalid solution, which is discarded from the population. They applied the selection
technique to an industrial case study and made a comparison between the proposed Trigger-Based Genetic Algorithm (TbGA) and
other techniques including random selection, greedy algorithms using
Trigger-based (Tb), Identical transition (It), and Identical state (Is). Hemmati \etal found ``TbGA'' to be more effective in
detecting real faults than the other selection techniques. Also,
they evaluated the fault detection of the selected subset in comparison with the original test suite. They were able to use the
technique to select only $27\%$ of the test suite maintaining the full fault-finding capabilities. Hemmati \etal~\cite{Hemmati2010C}
further discussed the concept of test case diversity in test case selection for model-based testing. They reported
that GAs are the best technique for STCS saving up to $80\%$ of test executions while maintaining $99\%$ of fault-detection rates.
Arrieta \etal~\cite{Arrieta2019} proposed a black-box test case selection approach for Simulink models. They defined six effectiveness
measures, four antipatterns based on output signals (Instability, Discontinuity, Growth, and Output minimum and maximum difference),
and two similarity metrics based on input and output signals similarities. They considered Test Execution Time (TET) as the cost measure
and applied it to all six measures and 15 of each of the pairs making up a total of 21 objective combinations. The empirical study
showed an increase by $18\%$ to $28\%$ over random selection in terms of Hypervolume quality indicator and that the proposed black-box
metrics outperformed the white-box metrics.

Some empirical studies were made to investigate the use of similarity-based approaches for MBTS.
Hemmati and Briand~\cite{Hemmati2010B} made an empirical study on industrial software to investigate different
similarity measures for test case selection in the context of model-based testing. The similarity measures used in the
study are Identical transition (\emph{It}) function, Hamming distance, Jaccard Index, Levenshtein distance, Needleman-Wunsch
(NW), and Smith-Waterman (SM). They reported that Jaccard Index gave the best fault-detection rate. They evaluated using
real faults and found that the fault detection rate increased by $50\%$ compared to coverage-based selection saving $77\%$
in execution cost. Another study by Hemmati \etal~\cite{Hemmati2011} made an empirical study on two industrial case studies
to understand the conditions in which similarity-based selection techniques can be beneficial to use. They implemented GA and
ART in the study using Jaccard Index and Needleman-Wunsch (NW) as similarity metrics. Then, they studied the test suites' properties in
order to identify the best situation to apply similarity-based techniques. That situation is when similar test cases
detect a common fault, and more importantly, when dissimilar test cases detect distinct faults. Also, they found that
when there are outliers (\ie a small number of test cases that are far away from the rest of the test cases), the fault-
detection rate of similarity-based test case selection drops significantly. This occurs because the approach selects
more test cases from a smaller group of test case clusters. They proposed a rank scaling approach to solve this scenario,
where all similarity values in the similarity matrix are sorted, and the rank of each value is its index in the sorted
array. The values in the similarity matrix are replaced by their rank. However, the rank scaling approach partially
solved the outliers problem. In a later study, Hemmati \etal~\cite{Hemmati2013} \label{pap:hemmati2013} introduced
320 variants of similarity-based test case selection (STCS) from different configurations of three
parameters of STCS, and
applied them to two industrial case studies. The three parameters of STCS are the representation of abstract test cases, the
similarity function, and the algorithm used for similarity minimisation. The best variant of STCS reduced the cost by $50\%$ to
$80\%$ and improved fault-detection rate up to $45\%$ over coverage-based selection and $110\%$ over random selection.

\subsubsection{Deep neural network}
\label{subsubsec:deep-neual-network}

Testing of Deep Neural Networks (DNNs) is a critical aspect of ensuring their reliability and accuracy in real-world
applications. Zhao \etal~\cite{Zhao2022} made an empirical study to find out the level of diversity of test input
selection for deep neural network (TIS-DNN) methods. The level of diversity here means whether some classes in the
original test set may have been missed by the selected subset. The test diversity is measured through the accuracy-based
performance measure $\mathit{AccEE}_{\mathit{avg}}$, which focuses on the average value of accuracy over all classes. Five TIS-DNN methods
used in the study are SRS, CSS, CES, DeepReduce and PACE on seven DNN datasets. The study showed that these methods
guarantee performance, but have a negative effect on the test diversity of the subset. Also, they discussed that there
is still great room for performance improvements in TIS-DNN methods. In another study, Aghababaeyan \etal~\cite{Aghababaeyan2023}
proposed a black-box strategy to test DNNs based on the diversity of inputs' features rather than the traditional neuron
coverage criteria. The approach can be applied to address multiple areas of testing DNNs, including test case selection.
The empirical study made using four image recognition datasets and five DNNs shows the superiority
of the \dbt technique over the neuron coverage approach in terms of fault-detection.

\subsection{Test Suite Quality Evaluation}
\label{subsec:test-suite-evaluation-diversity}

\Ctse techniques aim to judge on the quality of the current test suite and give insight to
testers about certain properties which can guide the tester to improve or fix some issues in the test suite.
Diversity of tests is one of the properties that testers generally aim to have in the test suites.
Nikolik~\cite{Nikolik2006} proposed a method for measuring the level a test suite executes on a program in
diverse ways with respect to control and data, and implemented a tool named \emph{The Diversity Analyzer}.
The proposed method measures conditional diversity, data diversity, standard deviation of diversity, and
test orthogonality. Conditional diversity measures the balance between the number of times branches are
true and false in the code. Data diversity measures the variation of the data inputs in the set of test cases.
Standard deviation of conditional diversity measures the control dispersion of the test cases. Orthogonality
measures the degree of linear dependencies in control and data between test cases. The experiments done using
the tool showed that more diverse test suites have higher fault-detection rates. Shi \etal~\cite{Shi2015}
proposed a black-box metric for \tse referred to as \emph{distance entropy} based on the
diversification concept. They represented the test cases as nodes in a complete graph where the weights of
the edges are the distance between each pair or node. The distance between the pairs of test cases is
measured using the learned distance metric. Then, they used a minimum spanning tree to measure the similarity of
the nodes in the graph calculating the distance entropy of the test set. Shi \etal showed that distance
entropy outperformed other simple metrics in equivalent time complexity.

Many empirical studies were made to investigate the importance of diversity in test suites~\cite{Xie2006B,Zhang2019,Flemstrom2016}
and some of its challenges~\cite{Neto2018}.
Xie and Memon~\cite{Xie2006B} made an empirical study to investigate the characteristics of ``\emph{good}'' GUI test
suite evaluating the testing cost and fault detection effectiveness. They found that the diversity of states and the
event coverage are the two major factors of fault-detection capabilities. In order to increase the diversity of GUI states,
testers need to increase the diversity of GUI states by developing a large number of small test cases to detect ``shallow''
faults (\ie faults that can be found by executing events in a different order). Another way of increasing diversity
is that, if more resources are available, testers can develop a small number of long test cases to reveal more ``deep''
faults (\ie faults that can be found through complex conditional statements).
Zhang and Xie~\cite{Zhang2019} conducted an empirical study to investigate the essential diversities to be used as testing
criteria for deep learning systems. They defined five metrics and leveraged metamorphic testing to reveal faulty behaviours.
They used the Jaccard distance to measure the diversity between two metamorphic relations. Also, they measured the diversity in
Neuron Coverage (\ie the number of activated neurons), the diversity in neuron output distribution with respect to major function
regions, the diversity in neuron output distribution with respect to corner-case regions, and the diversity in most activated neurons
and with respect to activated neuron distribution. Furthermore, they evaluated the quality of test suites using the defined metrics
and reported that diversity increases fault-detection.
Flemstrom \etal~\cite{Flemstrom2016} made an industrial case study on four real-world industrial projects in
the railway domain to investigate if test overlaps exist and how they are distributed over different test levels,
and reduce test effort. Test overlaps mean a high level of similarity between test cases, such that no additional
value can be had by including the redundant test cases. The results showed that most overlaps occurred within
the system integration test levels. Test cases in the system integration test consist of a series of ordered
test steps, which in turn consist of a stimulus (manual instructions) and an expected reaction.
Gomes de Oliveira Neto \etal~\cite{Neto2018} investigated some challenges of using diversity information for developers and
testers since results are usually many-dimensional. Also, it can be difficult to choose when and where to use
\dbt techniques due to their generality. They investigated: i) What are the trade-offs in using
different sources of diversity, such as diversity of test cases, or diversity of test requirements to optimize
large test suites? ii) How the visualization of test diversity data can help testers improve test suites? The
study made a comparison between three different techniques, which are \emph{manually selected test cases},
\emph{a diversity-based subset of test cases based on the Jaccard Index}, and a \emph{random subset of tests}.
The evaluation was performed based on the level of redundancy and the average percentage of fault detection (APFD).
They found that test similarity maps, based on pairwise diversity calculations, aided practitioners in the
industry in identifying problems in their test suites and helped them in making decisions to improve the test
suites. They concluded that the optimization and maintenance tasks can be improved using the visualization of
the diversity information.

Exploratory testing (ET) is a software testing approach that complements automated testing by leveraging business expertise,
and they are defined and executed on-the-fly by the testers~\cite{Leveau2022}, and can be considered as an evaluation technique.
The main goals of ET are to find new bugs not detected using manual or automated tests,
and improve the quality of the system under test. Leveau \etal~\cite{Leveau2020,Leveau2022} proposed an approach to help testers explore
any web application using Exploratory Testing. They used a prediction model based on $n$-gram language models, which tracks the
online interactions that the testers do and based on the prediction model gives probabilistic future interactions. Testers who
select lower probability interactions will increase the exploration diversity. Thus, achieving deeper investigation of the web
application. They made a controlled experiment and a case study to evaluate the approach, and reported that the approach helped
the tester not be stuck in repeating the same test interactions.

\subsection{Fault Localisation}
\label{subsec:fault-localization}

Fault localisation (FL) is the process of identifying the locations of faults in a program~\cite{Wong2016}. In this
section, we cover testing-based fault localisation, which utilises the testing information to localize the
faults~\cite{Hao2005}, and in particular \dbt techniques. Hao \etal~\cite{Hao2005}
proposed a technique named \emph{similarity-aware fault localisation} (SAFL) by using the theory of fuzzy
sets to evenly distribute the test cases. They evaluated their proposed approach against Dicing and TARANTULA testing-based fault
localisation approaches using two real-world programs, and found that SAFL is more effective when redundancy is present, and
competitive when it is not present. Hao \etal used SAFL within a more detailed description in~\cite{Hao2008}.
Zhao \etal~\cite{Zhao2011,Zhao2013} proposed a fault localisation framework to make use of execution similarities
of test cases to improve coverage-based fault localisation techniques. They count distinct execution paths
and store that information in a coverage vector to capture any execution similarity, and use the failing extent of
a coverage vector to estimate the execution similarity~\cite{Zhao2013}. They evaluated their framework on seven programs
from the Siemens benchmark and three UNIX utility programs. Gong \etal~\cite{Gong2012} presented a Diversity Maximisation
Speedup (DMS) for test case prioritisation to address the problem of test cases without manual labelling. DMS requires
testers to label much fewer test cases to achieve a competing fault localisation accuracy with other techniques thus
reducing the cost of the test. Xia \etal~\cite{Xia2016} used the same DMS approach to locate faults in single-fault and
multi-fault programs. They compared DMS with six test case prioritisation techniques on 12 $C$ programs and found that
DMS can reduce the cost while keeping the same accuracy of fault localisation. Later, You \etal~\cite{You2013} discussed a
number of modified similarity coefficients in fault localisation to evaluate the
significance of failing and passing test cases in similarity coefficients. The failing test cases provide more important
information for fault localisation, and therefore it is given more weight than passing test cases. They used Jaccard
similarity coefficient~\cite{Chen2002}, Tarantula similarity coefficient~\cite{Jones2005}, and Ochiai similarity coefficient
\cite{Meyer2004} in their approach modifying them with different weights. They evaluated their proposed approach on five
programs from the Siemens benchmark using 75 faulty versions and found it more effective and accurate.

Test cases have an important role in fault localisation, and adding more test cases to the test suite can help refine
the suspiciousness scores of the statements. However, increasing the number of test cases adds to the cost of execution
time and effort of the manual test oracle. Therefore, having a diverse and small number of test cases can improve fault
localisation while reducing the cost.
Liu \etal~\cite{Liu2017} proposed an approach of test case generation for fault localisation of Simulink models. They
developed a prediction model based on supervised learning that prevents adding more test cases that are unlikely
to improve fault localisation. Liu \etal~\cite{Liu2019} extended this work by adding a new test objective based on output diversity.
They evaluated the proposed approach on three industrial case studies and found that fault localisation has increased
significantly and that the prevention prediction model was able to reduce the size of the
generated test suite by more than half while maintaining the fault localisation capabilities.

\begin{tcolorbox}[title=Conclusions --- \rqproblems]
  The collected papers in this study handled a range of software testing problems including: \tdg,
  \tcp, \tsr, \tcs, \tse, and fault localisation.
  Test data generation is the most researched problem in software testing where \dbt techniques have
  been used with $37.9\%$ of the papers. \Ctcp comes next after \tdg in terms of the number of papers
  using \dbt techniques with $22.1\%$ of the papers. Followed by \tcs, \tse, \tsr, and fault localisation
  with $14.5\%$, $7.6\%$, $6.9\%$, and $6.2\%$, respectively.
  In \tdg, \dbt techniques achieved an overall higher mutation scores than random testing~\cite{Shahbazi2015B} with
  an increase of up to $30\%$~\cite{Bueno2014}, assisted in preventing the search process from stagnating in
  search-based approaches~\cite{Alshraideh2011,Albunian2020,Wang2023}, showed a significant increase
  in crash detection of between $28\%$ to $97\%$ \cite{Menendez2022}, and decreased the average generation time
  of valid tests in DL systems by $15.7\%$ to $47\%$~\cite{Jiang2023}. Furthermore, \dbt techniques
  increased the overall average percentage of fault detection compared to other \tcp
  techniqes~\cite{Miranda2018,Hemmati2015,Haghighatkhah2018A}, improved fault-detection rate up
  to $45\%$ over coverage-based selection and $110\%$ over random selection~\cite{Hemmati2013},
  and achieved $82.2\%$ \tsr maintaining both coverage and mutation score of the original test
  suites \cite{Chetouane2020}.
\end{tcolorbox}
\section{\rqsubjects}
\subsection*{\textbf{\rqsubjects}}
\label{sec:rqsubjects}

\begin{figure}[t]
    \centering
    \includegraphics[width=0.6\columnwidth]{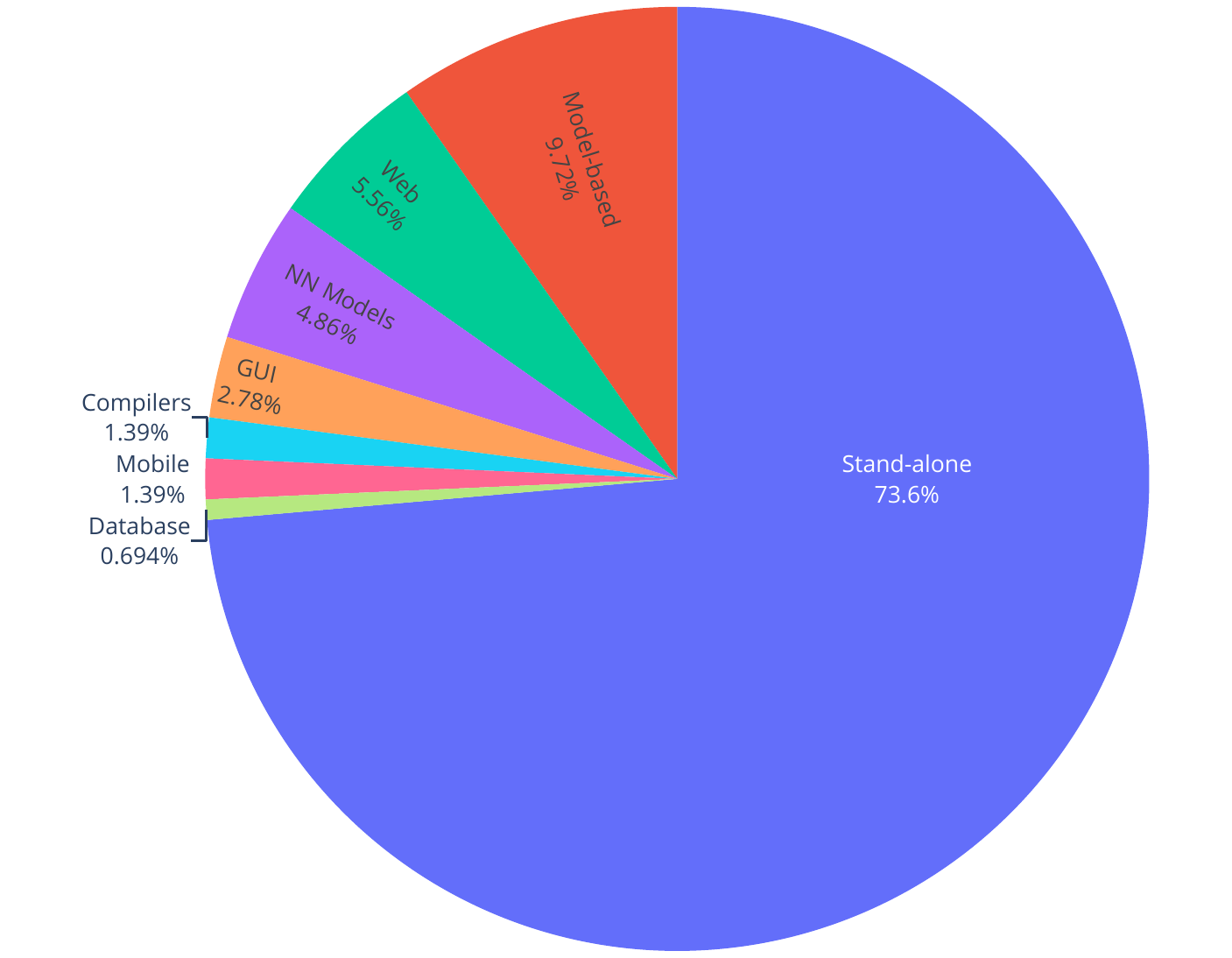}

    \vspace*{-1em}\caption{
        \label{fig:subjects-distribution}
        Distribution of the subject domains of the \dbt papers.
    }

\end{figure}

\dbt techniques have found applications in various domains, with the nature of each
domain influencing the specific approaches employed to incorporate diversity. In this section, we
highlight several key subject domains where \dbt has been explored.
Table \ref{table:applications-summary} lists the subject domains where DBT techniques
were applied, with a short description, total number of papers, and citation to these papers.
Figure~\ref{fig:subjects-distribution} shows the distribution of subject domains where \dbt
papers were applied.

\begin{table}[tbp]
  \centering
  \caption{
    \label{table:applications-summary}
    A list of the software application domains addressed using DBT techniques.
  }

  \begin{tabular}{p{3cm}@{~~~}p{6cm}@{~~~}rp{5cm}}
    \toprule
    {\bf Rank/App} & {\bf Description} & {\bf Total} & {\bf Papers}
    \\

    \midrule

    % The data in this file should be automatically output
% from code that processes raw results data.
\hspace{1mm} 1 Stand-Alone & Console programs and libraries that are written to perform any specific task on a local machine. & 106    &
\cite{Abd2019,Albunian2017,Albunian2020,Almulla2020,Alshraideh2011,Arafeen2013,Azizi2021,Beena2014,Benito2022,Bertolino2015,Boussaa2015,Brooks2009,Bueno2008,Bueno2014,Burdek2015,Cai2009,Cai2021,Cao2013,Cao2022,Carlson2011,Chen2011,Chen2021B,Chetouane2020,Coviello2018A,Coviello2018B,Cruciani2019,De2012,De2018,Neto2018,De2020,Dobslaw2020,Fang2014,Farzat2010,Feldt2008,Feldt2016,Feng2015,Fischer2016,Flemstrom2016,Flemstrom2018,Gong2012,Greca2022,Guarnieri2022,Haghighatkhah2018A,Haghighatkhah2018B,Hao2005,Hao2008,Hemmati2011,Henard2013,Henard2014,Huang2017B,Kichigin2009,Kifetew2013,Ledru2009,Ledru2012,Liu2017,Liu2019,Ma2018,Mahdieh2022,Marculescu2016,Masuda2021,Mei2013A,Mei2013B,Menendez2021,Menendez2022,Miranda2014,Miranda2018,Mondal2015,Nguyen2022,Nikolik2006,Noor2015,Nunes2015,Panchapakesan2013,Panichella2015B,Pei2014,Pizzoleto2020,Poulding2017,Reddy2020,Scalabrino2016,Semerath2018,Semerath2020,Shahbazi2014,Shahbazi2015B,Shahbazi2018,Shi2015,Shimari2022,Shin2016,Shin2018,Singh2013,Sondhi2022,Viggiato2023,Wang2009,Wang2014,Wang2015,Wu2012,Xia2016,Xiang2021,Xie2005,Xie2009,Yatoh2015,Yoo2009,You2013,Zhang2012,Zhao2011,Zhao2013,Zhao2015,Shimmi2022}
\\
\arrayrulecolor{lightgray}\hline
\hspace{1mm} 2 Model-Based & Systems represented as models like FSM or Simulink models. & 14    &
\cite{Arrieta2019,Asoudeh2013,Bueno2007,Cartaxo2007,Cartaxo2009,Coutinho2016,De2016,Hemmati2010A,Hemmati2010B,Hemmati2010C,Hemmati2013,Matinnejad2015A,Matinnejad2016,Matinnejad2019}
\\
\hline
\hspace{1mm} 3 Web & Applications that run on the web browser. & 8    &
\cite{Alshahwan2012,Alshahwan2014,Biagiola2019,Hemmati2015,Leveau2020,Lin2017,Marchetto2009,Wang2023}
\\
\hline
\hspace{1mm} 4 NN Models & The heart of deep learning algorithms which are inspired by the human brain, mimicking the way that biological neurons signal to one another. & 7    &
\cite{Aghababaeyan2023,Jiang2023,Joffe2019,Mosin2022,Xie2022,Zhang2019,Zhao2022}
\\
\hline
\hspace{1mm} 5 GUI & A program with a graphical user interface that has labels, buttons, text fields, and other widgets which a user can interact with. & 4    &
\cite{Xie2006B,Feng2012,He2015,Mariani2021}
\\
\hline
= 6 Compilers & A program that translates a programming language's source code into machine code, bytecode or another programming language. & 2    &
\cite{Chen2019C,Tang2022}
\\
\hline
=6 Mobile & Applications designed to run on a mobile device. & 2    &
\cite{Vogel2019,Vogel2021}
\\
\hline
\hspace{1mm} 8 Database  & Applications that process and manipulate data stored and work as a database management system. & 1    &
\cite{Rogstad2013}
\\
\arrayrulecolor{black}

    \bottomrule
  \end{tabular}

\end{table}

\subsection{Console Programs \& Libraries}
\label{subsec:stand-alone-diversity}

Stand-alone programs and libraries such as Defects4J and Siemens frameworks have received significant
attention in the context of \dbt
across various research areas. Notably, \dbt techniques have been applied in search-based
software testing~\cite{Bueno2007,Bueno2008,Bueno2014,Feldt2008,Xie2009,Shahbazi2015B,Albunian2020,Boussaa2015,Cai2021,Panchapakesan2013},
mutation testing~\cite{Shin2016,Shin2018,Semerath2018,Semerath2020,De2020,Pizzoleto2020},
continuous-integration environments~\cite{De2018,Haghighatkhah2018B,Azizi2021},
test case recommendation~\cite{Shimmi2022}, and machine translation systems~\cite{Cao2022}.

A number of papers have explored the incorporation of diversity into fuzzers, particularly in terms
of input generation. These include the utilisation of reinforcement learning to guide and control
the input generator~\cite{Reddy2020}, used feedback-driven fuzzing~\cite{Nguyen2022}, and exploration
of deep program semantics~\cite{Zhang2012}. Other approaches have aimed to enhance the diversity of
tests in general~\cite{Menendez2022}, or reduce the number of generated tests~\cite{Pei2014}.

Although the vast majority of the papers in the study used stand-alone programs or some libraries in
their experiments, a lot of the proposed approaches can be used in many other subject domains.

\subsection{Model-Based Systems}
\label{subsec:model-based-diversity}

\dbt techniques have been employed in the field of model-based testing. Noteworthy contributions
include the generation of test suites from finite state machines ~\cite{Asoudeh2013}, test suite reduction
for MBT~\cite{Coutinho2016}, test case selection for MBT~\cite{Cartaxo2007,Cartaxo2009,Hemmati2010A,Hemmati2010B,Hemmati2010C,Hemmati2013,De2016},
and the application of \dbt techniques to Simulink models~\cite{Matinnejad2015A,Matinnejad2016,Matinnejad2019,Arrieta2019,Liu2017}.

\subsection{Web Applications}
\label{subsec:web-diversity}

The realm of web testing has witnessed the utilisation of \dbt techniques, as discussed in
Section \ref{sec:rqproblems} and Section \ref{sec:rqartefacts}. Certain characteristics specific to web applications,
such as the document object model (DOM)~\cite{Biagiola2019}, HTML tag structure~\cite{Alshahwan2012, Alshahwan2014},
long interacting sequences~\cite{Marchetto2009}, exploration of
web applications~\cite{Leveau2020,Leveau2022}, and challenges related to string-matching-based rules in crawling-based
web application testing~\cite{Lin2017}, influence the application of \dbt techniques. Another
work used parallel evolutionary to generate test cases faster for web applications~\cite{Wang2023}.

\subsection{Database Applications}
\label{subsec:database-diversity}

\dbt techniques have been applied to database applications in one paper, which
focused on the selection of a diverse subset of test cases for large-scale databases~\cite{Rogstad2013}.

\subsection{Graphical User Interface}
\label{subsec:fuzzers-diversity}

Graphical User Interface (GUI) are applications that contain widgets (e.g. Buttons, Text fields, etc.)
where a user interacts with. GUI can be applied in desktop, web, or mobile applications. However, some
researchers focused on the characteristics of GUI components in testing and tried to investigate the
characteristics of what makes a GUI test suite effective~\cite{Xie2006B}, generate GUI test cases
\cite{Feng2012,He2015}, or investigated the reuse of GUI tests based on semantic similarities~\cite{Mariani2021}.

\subsection{Compilers}
\label{subsec:compiler-diversity}

Compilers are programs that translate source code written in some programming language into
machine code, byte-code, or the source code of some other programming language. A few studies have explored
\dbt techniques in compiler testing, including history-guided bug revealing techniques~\cite{Chen2019C},
and the detection of compiler warning defects~\cite{Tang2022}.

\subsection{Mobile Applications}
\label{subsec:mobile-diversity}

Although limited, \dbt techniques have also been applied to mobile application testing. Notably,
Vogel et al.~\cite{Vogel2019,Vogel2021} investigated diversity in the context of test data generation for
mobile applications, leveraging a modified version of the heuristic NSGA-II algorithm to introduce diversity
and achieve improved coverage, albeit at higher computational costs.

\subsection{Neural Network Models}
\label{subsec:neural-network-diversity}

Recent interest has emerged in employing \dbt techniques within the realm of Neural Networks (NNs).
For instance, NNs have been utilised to construct diverse search spaces based on the diversity of salient
features~\cite{Joffe2019}. Also, NNs were the target
of testing utilising diversity of neuron coverage for testing DL models~\cite{Zhang2019,Zhao2022}, or through
population diversity to generate diverse test cases~\cite{Braiek2019}. Furthermore, Aghababaeyan et al.~\cite{Aghababaeyan2023}
and Jiang et al.~\cite{Jiang2023} used inputs' feature diversity to test DNNs and showed significant
improvements over the traditional coverage-guided, greedy, or random approaches in terms of fault-detection
and average generation time.

In test input prioritisation for DL models, Xie et al.~\cite{Xie2022} found the diversity-based approach better
in terms of the average number of detected violations than the neuron coverage-based approach. Similarly, Mosin et al.
\cite{Mosin2022} used a similarity-based approach for input prioritisation and found that it was much more efficient
compared to white-box and data-box approaches, but with lower effectiveness (APFD) than white-box and comparable
effectiveness to data-box.

These diverse subject domains demonstrate the broad range of contexts in which \dbt
techniques have been investigated. The exploration of diversity across these domains reflects ongoing efforts
to enhance the effectiveness and efficiency of software testing practices.

\begin{tcolorbox}[title=Conclusions --- \rqsubjects]
  The subjects reported in the collected papers were stand-alone programs and libraries, web applications,
  database applications, mobile applications, model-based applications, compilers, and neural
  network models. Almost three-quarters implemented their approaches on console programs and libraries
  written in different
  languages such as Java, C, Python, etc. Model-based applications come next in the popularity of reported
  studies. Moreover, a few studies reported the use of diversity in web applications, but one might have
  expected such systems to receive more attention since they are widely used in both industry and academia.
  Also, only two
  papers applied diversity in compilers, two papers used diversity in mobile applications, and only one
  paper used diversity in databases. Finally, since 2019, there has been a significant increase of \dbt
  techniques for neural-network models.
\end{tcolorbox}
\section{\rqtools}
\label{sec:rqtools}

\subsection*{\textbf{\rqtools}}

\begin{figure}[t]
    \centering
    \includegraphics[width=1.\columnwidth]{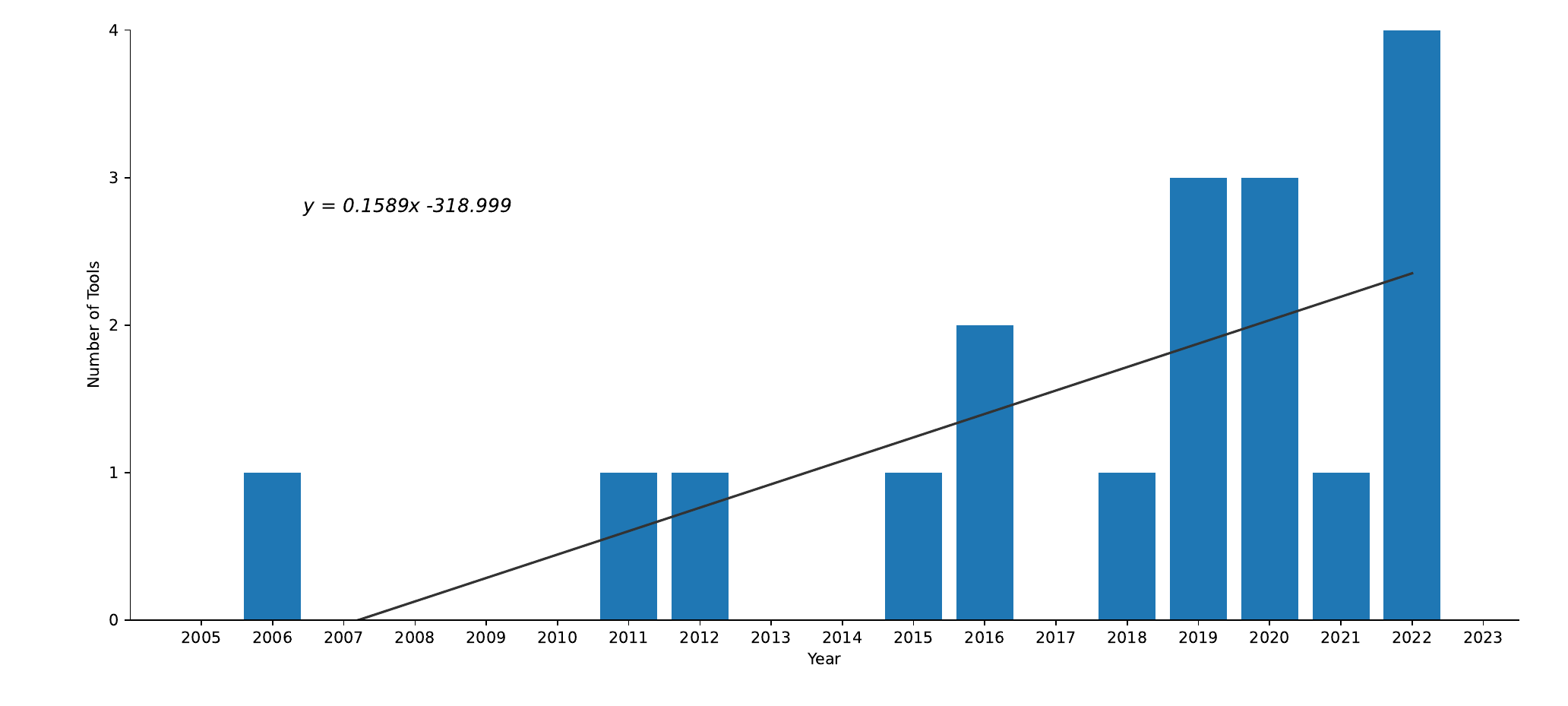}
    \vspace{-3em}
    \caption{
        \label{fig:tools-per-year}
        The number of diversity-based tools developed per year.
    }
    
\end{figure}

In this section, we present the tools that were developed based on \dbt techniques. Table \ref{table:tools-summary}
lists the tools encountered in the study. Figure \ref{fig:tools-per-year} shows the number of developed tools
since 2005. The number of tools started to increase in the last 5 years, which shows a level of maturity in
the field of applying diversity in software testing. Also, the trend line shows an increase of \dbt tools
in the field, with the highest number of \dbt tools developed in 2022. Note also that 2022 was the year
with the most publications

\begin{table}[tbp]
    \centering
    \caption{
        \label{table:tools-summary}
        A summary of the tools encountered in the study.
    }

    \begin{tabular}{p{2.5cm}p{4cm}lll}
        \toprule

            {\bf Tool} & {\bf Problem} & {\bf Application} & {\bf Proposed} & {\bf Used in} 
            \\

        \midrule

        % The data in this file should be automatically output
% from code that processes raw results data.
The Diversity Analyzer  & Test suite evaluation & General-purpose Applications & \cite{Nikolik2006} & \cite{Nikolik2006B,Nikolik2008,Nikolik2012}  \\
% used in "Test Suite Oscillations" , "Software quality assurance economics" , "The π measure". All by the same author
\arrayrulecolor{lightgray}\hline
DART  & Test case selection & Database Application & \cite{Rogstad2011} & \cite{Rogstad2016A,Rogstad2016B,Rosero2017}\\
\hline
% used in "Cost-effective strategies for the regression testing of database applications: Case study and lessons learned"
% , "Clustering Deviations for Black Box Regression Testing of Database Applications"
% , "An Approach for Regression Testing of Database Applications in Incremental Development Settings"
SimFuzz  & Test data generation & General-purpose Applications & \cite{Zhang2012} & \cite{Sadeghiyan2017,Alqahtani2022}\\
\hline
% used in "A study on the use of vulnerabilities databases in software engineering domain" 
% , "A New View on Classification of Software Vulnerability Mitigation Methods" , "Msc thesis: An Evaluation of Free Fuzzing Tools"
SIMTAC  & Test case prioritisation & C \& Access Control & \cite{Bertolino2015} & -\\
\hline
SimFL  & Fault localisation & Simulink Models & \cite{Liu2016B} & \cite{Liu2019}\\
\hline
% used in \cite{Liu2019}
SimCoTest  & Test data generation \& Test case prioritisation & Simulink Models & \cite{Matinnejad2016B} & \cite{Matinnejad2019,Yang2022}\\
\hline
% used in \cite{Matinnejad2019} , "Improve Model Testing by Integrating Bounded Model Checking and Coverage Guided Fuzzing"
% , "Msc thesis: Evaluation of Automated Test Generation for Simulink: A Case Study in the Context of Propulsion Control Software"
FAST  & Test case prioritisation & C \& Java programs & \cite{Miranda2018} & \cite{Cruciani2019,Greca2022}\\
\hline
% used in \cite{Greca2022} , \cite{Cruciani2019}
HiCOND  & Test data generation & Compilers & \cite{Chen2019C} & \cite{Rabin2021,Tang2022}\\
\hline
% used in \cite{Tang2022}, "Configuring Test Generators using Bug Reports: A Case Study of GCC Compiler and Csmith"
DIG  & Test data generation & Web Applications & \cite{Biagiola2019} & \cite{Zheng2021} \\
\hline
% used in "Automatic Web Testing Using Curiosity-Driven Reinforcement Learning" , 
\textsc{Sapinez}$^{\mathit{DIV}}$  & Test data generation & Mobile Applications & \cite{Vogel2019} & \cite{Vogel2021} \\
\hline
% used in \cite{Vogel2021}
OutGen  & Test data generation & General-purpose Applications & \cite{Benito2022} & - \\
\hline
SiMut  & Test data generation & Java programs & \cite{Pizzoleto2020} & - \\
\hline
RLCheck  & Test data generation & Java programs & \cite{Reddy2020} & \cite{Nguyen2022,Tsai2022} \\
\hline
% used in "DeepRNG: Towards Deep Reinforcement Learning-Assisted Generative Testing of Software" ,  \cite{Nguyen2022}
DIPROM  & Test data generation & Compilers & \cite{Tang2022} & - \\
\hline
\textsc{BeDivFuzz}  & Test data generation & Java programs & \cite{Nguyen2022} & - \\
\hline
TSE  & Test data generation & Java programs & \cite{Shimmi2022} & - \\
\hline
\textsc{Fastazi}  & Test case prioritisation & Java programs & \cite{Greca2022} & - \\
\hline
\textsc{Metallicus}  & Test adaptation & Python programs & \cite{Sondhi2022} & - \\
\arrayrulecolor{black}

        \bottomrule
    \end{tabular}
\end{table}

Numerous research papers have contributed to the development of \dbt tools aimed at generating test
data for various applications. Many of these tools are categorized as fuzzers, such as ``SimFuzz''~\cite{Zhang2012},
``RLCheck''~\cite{Reddy2020}, and ``BeDivFuzz''~\cite{Nguyen2022}. Furthermore, ``OutGen''~\cite{Benito2022} caters
for a broad range of programs and software systems, focusing on the generation of diverse output data.
In compiler testing, there is ``HiCOND''~\cite{Chen2019C}
a test-program generation tool that leverages historical data for compiler testing, and ``DIPROM''~\cite{Tang2022},
which constructs warning-sensitive programs to detect compiler warning defects.
Moreover, in test case recommendation,
Test Suite Evolver (TSE)~\cite{Shimmi2022} augments an existing test suite to cope with changes made in the
source code, or add more test cases to an incomplete test suite by exploiting structural similarity between methods.

Specifically targeting web applications, ``DIG''~\cite{Biagiola2019} (\textbf{DI}versity-based \textbf{G}enerator)
has been introduced for generating test cases. For mobile applications, \textsc{Sapienz}$^{\mathit{DIV}}$~\cite{Vogel2019}
used diversity to enhance the mobile application testing tool \textsc{Sapienz} resulting in improved coverage albeit
at the cost of increased processing time.
In another type of application, ``SimCoTest''~\cite{Matinnejad2016B}, is a tool for generating tests for Simulink
models. Additionally, ``SiMut''~\cite{Pizzoleto2020} is notable for its ability to generate a reduced set of mutants,
offering efficiency gains in mutation testing.

Several additional papers have contributed to the realm of \dbt tools in the context of test case prioritisation.
One such tool, ``SIMTAC'' (Similarity Testing for Access-Control)~\cite{Bertolino2015}, focuses on reordering test cases for
Access Control systems based on their similarity. Similarly, ``SimCoTest''~\cite{Matinnejad2016B} also addresses test
case prioritisation, particularly within Simulink models, while also encompassing test data generation capabilities, as
previously mentioned. Furthermore, two tools, namely ``FAST'' (Fast and Scalable Test Case prioritisation)~\cite{Miranda2018}
and ``Fastazi''~\cite{Greca2022}, utilise diversity to prioritise test cases within test suites for Java applications.
FAST demonstrates improved efficiency without compromising effectiveness, while Fastazi combines file-based test case
selection with similarity-based test case prioritisation.

Other areas in software testing had very few \dbt tools. In 2006, Nikolik~\cite{Nikolik2006} introduced the
first \dbt tool named ``The Diversity Analyzer'' for quantifying the extent to which a test suite exercises
a program in diverse manners with regard to control and data. In the domain of test suite reduction,
``DART'' (Database Regression Testing)~\cite{Rogstad2011} employs a similarity-based regression test reduction approach
for large-scale database applications. Another tool, ``SimFL'' (Simulink Fault Localisation)~\cite{Liu2016B}, focuses on fault
localisation within Simulink models. Lastly, \textsc{Metallicus}~\cite{Sondhi2022} is a \dbt test recommendation
tool that generates a corresponding test suite given a query function.

These \dbt tools showcased in the literature contribute to various aspects of software testing, ranging from
test case prioritisation and reduction to fault localisation and test suite generation. Their adoption demonstrates the
ongoing efforts to leverage diversity as a means to enhance the effectiveness, efficiency, and quality of software
testing practices.

\begin{tcolorbox}[title=Conclusions --- \rqtools]
  The trend line shows an increase of \dbt tools in software testing with a notable increase in the past 5 years.
  The highest number of developed \dbt tools was in 2022, which also has the highest number of \dbt
  papers published from our collection of papers. Most of the reported \dbt tools ($77.78\%$) were developed to
  deal with the problems of test data generation and test case
  prioritisation. One tool supports both test data generation and test case prioritisation and 10 more tools deal with
  the problem of test data generation, while 3 more tools were developed for test case prioritisation. Only one \dbt tool was
  developed for test suite quality evaluation, one in test case selection, and one for test adaptation.
  The tool developed for test case selection is only for database applications.
\end{tcolorbox}
\section{Future Research Directions}
\label{sec:futre-research-directions}

We discuss a number of the challenges in \dbt and future research that can be
done to address these challenges.

\subsection{Areas of Applying \dbt Techniques}

Test data generation is the most researched area where \dbt techniques were
implemented as we showed in Section~\ref{sec:rqproblems}. Although \tsr, \tcp, and \tcs are closely
related problems that deal with the cost of large regression test suites, many authors have
applied \dbt techniques in \tcp and \tcs, but there are very few papers on \tsr. Since \dbt techniques
showed great promise in \tcp~\cite{Arafeen2013,Miranda2018,Ledru2012,Huang2017B}, it would make sense to use them to solve \tsr as well. More
work needs to be performed in \tsr, and more extensive investigations need to be done to
compare \dbt techniques with other reduction techniques reported in the literature.

\subsection{Diversity-Based Subjects}

Most reported papers in this study used \dbt techniques on stand-alone
programs and in Model-Based testing. This was discussed in Section \ref{sec:rqsubjects},
and we showed that there was little work on the use of \dbt techniques in web, database and mobile
testing.

Although web applications are widely used, there are fewer paper
applying \dbt techniques on them. As discussed in Section \ref{subsec:web-diversity},
all of these address \tdg
\cite{Marchetto2009, Alshahwan2012, Alshahwan2014, Lin2017, Selay2018, Biagiola2019, Leveau2020,Leveau2022,Wang2023},
and none of the reported papers dealt with evaluating the quality of the current test suite or
reducing the cost of test suites for web applications.

The limited memory and processing power of mobile devices make it difficult to use
an extensive and more demanding techniques. This can be an
obstacle for using \dbt in test data generation for Mobile applications as was
reported by Vogel \etal~\cite{Vogel2021}. The modified tool \textsc{Sapinez}$^{\mathit{DIV}}$
has more computational cost than \textsc{Sapinez}. However, applying \tcp or \tsr for
mobile test suites could be very appealing. A scalable prioritization approach such as the
one proposed by Miranda \etal~\cite{Miranda2018}, may be useful.
Further investigations are needed to apply \dbt techniques in mobile applications.

\subsection{Tool Support}

As discussed in Section~\ref{sec:rqtools}, there has been a spike on the number of \dbt
tools developed since 2019, which indicates a possible maturity in the field.
The trend line of Figure~\ref{fig:tools-per-year} also shows an increase of \dbt tools developed.
However, most of these tools were developed for test data generation. There is a need for more
tools in other areas, especially for test case selection, test suite reduction, and
\tse.

Furthermore, there are many Simulink \dbt tools developed for a variety of areas, such as \tdg~\cite{Matinnejad2016B},
\tcp~\cite{Matinnejad2016B}, and FL~\cite{Liu2016B}. However, other applications, such as web and mobile,
have very little tool support for \dbt techniques. Only a single tool for Web applications was
developed~\cite{Biagiola2019}, and only a single tool for Mobile applications was developed~\cite{Vogel2019}.
These tools are only for test data generation, which leaves other areas such as \tcp, \tsr, and \tcs open
for \dbt tools to be developed for both Web applications and Mobile applications. For Database
applications, DART~\cite{Rogstad2011} is for \tcs. There is no available \dbt techniques
or tools to generate test cases for such applications.

\subsection{Guidelines for Selecting The Diversity Parameters}

We refer to the similarity metric and the testing artefact 
as the \emph{diversity parameters}.
Many similarity metrics were discussed in Section \ref{sec:rqmetrics}, and there was no clear indication
of which similarity metric would perform best. In Section \ref{sec:rqartefacts}, many \artefact{s} are described
that can be used as a basis for applying \dbt techniques, and most of the papers used
test script or input diversity. However, there is a lack of guidelines for selecting these
parameters. Currently, it is primarily the role of the researcher to select
based on intuition or experience. The nature of the application or the type of the testing \artefact{}
can guide the selection of the similarity metric that can be most appropriate. Also, is there a
correlation between the software testing problem to be solved and these diversity parameters?
The existence of such guidelines could help researchers and practitioners identify the most suitable
the similarity metric and \artefact{} to use for a specific problem or application.

\subsection{Exploring Other Diversity Artefacts}

Most current works focused on input and test script diversity as shown in Section \ref{sec:rqartefacts}.
Other \artefact{s} did not receive much attention, and some \artefact{s} are used with
other diversity \artefacts, which makes it difficult to identify their importance and contribution to
the effectiveness of the \dbt technique. Non-functional properties are prime
examples of such \artefact{s}. Feldt \etal~\cite{Feldt2008}, used non-functional
properties such as memory and processing time as one of 11 variation points in their model.
However, since there were several other factors, it is difficult to determine how useful
these properties can be. Shimari \etal~\cite{Shimari2022} used the run time information
obtained when executing the SUT with the test cases to develop a test case
selection technique for an Industrial Simulator.
Apart from their work, no other works investigated how non-functional properties would be beneficial.
The same can be said about the program's state diversity, which is not investigated at all.
\section{Conclusions}
\label{sec:conclusions}

In this study we covered \numpapers{} papers that described the use of \dbt techniques
to solve a software testing problem. Our survey discussed \dbt publication trends
showing the number of DBT publications per year, and we presented the prominent venues,
papers and authors in the field. We presented \nummetrics similarity metrics there
are used in the literature to calculate the level of similarity between different testing
\artefact{s}, and found that generic similarity metrics are the most commonly used
metrics. The most popular similarity metrics were Euclidean distance, Jaccard distance,
and Edit distance used in 29, 28, and 27 papers, respectively.
the Euclidean distance is the most popular one used in 29 of the papers in
our study. However, there is no clear evidence that a particular metric performs best,
and it is clear that different factors could affect the results.

The \dbt papers attempted to address the problems of \tdg,
\tcp, \tcs, \tsr, \tse,
and fault localization. Test data generation and test case prioritization were the most
researched areas with $37.9\%$ of the collected papers dealing with \tdg, then $22.1\%$ dealing
with \tcp. Also, we presented the various testing \artefact{s} used as a basis
for \dbt techniques, such as inputs, output, test scripts, test executions, features,
etc. Of these, $18.2\%$ used test scripts, then $14.9\%$ used input diversity as \artefact{s}
for applying \dbt techniques.
Furthermore, we explored the various subject domains where \dbt techniques were utilised.
Stand-alone programs and libraries written in languages such as Java, C++, Python, and so on were the
most used subject domain for the collected papers followed by model-based applications. Relatively few papers
dealt with Web applications, Database applications, Mobile applications, Neural-Networks, and
Compilers. There has been a clear increase in the use of \dbt techniques for deep neural networks
since 2019. Moreover, we presented the \dbt tools reported in the collected papers and
found that $77.78\%$ of the tools presented in the study support test data generation and test case
prioritization. Finally, we discussed some challenges and future work in \dbt that can be
addressed by researchers in the future.

%Bibliography
\bibliographystyle{unsrt}  
\bibliography{references}

\end{document}